\documentclass[aps,pre,onecolumn,showpacs,preprintnumbers,amsmath,amssymb,footinbib,superscriptaddress]{revtex4-2}

\usepackage{graphicx}
\usepackage{dcolumn}
\usepackage{bm}
\usepackage{amsmath}
\usepackage{amssymb}
\usepackage{natbib}
\setcitestyle{numbers,square}
\usepackage{textgreek}
\usepackage{subcaption}

\DeclareMathOperator{\erfc}{erfc}
\DeclareMathOperator{\sech}{sech}

\newcommand\ddfrac[2]{\frac{\displaystyle #1}{\displaystyle #2}}
\begin{document}
	
\title{Replica analysis of the lattice-gas Restricted Boltzmann Machine partition function}

\author{David C. Hoyle}

\affiliation{Oakthorpe Consulting, Manchester, UK.}
\email{david.hoyle@oakthorpe-consulting.com}

\affiliation{Division of Pharmacy \& Optometry, School of Health Sciences, University of Manchester, UK.}
\email{david.hoyle@manchester.ac.uk}

\date{\today}

\begin{abstract}		
We study the expectation value of the logarithm of the partition function of large binary-to-binary lattice-gas Restricted Boltzmann Machines (RBMs) within a replica-symmetric ansatz, averaging over the disorder represented by the parameters of the RBM Hamiltonian. Averaging over the Hamiltonian parameters is done with a diagonal covariance matrix. Due to the diagonal form of the parameter covariance matrix not being preserved under the isomorphism between the Ising and lattice-gas forms of the RBM, we find differences in the behaviour of the quenched log partition function of the lattice-gas RBM compared to that of the Ising RBM form usually studied. We obtain explicit expressions for the expectation and variance of the lattice-gas RBM log partition function per node in the thermodynamic limit. We also obtain explicit expressions for the leading order finite size correction to the expected log partition function per node, and the threshold for the stability of the replica-symmetric approximation. We show that the stability threshold of the replica-symmetric approximation is equivalent, in the thermodynamic limit, to the stability threshold of a recent  message-passing algorithm used to construct a mean-field Bethe approximation to the RBM free energy. Given the replica-symmetry assumption breaks down as the level of disorder in the spin-spin couplings increases, we obtain asymptotic expansions, in terms of the variance controlling this disorder, for the replica-symmetric log partition function and the replica-symmetric stability threshold. We confirm the various results derived using simulation.
\end{abstract}
	
	
\maketitle

\section{\label{sec:Introduction} Introduction}
Restricted Boltzmann Machines (RBMs) provide an archetypal machine learning model that have gained particular prominence in early studies of deep learning neural networks \cite{Hinton2002, Fischer2014}. An RBM provides a model for the joint distribution of observed features, represented by nodes in a visible layer, and latent features, represented by nodes in a hidden layer. Variants of RBMs have visible and hidden features as continuous or binary variables. Estimating the partition function, $Z$, of the RBM Hamiltonian then becomes an important task, for example in training and assessing RBM models. It is common to need to evaluate $\log Z$ - equivalently the negative of the Helmholtz free energy. Computational approaches to estimating $\log Z$ for RBMs are numerous and have recently been extensively reviewed by Krause et al.\cite{Krause2020}. 

When visible and hidden layer features are binary, the logarithm of the joint probability is equivalent to the Hamiltonian of a bi-partite spin-glass model, and so unsurprisingly, analytical studies of the log of the RBM partition function have made use of statistical physics techniques that have been previously applied to the study of spin-glasses \cite{Nishimori2001, MontanariSen2022}, including the use of the replica trick to study the paramagnetic, spin-glass and ferromagnetic phases of different variants of RBMs \cite{Barra2011, Barra2017, Barra2017b, Agliari2019}. The work of Barra and collaborators \cite{Barra2017, Barra2017b, Agliari2019} in particular has focused on understanding this phase behaviour. Spherical versions of the bi-partite Sherrington-Kirkpatrick (SK) model have also been studied recently by Baik and Lee \cite{Baik2020}. The extensive literature on the application of statistical physics techniques to the analysis of RBMs has been recently and succinctly reviewed by Decelle and Furtlehner \cite{DecelleFurtlehner2021}. Less study has been made of the free energy for binary-to-binary RBMs with inhomogeneous external fields. Barra et al. \cite{Barra2017b} consider generic forms of the prior distributions acting on the variables at the visible and hidden nodes, which in one particular limit, become a distribution of a binary variable, but those generic priors are homogeneous across nodes. Similarly, less work within the statistical physics literature has been done on RBMs using the lattice-gas form of the RBM Hamiltonian, a notable exception to this being the work of Tubiana and Monasson \cite{TubianaMonasson2017} on emergent compositional representations. 

\section{Lattice-gas RBMs}
In this paper we study binary-to-binary RBMs, but in keeping with a lot of the machine learning literature where the variables at each node are used to represent presence or absence of a state or feature, we will use a lattice-gas form of the RBM Hamiltonian, where the visible and hidden node variables take values in $\{0,1\}$, rather than the Ising model form where variables take values in $\{-1,1\}$. We use $n_{i},\;i=1,\dots,N$, to denote the $N$ units in the visible layer, and $m_{a},\;a=1,\ldots,M$, to denote the $M$ units in the hidden layer. We set $M=\alpha N$ and will keep $\alpha$ fixed when we go to the thermodynamic limit $N\rightarrow\infty$. The variables, $n_{i}$ and $m_{a}$ are binary, that is, $n_{i},m_{a}\in \{0,1\}\,\;\forall i,a$. The Hamiltonian for the RBM can be written in terms of these binary variables as,
\begin{equation}
H({\bm n}, {\bm m})  \;=\; -\sum_{i,a}n_{i}J_{ai}m_{a}\;-\;\sum_{i=1}^{N}n_{i}v_{i}\;-\;\sum_{a=1}^{M}m_{a}h_{a}\;\;.
\label{eq:B.1}
\end{equation}

\noindent Here we have used the vector ${\bm n}$ of binary variables to denote the state of the whole visible layer, i.e., ${\bm n}=(n_{1},\ldots,n_{N})$. Likewise, we have used ${\bm m}=(m_{1},\ldots,m_{M})$ to denote the state of the hidden layer. The Hamiltonian in Eq.(\ref{eq:B.1}) is isomorphic to an Ising model formulation of the RBM, with the mapping between the Ising model formulation and this lattice-gas formulation corresponding to the trivial transformations $\sigma_{i}=2n_{i}-1$, $s_{a}=2m_{a}-1$. For example, Huang and Toyoizumi study an Ising RBM Hamiltonian of the form \cite{HuangToyoizumi2015},
\begin{equation}
H\;=\; -\sum_{i,a}\sigma_{i}w_{ai}s_{a}\;-\;\sum_{i=1}^{N}\sigma_{i}\phi_{i}\;-\;\sum_{a=1}^{M}s_{a}\psi_{a}\;\;.
\label{eq:B.1b}
\end{equation}

\noindent The Hamiltonian in Eq.(\ref{eq:B.1b}) is identical (up to a global constant) to the Hamiltonian in Eq.(\ref{eq:B.1}). The mapping between the external fields and interaction couplings in the lattice-gas and Ising formulations is trivial, given the mappings between $n_{i}$ and $\sigma_{i}$, and between $m_{a}$ and $s_{a}$. For example, 
\begin{equation}
w_{ai} \; = \; \frac{1}{4}J_{ai}\;\;\;,\;\;\; 
\phi_{i} \; = \; \frac{1}{2}v_{i}\;+\;\frac{1}{4}\sum_{a=1}^{M}J_{ai}\;\;.
\label{eq:B.1c}
\end{equation}

A specific instance of the interaction weights $J_{ai}$ and the external fields $v_{i}, h_{a}$ represents a specific joint distribution of visible and hidden features, and consequently we are interested in calculating the partition function $Z=\sum_{{\bm n},{\bm m}}e^{-H({\bm n},{\bm m})}$, and then also in calculating $\log Z$. For a specific instance of the parameters, algorithms exist to estimate $\log Z$, such as that developed by Huang and Toyoizumi to study the Hamiltonian in Eq.(\ref{eq:B.1b}) \cite{HuangToyoizumi2015}, or the algorithms reviewed by Krause et al. \cite{Krause2020}.

To study $\log Z$ analytically it is typical to consider the behaviour of the quenched average of $\log Z$. That is, we are interested in the behaviour of $\log Z$, averaged over the disorder represented by the random variables $J_{ai}, v_{i}, h_{a}$. The quenched average is often evaluated through the use of replicas and a replica-symmetric ansatz \cite{MezardEtAl1987, Nishimori2001, MontanariSen2022}. Typically, we study the case where the interaction weights $J_{ai}$ and the external fields $v_{i}, h_{a}$ are independently Gaussian distributed. That is,
\begin{eqnarray}
J_{ai} &\sim& {\mathcal N}\left (\mu_{J}, \sigma^{2}_{J} \right)\;\;\forall i,a\;\;, \label{eq:B.2a} \\
v_{i} &\sim& {\mathcal N}\left (\mu_{v}, \sigma^{2}_{v} \right)\;\;\forall i\;\;, \label{eq:B.2b} \\
h_{a} &\sim& {\mathcal N}\left (\mu_{h}, \sigma^{2}_{h} \right)\;\;\forall a\;\;. \label{eq:B.2c}
\end{eqnarray} 

\noindent To ensure comparability of systems of different sizes but similar total disorder, we introduce the scaled variance, $\sigma^{2}_{J}=\tilde{\sigma}^{2}_{J}/N$.

When studying RBMs in the Ising formulation it is also common, just as in Eq.(\ref{eq:B.2a}) - Eq.(\ref{eq:B.2c}), to consider the case where we have a diagonal covariance matrix controlling the joint distribution of the parameters $w_{ai}, \phi_{i}, \psi_{a}$ in Eq.(\ref{eq:B.1b}), i.e., the distribution of $w_{ai}$ is independent from the distribution of $\phi_{i}$, and so on. However, it is clear from the mapping in Eq.(\ref{eq:B.1c}) that despite the Ising and lattice-gas RBM Hamiltonians being isomorphic to each other, if the Hamiltonian parameters in the lattice-gas formulation of the RBM have a diagonal covariance then the parameters in the Ising formulation will not, and vice-versa. For example, with the distributions in Eq.(\ref{eq:B.2a}) - Eq.(\ref{eq:B.2c}) and the relations in Eq.(\ref{eq:B.1c}), the covariance between $w_{ai}$ and $\phi_{i}$ is $\tilde{\sigma}^{2}_{J}/16N$. Whilst this covariance is weak, being ${\mathcal O}\left ( N^{-1}\right )$, there are $M\times N$ such covariance matrix elements overall. Consequently, the behaviour of the expectation $\mathbb{E}\left (\log Z\right )$ evaluated with a diagonal parameter covariance matrix can be qualitatively different for the Hamiltonian in Eq.(\ref{eq:B.1}) compared to that of an Ising RBM Hamiltonian of the form in Eq.(\ref{eq:B.1b}). Similar observations about the non-equivalence of the Ising and lattice-gas disorder averaged free energies have been made by Russo \cite{Russo1998} in the context of the ordinary SK model, and also by Tsodyks and Feigel'man \cite{TsodyksFeigelman1988} in the context of the Hopfield model. This raises several questions; firstly, whether a replica-symmetric approximation to the quenched RBM log partition function is still accurate when a diagonal covariance matrix for the Hamiltonian parameters is used with a lattice-gas RBM Hamiltonian; secondly, what is the replica-symmetric quenched RBM log partition function
for the lattice-gas RBM with inhomogeneity in all the Hamiltonian parameters and a diagonal parameter covariance, and can we derive simple (quick to calculate) but accurate approximate expressions for that quenched log partition function; thirdly, whether the same qualitative behaviour of the Ising RBM quenched log partition function is observed in the lattice-gas RBM quenched log partition function; and fourthly what is the behaviour of the stability threshold of the replica-symmetric ansatz in the lattice-gas RBM formulation. These are the questions we address in this paper.

\section{The replica partition function}
To derive $\mathbb{E}(\log Z)$ for lattice-gas RBMs our analysis, to start with, largely follows that of Nishimori \cite{Nishimori2001}. The expectation of the log partition function, averaged over the disorder of the interactions and external fields is given by,
\begin{equation}
\mathbb{E}\left ( \log Z\right) \;= \; \int_{-\infty}^{\infty}\prod_{a,i}dJ_{ai} P(J_{ai}) \int_{-\infty}^{\infty} \prod_{i}dv_{i}P(v_{i}) \int_{-\infty}^{\infty} \prod_{a}dh_{a}P(h_{a}) \log \sum_{{\bm n}}\sum_{{\bm m}}e^{-H({\bm n},{\bm m})}\;\;.
\label{eq:B.3}
\end{equation} 

\noindent The expectation $\mathbb{E}\left ( \log Z\right)$ is then consequently a function of $N,M,\tilde{\sigma}^{2}_{J}, \mu_{v}, \mu_{h}, \sigma^{2}_{v}, \sigma^{2}_{h}$. Obviously, the designation of the labels,`visible layer' and 'hidden layer' is just one of convenience  and relates to how the RBM might be used in processing an input pattern. However, from the perspective of calculating $\mathbb{E}\left ( \log Z\right)$ which layer is considered the visible layer and which is considered the hidden layer cannot affect its value. Consequently, we expect the value of  $\mathbb{E}\left ( \log Z\right)/N$ to be invariant under the transformation,
\begin{equation}
N\rightarrow M\;,\;
\alpha\rightarrow\alpha^{-1}\;,\;
\tilde{\sigma}^{2}_{J}\rightarrow\alpha\tilde{\sigma}^{2}_{J}\;,\;
\mu_{v}\rightarrow\mu_{h}\;,\;
\mu_{h}\rightarrow\mu_{v}\;,\;
\sigma^{2}_{v}\rightarrow\sigma^{2}_{h}\;,\;
\sigma^{2}_{h}\rightarrow\sigma^{2}_{v}\;\;.
\label{eq:B.4}
\end{equation}

\noindent To evaluate the expectation in Eq.(\ref{eq:B.3}) above we use the familiar replica trick,
\begin{equation}
\log Z \;=\; \lim_{n\rightarrow 0} \frac{Z^{n}-1}{n}\;\;,
\nonumber
\end{equation}

\noindent and so introduce $n$ copies, or replicas of the original partition function, indexed by $\nu=1,\ldots,n$. So we have,

\begin{equation}
\mathbb{E}\left ( Z^{n}\right) \;= \; \int_{-\infty}^{\infty}\prod_{a,i}dJ_{ai} P(J_{ai}) \int_{-\infty}^{\infty} \prod_{i}dv_{i}P(v_{i}) \int_{-\infty}^{\infty} \prod_{a}dh_{a}P(h_{a})\sum_{{\bm n}^{(1)}, {\bm m}^{(1)}}\ldots\sum_{{\bm n}^{(n)}, {\bm m}^{(n)}}e^{-\sum_{\nu}H({\bm n}^{(\nu)}, {\bm m}^{(\nu)})}\;\;.
\label{eq:B.6}
\end{equation}

\noindent The integrals over the interaction weights, $J_{ai}$, and external fields $v_{i}, h_{a}$ are Gaussian and easily evaluated. As it is common to only consider the zero-mean case, $\mathbb{E}\left (w_{ai}\right ) = 0$, in the Ising RBM formulation and this leads to zero-mean $\mathbb{E}\left (J_{ai}\right ) = 0$ in the lattice-gas formulation, we will predominantly only consider the $\mu_{J}=0$ case. With this simplification of $\mu_{J}=0$ we obtain,
\begin{multline}
\mathbb{E}\left ( Z^{n}\right) \; = \;\sum_{{\bm n}^{(1)}, {\bm m}^{(1)}}\ldots\sum_{{\bm n}^{(n)}, {\bm m}^{(n)}} \left [
\prod_{i}\exp \left ( \mu_{v}\sum_{\nu}n^{(\nu)}_{i}\;+\;\frac{\sigma_{v}^{2}}{2}\left ( \sum_{\nu}n^{(\nu)}_{i}\right )^{2}\right ) \right . \\
\times\; \left . \prod_{a}\exp \left ( \mu_{h}\sum_{\nu}m^{(\nu)}_{a}\;+\;\frac{\sigma_{h}^{2}}{2}\left ( \sum_{\nu}m^{(\nu)}_{a}\right )^{2}\right ) \times \prod_{i,a}\exp \left ( \frac{\sigma_{J}^{2}}{2}\left ( \sum_{\nu}n^{(\nu)}_{i}m^{(\nu)}_{a}\right )^{2}\right ) \right ]\;\;.
\label{eq:B.7}
\end{multline}

\noindent We can write,
\begin{equation}
\prod_{i}\exp \left ( \frac{\sigma_{v}^{2}}{2}\left ( \sum_{\nu}n^{(\nu)}_{i}\right )^{2}\right )  \; = \; 
\exp\left ( \frac{1}{2}\sigma^{2}_{v}\sum_{\nu,\nu'}\sum_{i}n^{(\nu)}_{i}n^{(\nu')}_{i}\right )
\; = \; \exp\left (\frac{1}{2}\sigma^{2}_{v} \sum_{\nu,\nu'}x_{\nu\nu'}\right )\;\;,
\nonumber 
\end{equation}

\noindent where we have defined $x_{\nu\nu'}=\sum_{i}n^{(\nu)}_{i}n^{(\nu')}_{i}$. Similarly, if we define, $y_{\nu\nu'}=\sum_{a}m^{(\nu)}_{a}m^{(\nu')}_{a}$, we can write the other factors in Eq.(\ref{eq:B.7}) as,

\begin{eqnarray}
\prod_{a} \exp \left ( \frac{\sigma_{h}^{2}}{2}\left ( \sum_{\nu}m^{(\nu)}_{a}\right )^{2}\right ) & = & \exp\left (\frac{1}{2}\sigma^{2}_{h} \sum_{\nu,\nu'}y_{\nu\nu'}\right )\;\;, \nonumber \\
\prod_{i,a}\exp \left ( \frac{\sigma_{J}^{2}}{2}\left ( \sum_{\nu}n^{(\nu)}_{i}m^{(\nu)}_{a}\right )^{2}\right ) & = &
\exp \left ( \frac{1}{2}\sigma^{2}_{J} \sum_{\nu,\nu'}x_{\nu\nu'}y_{\nu\nu'}\right )\;\;.
\label{eq:B.10b}
\end{eqnarray}

\noindent We further linearize the exponent in Eq.(\ref{eq:B.10b}) by writing,
\begin{multline}
\exp \left ( \frac{1}{2}\sigma^{2}_{J} x_{\nu\nu'}y_{\nu\nu'}\right ) \;=\; 
\frac{2}{\pi}\sigma^{-2}_{J}\int_{-\infty}^{\infty}dz_{\nu\nu'}\int_{-\infty}^{\infty}d\tilde{z}_{\nu\nu'}\left [ \exp \left (-2\sigma^{-2}_{J}\left ( z^{2}_{\nu\nu'}\;+\;\tilde{z}^{2}_{\nu\nu'}\right)\right ) \right . \\
\times \exp \left (z_{\nu\nu'}\left( x_{\nu\nu'} \;+\; y_{\nu\nu'}\right)\right ) \exp \left (\mathrm{i}\tilde{z}_{\nu\nu'}\left( x_{\nu\nu'} \;-\; y_{\nu\nu'}\right)\right ) \big ]\;\;.
\label{eq:B.11}
\end{multline}

\noindent Using the representation in Eq.(\ref{eq:B.11}) above we arrive at,
\begin{equation}
\mathbb{E}\left ( Z^{n}\right) \; = \; \left ( \frac{2}{\pi\sigma^{2}_{J}} \right )^{n^{2}}
\int_{-\infty}^{\infty}\prod _{\nu,\nu'} dz_{\nu\nu'}
\int_{-\infty}^{\infty}\prod _{\nu,\nu'} d\tilde{z}_{\nu\nu'}\,\exp \left (-2\sigma^{-2}_{J}\sum_{\nu,\nu'}\left ( z^{2}_{\nu\nu'}\;+\;\tilde{z}^{2}_{\nu\nu'}\right)\right ) Z_{1}^{N}Z_{2}^{M}\;\;.
\label{eq:B.12}
\end{equation}

\noindent Here, $Z_{1}$ and $Z_{2}$ are defined as,
\begin{equation}
Z_{1} \; = \; \sum_{n_{1},\ldots,n_{n}\in \{0,1\}} \exp\left ( \mu_{v}\sum_{\nu}n_{\nu}\;+\;\sum_{\nu,\nu'}\left( \frac{1}{2}\sigma^{2}_{v}+z_{\nu\nu'}+\mathrm{i}\tilde{z}_{\nu\nu'}\right)n_{\nu}n_{\nu'}\right)\;\;,
\label{eq:B.13a}
\end{equation}
\begin{equation}
Z_{2} \; = \; \sum_{m_{1},\ldots,m_{n}\in \{0,1\}} \exp\left ( \mu_{h}\sum_{\nu}m_{\nu}\;+\;\sum_{\nu,\nu'}\left( \frac{1}{2}\sigma^{2}_{h}+z_{\nu\nu'}-\mathrm{i}\tilde{z}_{\nu\nu'}\right)m_{\nu}m_{\nu'}\right)\;\;.
\label{eq:B.13b}
\end{equation}

\noindent On writing $M\;=\;\alpha N$ the integrand in Eq.(\ref{eq:B.12}) becomes,
\begin{equation}
\exp\left ( N\left [ \log Z_{1}\;+\;\alpha\log Z_{2}\;-\;2\tilde{\sigma}^{-2}_{J}\sum_{\nu,\nu'}\left ( z^{2}_{\nu\nu'}\;+\;\tilde{z}^{2}_{\nu\nu'}\right)\right ]\right )\;\equiv\; \exp\left ( NS\right )\;\;,
\label{eq:B.14}
\end{equation}

\noindent where the right-hand side of Eq.(\ref{eq:B.14}) simply defines the action $S$ which is a function of the various replica variables $z_{\nu\nu'},\tilde{z}_{\nu\nu'}$. The extension of the derivation above to the $\mu_{J} \neq 0$ case is straightforward and for completeness we give it in \ref{sec:AppendixA1}. 

The derivation of Eq.(\ref{eq:B.13a}), Eq.(\ref{eq:B.13b}) and Eq.(\ref{eq:B.14}) has not yet made explicit use of the fact that the variables, $n_{i}, m_{a}$, at the visible and hidden nodes, take values in the set $\{0,1\}$, and so the derivation of Eq.(\ref{eq:B.13a}), Eq.(\ref{eq:B.13b}) and Eq.(\ref{eq:B.14}) and subsequent analyses and results can be trivially modified to the Ising RBM case, where the visible and hidden node variables take values in $\{-1,1\}$. Again for completeness we given the derivation and subsequent analyses in \ref{sec:AppendixIsing}.

As we will study the behaviour of $\mathbb{E}\left ( \log Z\right)$ for large values of $N$, and ultimately go to the thermodynamic limit $N\rightarrow\infty$, this suggests evaluating the multi-dimensional integral in Eq.(\ref{eq:B.12}) via saddle-point expansion. As is common in replica calculations, we will initially assume that the dominant saddle-points are replica-symmetric. Location of the replica-symmetric saddle-points is performed in the next section.

\section{Replica symmetric saddle-points}
To evaluate the integration in Eq.(\ref{eq:B.12}) via saddle-point expansion, we make the assumption that the dominant saddle-points display replica symmetry, that is, at the replica-symmetric saddle-points we have,
\begin{eqnarray}
z_{\nu\nu} & = &z_{0}\;\;,\;\tilde{z}_{\nu\nu}\;=\;\tilde{z}_{0}\;\;,\forall \nu\;\;, \label{eq:B.15a} \\
z_{\nu\nu'} & = &z_{1}\;,\; \tilde{z}_{\nu\nu'}\;=\;\tilde{z}_{1}\;\;,\forall \nu'\neq\nu\;\;. \label{eq:B.15b}
\end{eqnarray}

\noindent With this ansatz the partition functions $Z_{1}, Z_{2}$ in Eq.(\ref{eq:B.13a}) and Eq.(\ref{eq:B.13b}) are easily evaluated as,
\begin{equation}
Z_{1} \; = \; \frac{1}{\sqrt{2\pi}}\int_{-\infty}^{\infty}dt\,e^{-\frac{1}{2}t^{2}} \left (1+\exp\left (\mu_{v}+z_{0}+\mathrm{i}\tilde{z}_{0}-z_{1}-\mathrm{i}\tilde{z}_{1}\;+\;t\left ( \sigma^{2}_{v}+2z_{1}+2\mathrm{i}\tilde{z}_{1}\right )^{\frac{1}{2}} \right) \right )^{n} \;\;,
\nonumber 
\end{equation}

\begin{equation}
Z_{2} \; = \; \frac{1}{\sqrt{2\pi}}\int_{-\infty}^{\infty}dt\,e^{-\frac{1}{2}t^{2}} \left (1+\exp\left (\mu_{h}+z_{0}-\mathrm{i}\tilde{z}_{0}-z_{1}+\mathrm{i}\tilde{z}_{1}\;+\;t\left ( \sigma^{2}_{h}+2z_{1}-2\mathrm{i}\tilde{z}_{1}\right )^{\frac{1}{2}} \right) \right )^{n} \;\;,
\label{eq:B.17}
\end{equation}

\noindent from which we have (under replica symmetry),
\begin{equation}
\log Z_{1}\;=\;nI_{0,1}\left (\mu_{v}, \sigma^{2}_{v}, -z_{0}-\mathrm{i}\tilde{z}_{0} + z_{1}+\mathrm{i}\tilde{z}_{1}, z_{1}+\mathrm{i}\tilde{z}_{1} \right)\;+\; {\mathcal O}(n^{2})\;\;,
\nonumber 
\end{equation}
\begin{equation}
\log Z_{2}\;=\;nI_{0,1}\left (\mu_{h}, \sigma^{2}_{h}, -z_{0}+\mathrm{i}\tilde{z}_{0} + z_{1}-\mathrm{i}\tilde{z}_{1}, z_{1}-\mathrm{i}\tilde{z}_{1} \right)\;+\; {\mathcal O}(n^{2})\;\;,
\label{eq:B.19}
\end{equation}

\noindent where $I_{0,s}$ is given by,
\begin{equation}
I_{0,s}(\mu, \sigma^{2},x,y)\;=\;\frac{1}{\sqrt{2\pi}}\int_{-\infty}^{\infty}dt
\, e^{-\frac{1}{2}t^{2}} \left [ \log \left ( 1+\exp\left (\mu-x\;+\;t\left ( \sigma^{2}+2y\right )^{\frac{1}{2}}\right ) \right ) \right ] ^{s}\;\;.
\label{eq:B.23}
\end{equation}

\noindent Defining $\tilde{z}_{0}=\mathrm{i}\hat{z}_{0}$, $\tilde{z}_{1}=\mathrm{i}\hat{z}_{1}$, we find that in the replica limit $n\rightarrow 0$, the replica-symmetric saddle-point solutions satisfy,
\begin{flalign}
z_{0} &= \; \frac{1}{4}\tilde{\sigma}^{2}_{J}I_{1}(\mu_{v}, \sigma^{2}_{v}, \Delta z - \Delta \hat{z}, z_{1} - \hat{z}_{1})\;+\;\frac{1}{4}\tilde{\sigma}^{2}_{J}\alpha I_{1}(\mu_{h}, \sigma^{2}_{h}, \Delta z + \Delta \hat{z}, z_{1} + \hat{z}_{1})\;\;, \nonumber \\
z_{1} &= \; \frac{1}{4}\tilde{\sigma}^{2}_{J}I_{2}(\mu_{v}, \sigma^{2}_{v}, \Delta z - \Delta \hat{z}, z_{1} - \hat{z}_{1})\;+\;\frac{1}{4}\tilde{\sigma}^{2}_{J}\alpha I_{2}(\mu_{h}, \sigma^{2}_{h}, \Delta z + \Delta \hat{z}, z_{1} + \hat{z}_{1})\;\;, \nonumber   \\
\hat{z}_{0} &= \; \frac{1}{4}\tilde{\sigma}^{2}_{J}I_{1}(\mu_{v}, \sigma^{2}_{v}, \Delta z - \Delta \hat{z}, z_{1} - \hat{z}_{1})\;-\;\frac{1}{4}\tilde{\sigma}^{2}_{J}\alpha I_{1}(\mu_{h}, \sigma^{2}_{h}, \Delta z + \Delta \hat{z}, z_{1} + \hat{z}_{1})\;\;, \nonumber \\
\hat{z}_{1} &= \; \frac{1}{4}\tilde{\sigma}^{2}_{J}I_{2}(\mu_{v}, \sigma^{2}_{v}, \Delta z - \Delta \hat{z}, z_{1} - \hat{z}_{1})\;-\;\frac{1}{4}\tilde{\sigma}^{2}_{J}\alpha I_{2}(\mu_{h}, \sigma^{2}_{h}, \Delta z + \Delta \hat{z}, z_{1} + \hat{z}_{1})\;\;,\nonumber \\
\label{eq:B.20d}
\end{flalign}

\noindent where $\Delta z = z_{1} - z_{0}$ and $\Delta \hat{z} = \hat{z}_{1} - \hat{z}_{0}$, and the functions $I_{1}$ and $I_{2}$ are defined via the integrals,
\begin{flalign}
I_{k}(\mu, \sigma^{2}, x, y) &=\; \frac{1}{\sqrt{2\pi}} \int_{-\infty}^{\infty} dt\, e^{-\frac{1}{2}t^{2}}  \left ( \frac{\exp\left (\mu - x + t\left ( \sigma^{2} + 2y\right )^{\frac{1}{2}} \right)}{1+\exp\left (\mu - x + t\left ( \sigma^{2} + 2y\right )^{\frac{1}{2}}\right) } \right ) ^{k} \nonumber \\
&=\; \frac{1}{\sqrt{2\pi}}\frac{1}{\sqrt{\sigma^{2} + 2y}}\int_{-\infty}^{\infty}dt\,
\exp\left ( -\frac{(t-\mu + x)^{2}}{2(\sigma^{2} + 2y)}\right )\left ( \frac{e^{t}}{1+e^{t}}\right )^{k} \;\;. \label{eq:B.21}
\end{flalign}

\noindent The system of equations in Eq.(\ref{eq:B.20d}) are easily solved by iteration. For the values of $\mu_{v}, \sigma^{2}_{v}, \mu_{h}, \sigma^{2}_{h}, \tilde{\sigma}_{J}^{2}$ we have studied, we have not observed more than a single solution to Eq.(\ref{eq:B.20d}) for a given set of parameter values. This suggests that Eq.(\ref{eq:B.20d}) possesses only a single fixed point, although we have not been able to analytically prove this.

In terms of the saddle-point values of $z_{0},\hat{z}_{0},z_{1},\hat{z}_{1}$, the replica-symmetric approximation to $\mathbb{E}\left ( \log Z\right)$ then becomes,
\begin{multline}
\lim_{\substack{N\rightarrow\infty\\M=\alpha N}} \frac{1}{N}\mathbb{E}\left (\log Z \right ) \; = \; \frac{2}{\tilde{\sigma}^{2}_{J}}\left( \hat{z}_{0}^{2}-z_{0}^{2}\;-\;\hat{z}_{1}^{2}+z_{1}^{2}\right ) \; + \; I_{0,1}\left( \mu_{v},\sigma^{2}_{v},\Delta z -\Delta \hat{z}, z_{1} -\hat{z}_{1}\right ) \\
\; + \; \alpha I_{0,1}\left ( \mu_{h}, \sigma^{2}_{h}, \Delta z + \Delta \hat{z}, z_{1} + \hat{z}_{1} \right)\;\;.
\label{eq:B.22}
\end{multline}

\noindent Again, with minimal change to the definition of the forms of the integrands defining $I_{0,s}$ and $I_{k}$ in Eq.(\ref{eq:B.23}) and Eq.(\ref{eq:B.21}), the same derivation can be used to obtain the leading order replica-symmetric approximation to $\mathbb{E}(\log Z)$ for the $\mu_{J}\neq 0$ case. This is done in \ref{sec:AppendixA1}. From the form of the saddle-point equations in Eq.(\ref{eq:B.20d}) it is easy to verify that the leading order approximation to $\mathbb{E}\left ( \log Z\right)$ given in Eq.(\ref{eq:B.22}) is invariant under the transformation given in Eq.(\ref{eq:B.4}).

Similarly, by considering the ${\mathcal O}(n^{2})$ contribution to $\mathbb{E}\left ( Z^{n}\right )$ we can obtain an expression for the variance of $\log Z$, taken over the disorder of the lattice-gas RBM Hamiltonian parameters. Specifically, we have,
\begin{equation}
{\rm Var}\left (\log Z \right) \; = \; \lim_{n\rightarrow 0}\frac{\partial^{2}\mathbb{E}\left ( Z^{n}\right)}{\partial n^{2}}\;-\;\left ( \lim_{n\rightarrow 0}\frac{\partial\mathbb{E}\left ( Z^{n}\right)}{\partial n}\right )^{2}\;\;,
\nonumber 
\end{equation}
\noindent from which we find,
\begin{equation}
\lim_{\substack{N\rightarrow\infty\\M=\alpha N}}
\frac{1}{N}{\rm Var}\left (\log Z \right)
\; = \; I_{0,2}(v)-
I^{2}_{0,1}(v) +  \alpha  I_{0,2}(h) -
\alpha I^{2}_{0,1}(h) - \alpha\tilde{\sigma}^{2}_{J} I_{2}(v)\,I_{2}(h)\;\;,
\label{eq:B.35}
\end{equation} 

\noindent where we have used an obvious shorthand notation, $v \equiv \left ( \mu_{v}, \sigma^{2}_{v}, \Delta z - \Delta \hat{z}, z_{1} - \hat{z}_{1}\right)$ and $h\equiv \left (  \mu_{h}, \sigma^{2}_{h}, \Delta z + \Delta \hat{z}, z_{1} + \hat{z}_{1} \right )$.

\section{Simulation Results}
\begin{figure*}
	\begin{minipage}[h]{0.485\textwidth}
	\begin{center}	
		\includegraphics*[bb=0 0 432 288, scale=0.5]{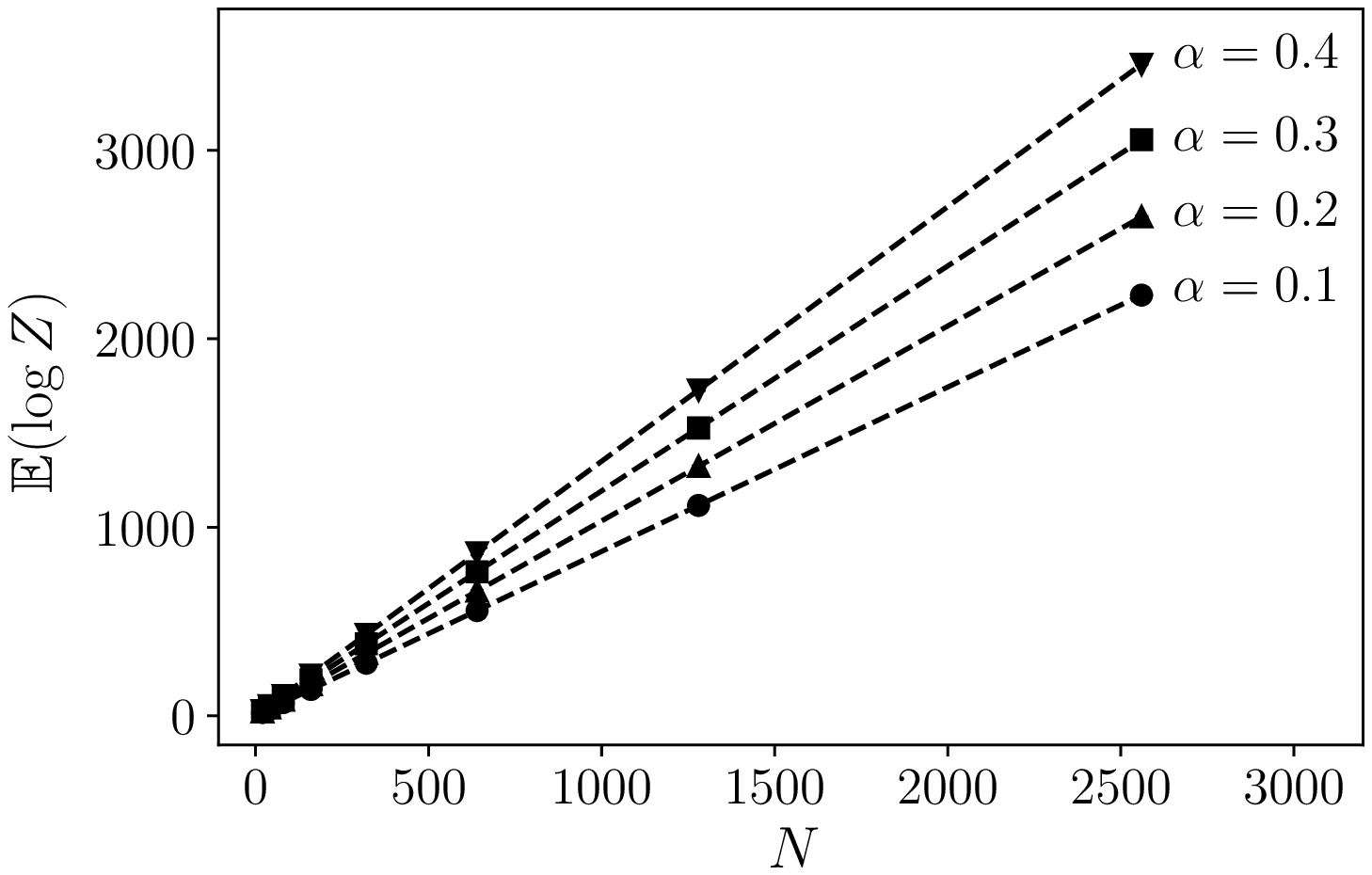}
	\end{center}
    \end{minipage}
	\begin{minipage}[h]{0.495\textwidth} 
		\begin{subfigure}[h]{0.475\textwidth}
		\includegraphics[bb=0 0 432 288, scale=0.27]{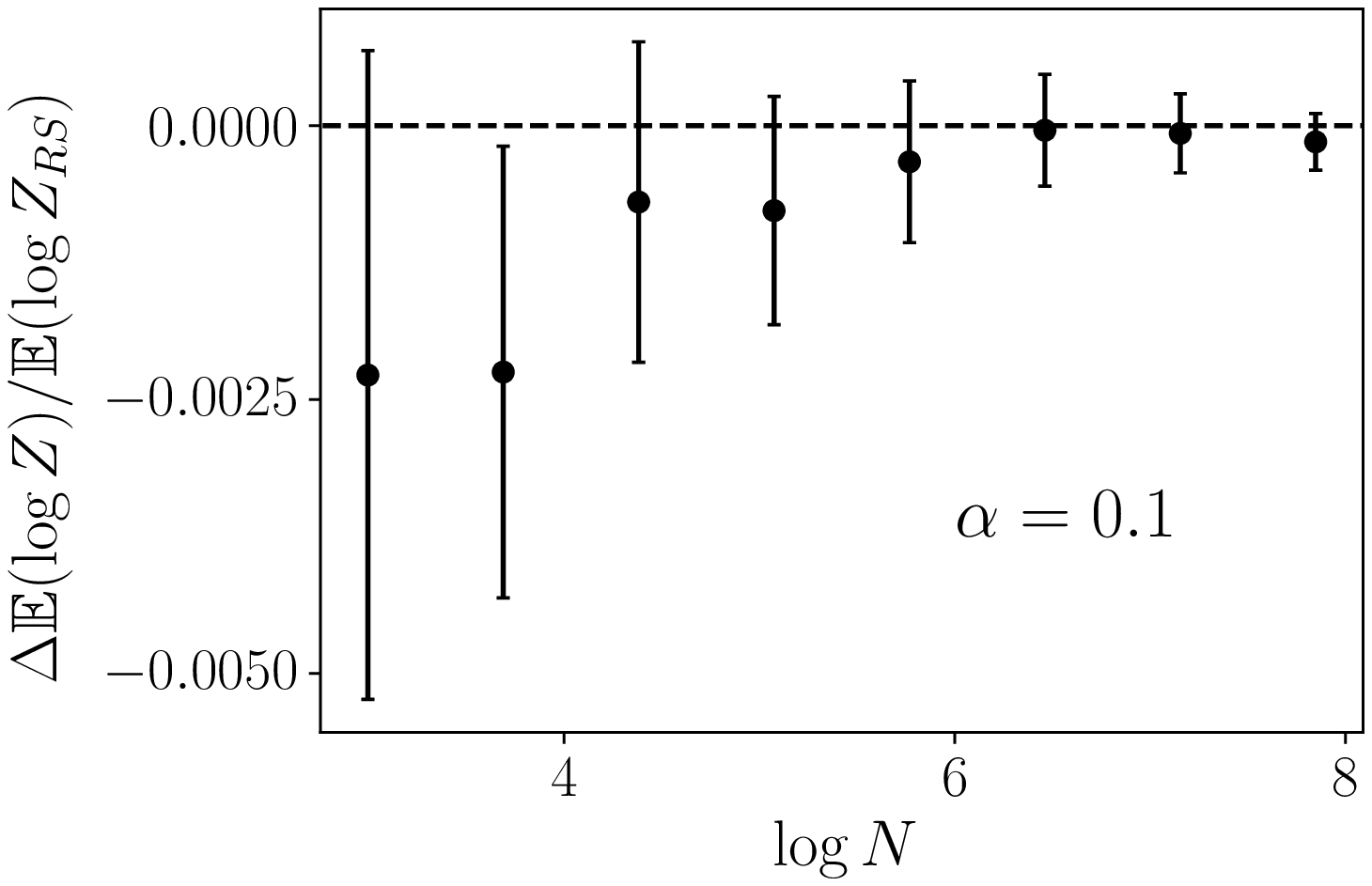}
		\end{subfigure}
	    \begin{subfigure}[h]{0.475\textwidth}
		\includegraphics[bb=0 0 432 288, scale=0.27]{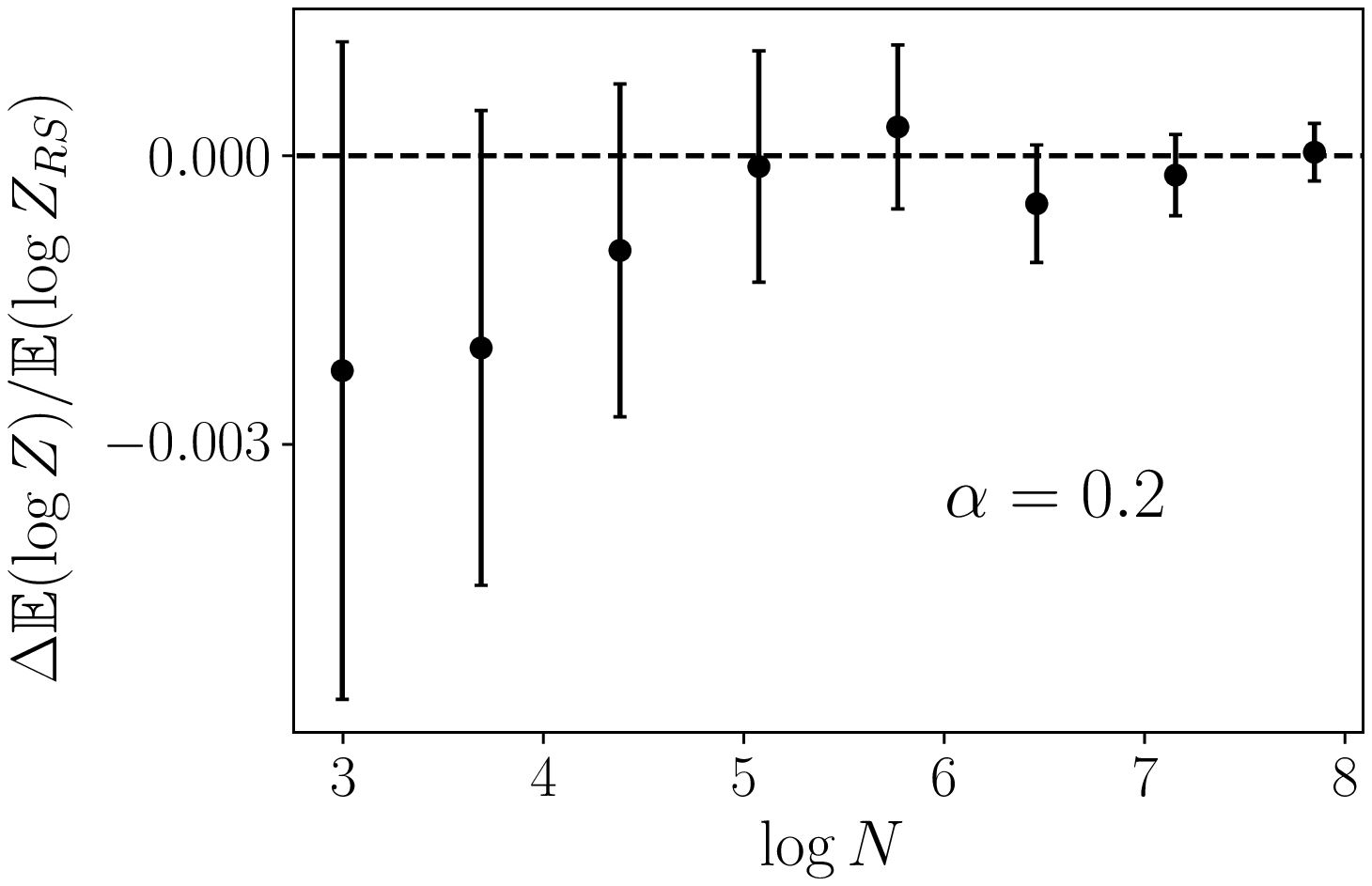}
	    \end{subfigure} 
		\begin{subfigure}[h]{0.475\textwidth}
		\includegraphics[bb=0 0 432 288, scale=0.27]{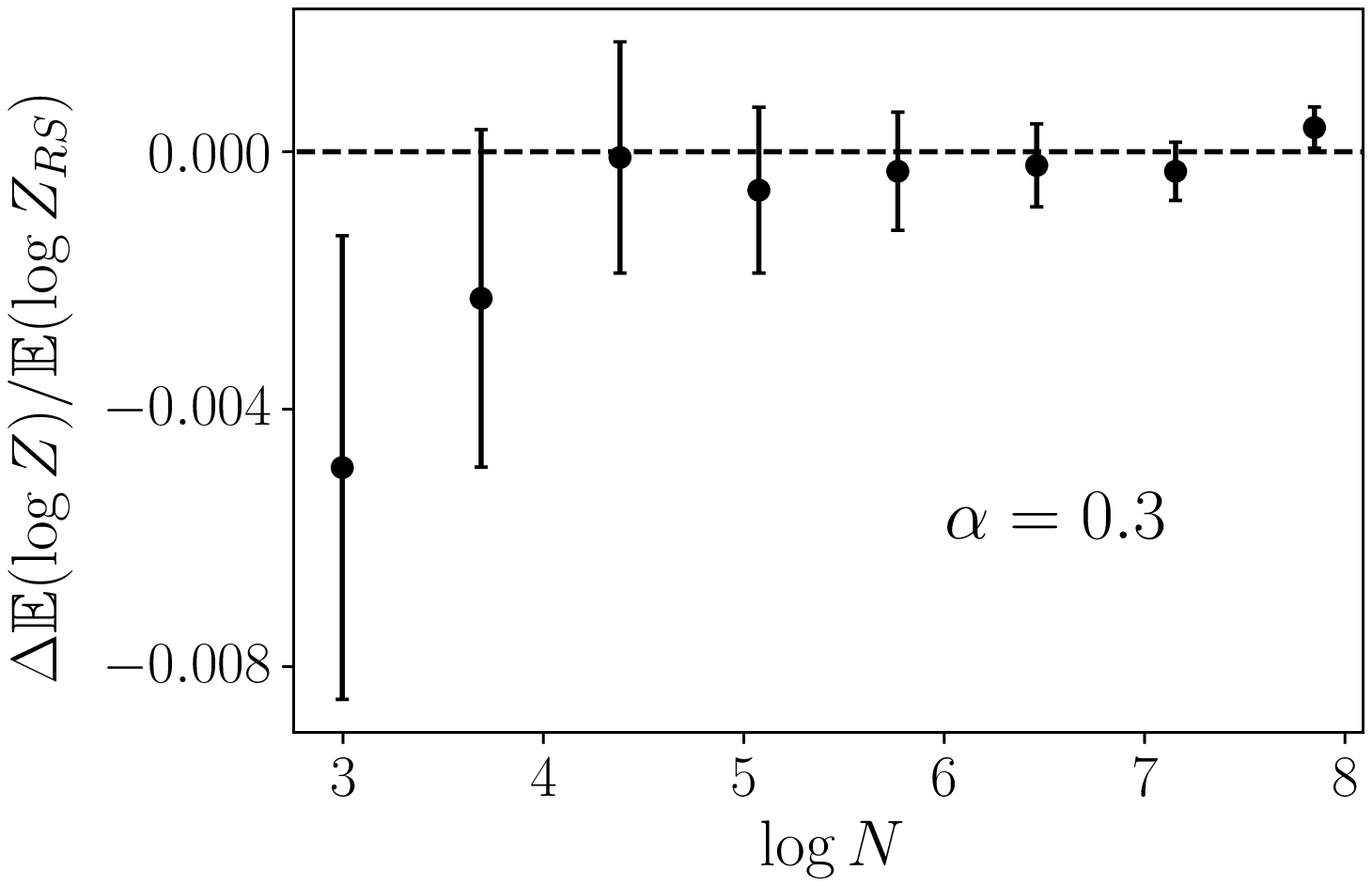}
	    \end{subfigure} \hspace{0.7mm}
		\begin{subfigure}[h]{0.475\textwidth}
		\includegraphics[bb=0 0 432 288, scale=0.27]{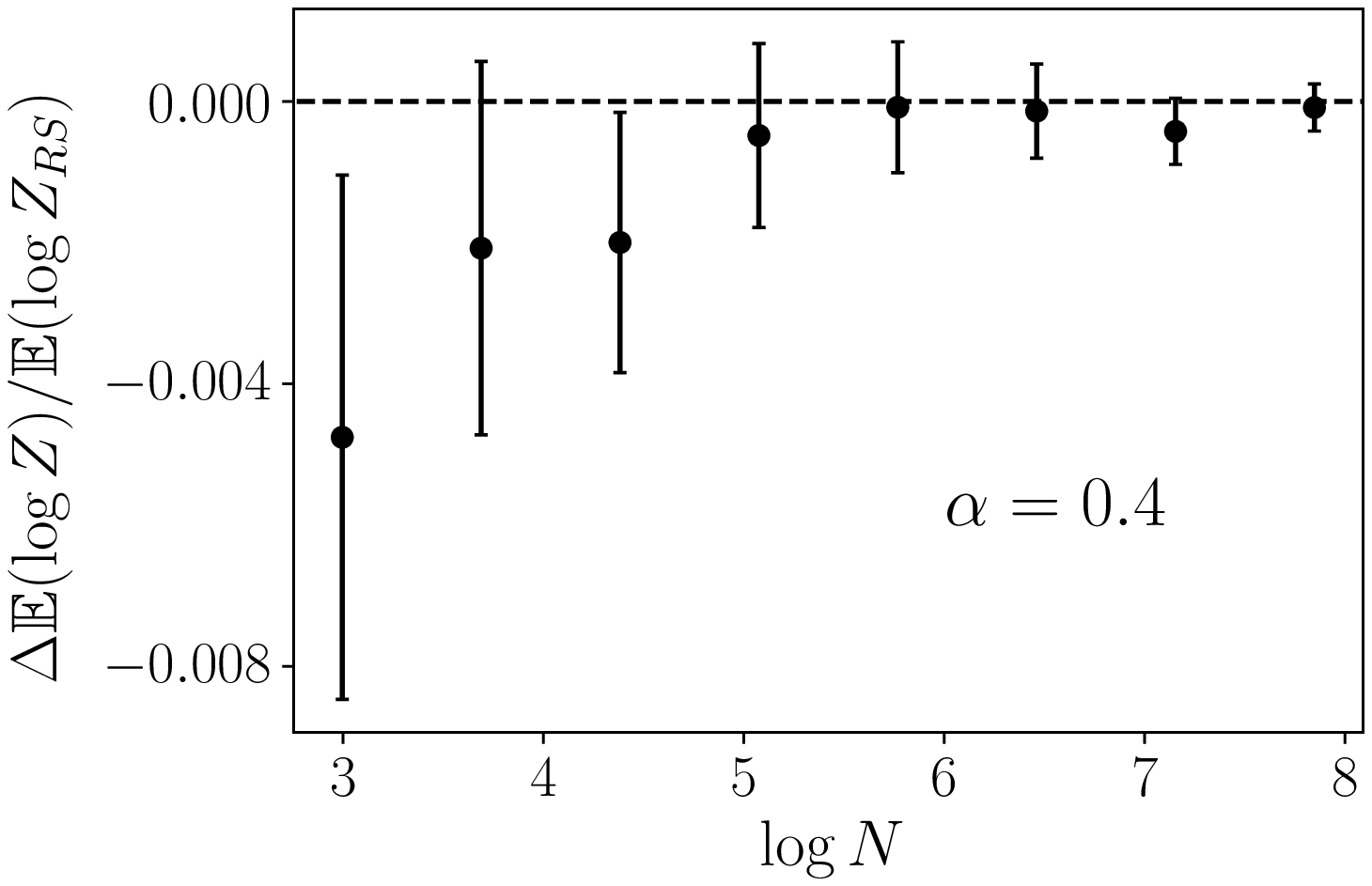}
	\end{subfigure}
    \end{minipage}

    \vspace{0.25cm}
	\begin{minipage}[h]{.485\textwidth}
	\begin{center}	
		\includegraphics*[bb=0 0 432 288, scale=0.5]{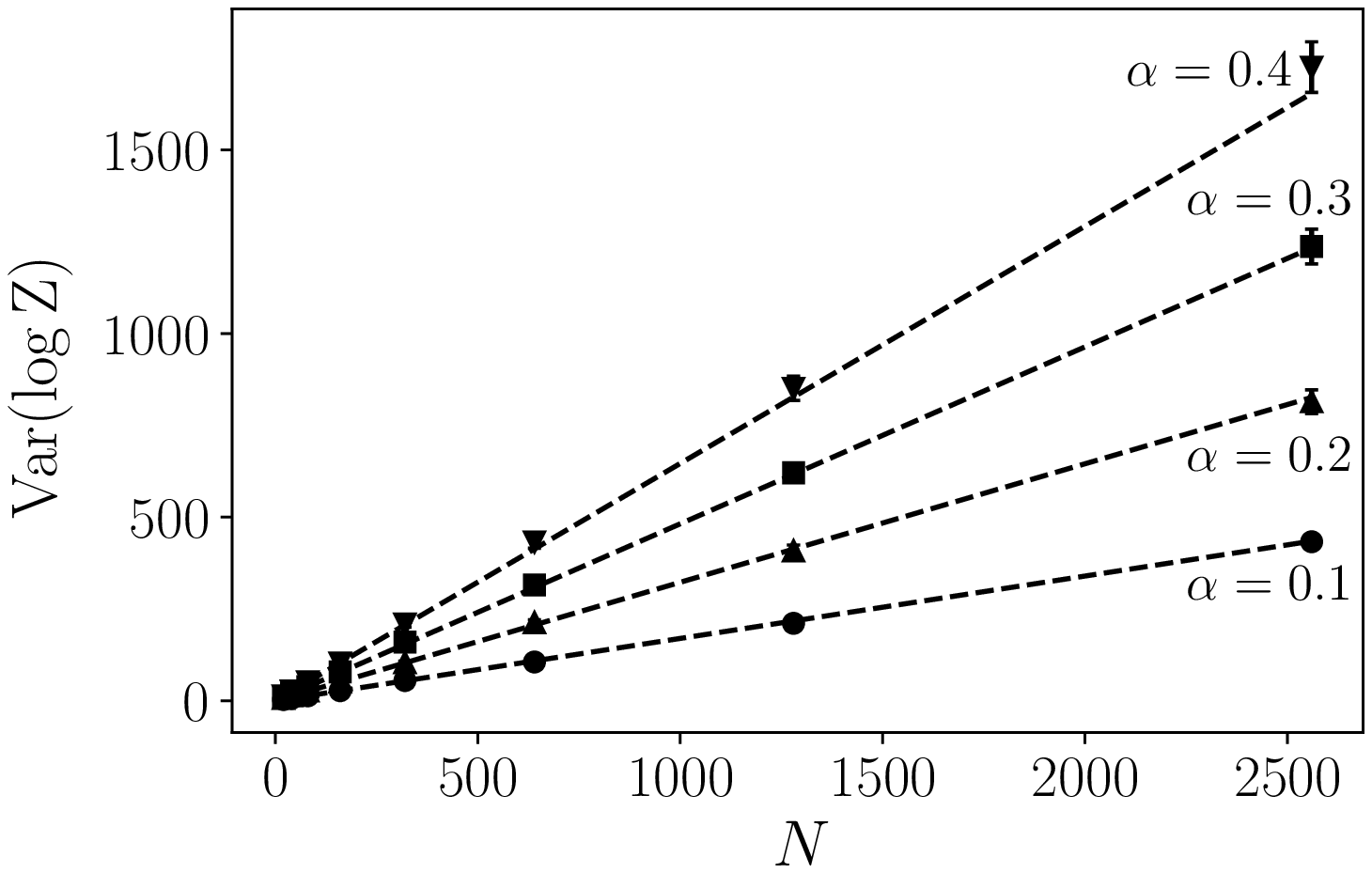}
	\end{center}
    \end{minipage}%
    \hspace{0.025cm}
    \begin{minipage}[h]{.485\textwidth} 
	    \begin{subfigure}[h]{0.475\textwidth}
		    \includegraphics[bb=0 0 432 288, scale=0.27]{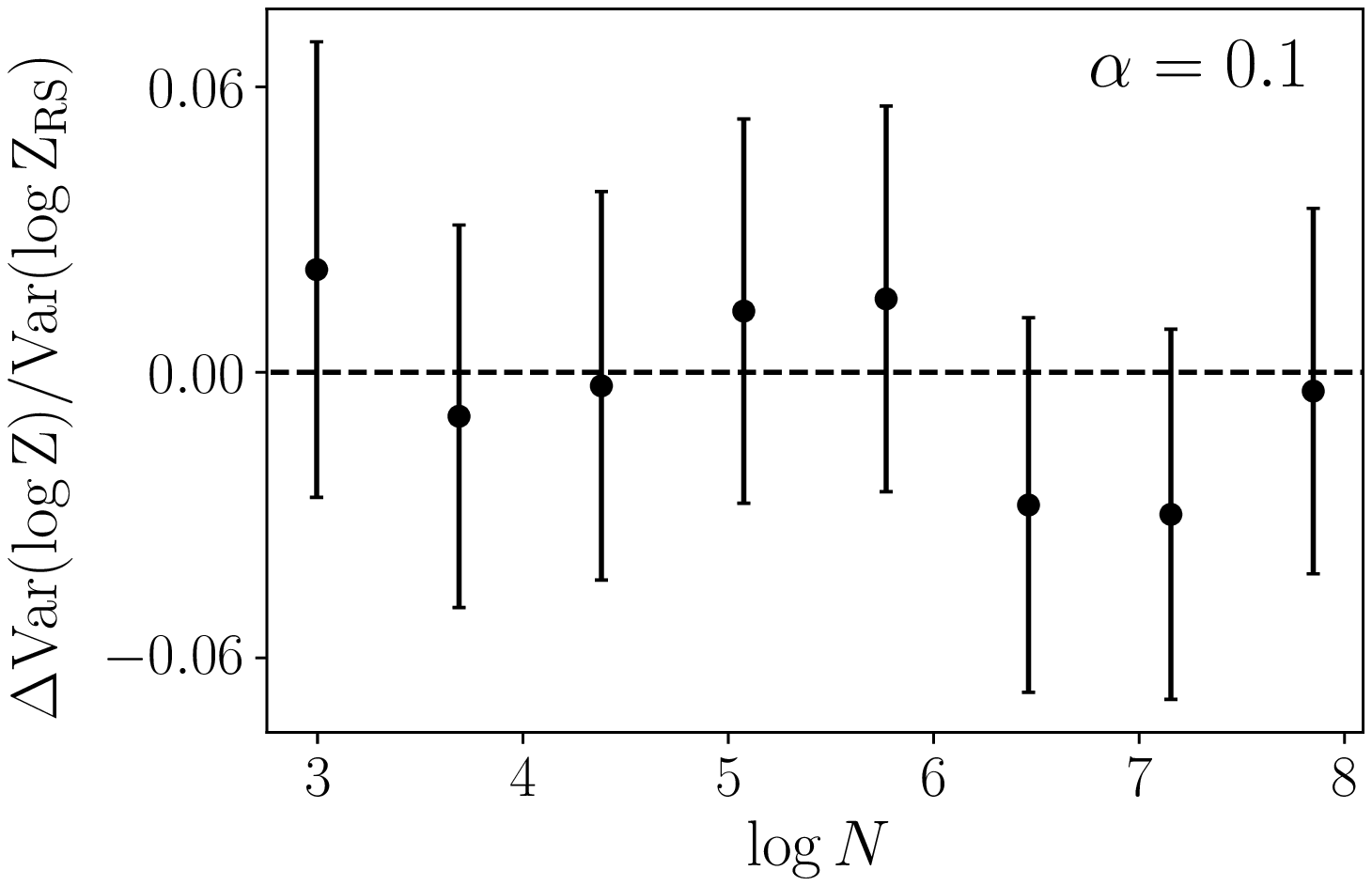}
	    \end{subfigure}
	    \begin{subfigure}[h]{0.475\textwidth}
		    \includegraphics[bb=0 0 432 288, scale=0.27]{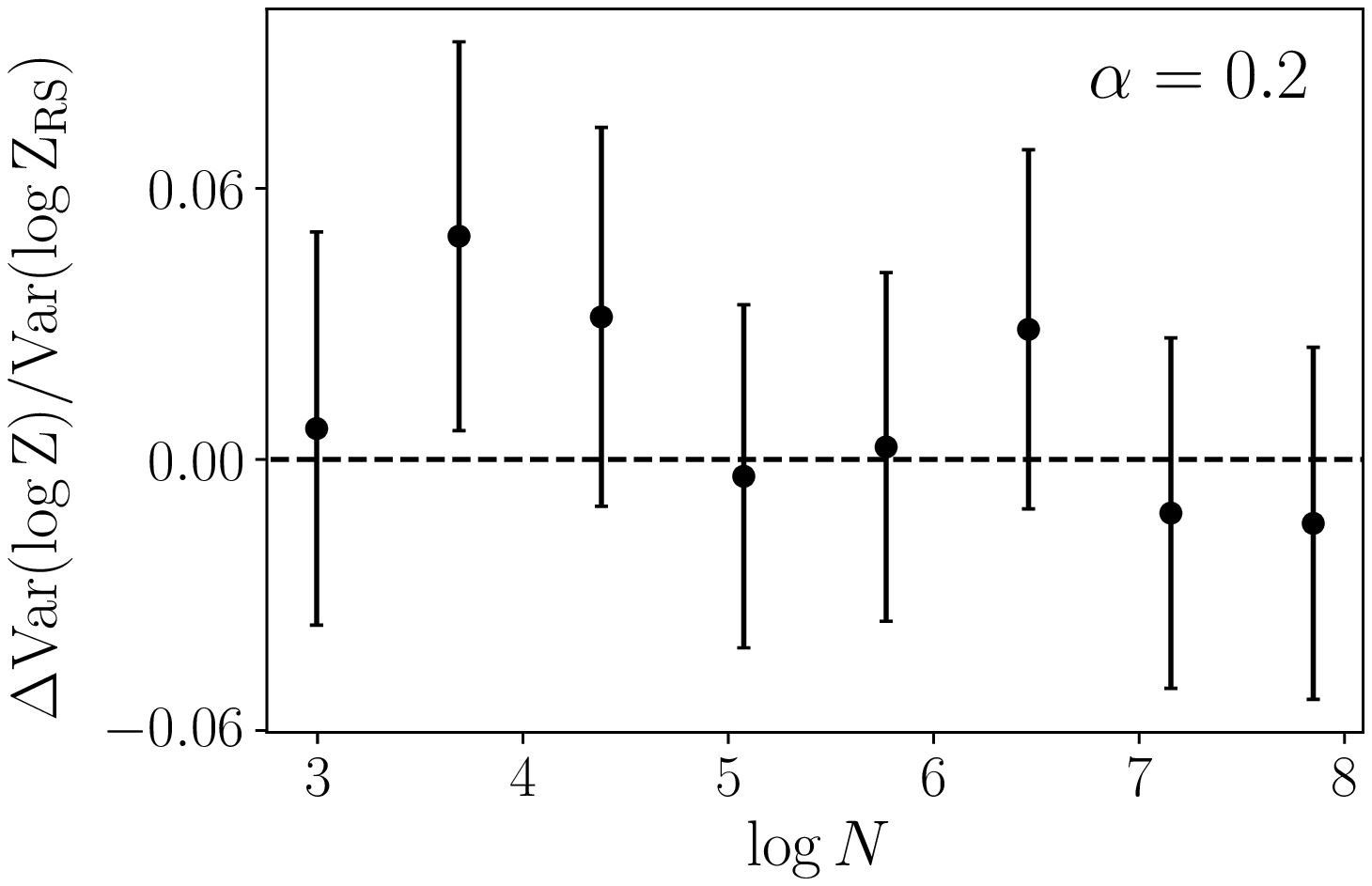}
	    \end{subfigure}
	    \begin{subfigure}[h]{0.475\textwidth}
		    \includegraphics[bb=0 0 432 288, scale=0.27]{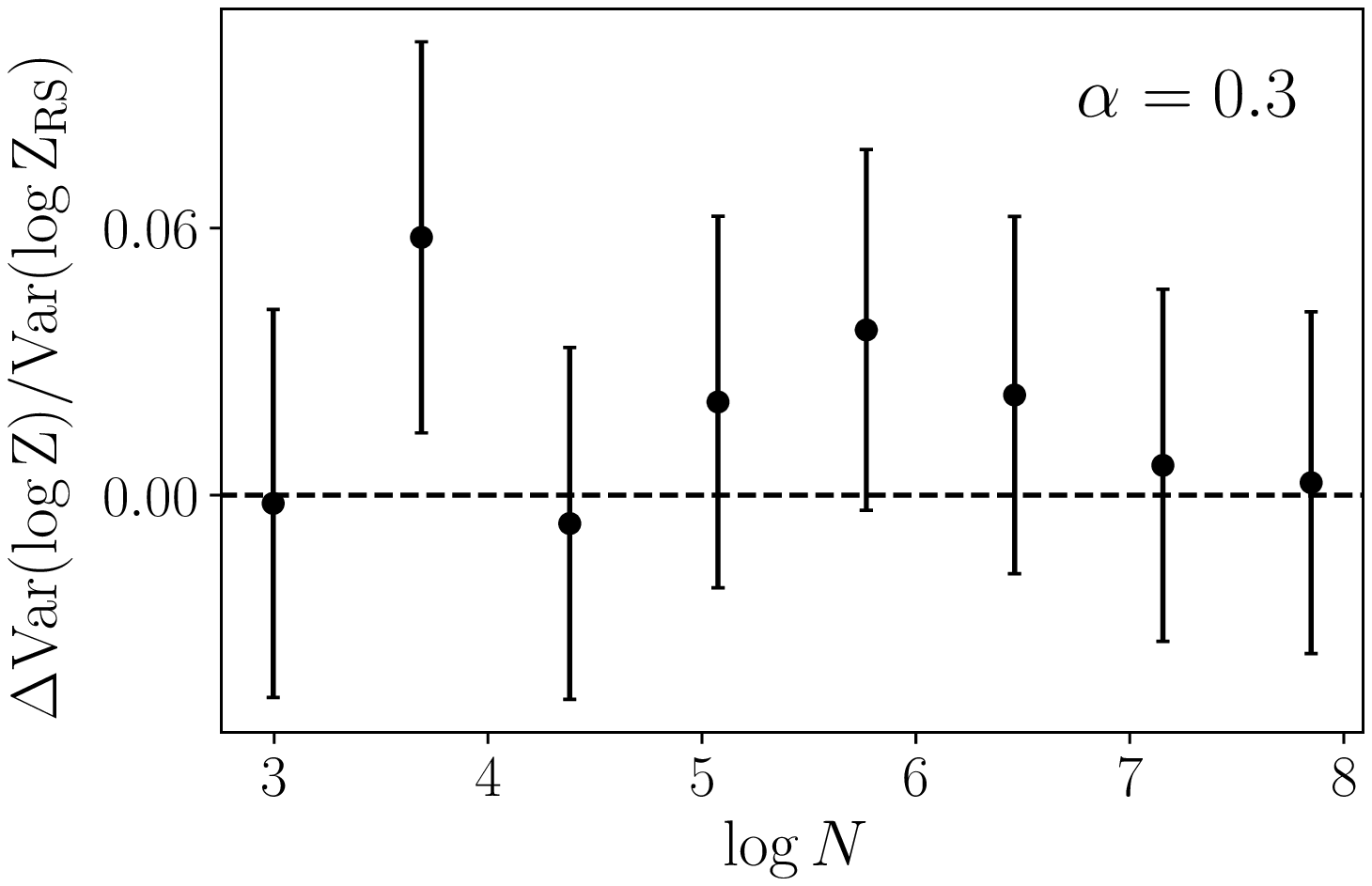}
	    \end{subfigure} \hspace{0.7mm}
	    \begin{subfigure}[h]{0.475\textwidth}
		    \includegraphics[bb=0 0 432 288, scale=0.27]{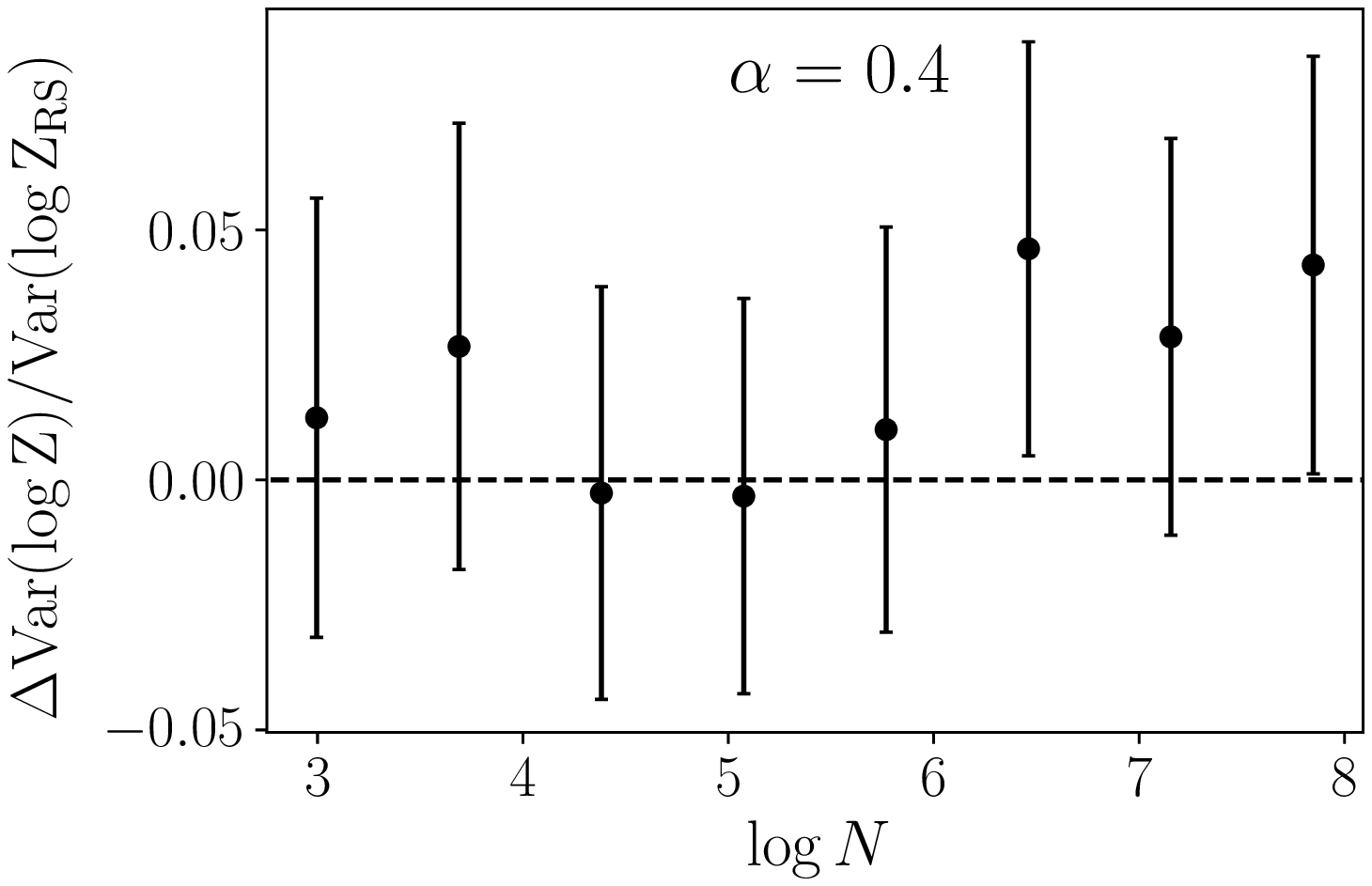}
	    \end{subfigure}
    \end{minipage}
    \caption{Plots of estimates of $\mathbb{E}\left(\log Z\right)$ (upper row) and ${\rm Var}\left(\log Z\right)$ (lower row). In the main (left hand) plot of both rows the solid symbols show the simulation average (over 5000 RBMs) calculated using the Huang \& Toyozuimi message-passing algorithm, whilst the dashed lines show the leading order replica-symmetric approximation. Error bars (typically less than the size of the symbols correspond to 95\% confidence intervals of the simulation estimates. The right-hand subplots in each row show the fractional differences, with 95\% confidence intervals, between the two methods of calculation.}
    \label{fig:Figure1}
\end{figure*}

To test the accuracy of the leading order expressions in Eq.(\ref{eq:B.22}) and Eq.(\ref{eq:B.35}) for the mean and variance of $\log Z$, we can compare to simulation estimates of the same quantities. We evaluate $\log Z$ of a specific RBM instance (with values of $J_{ai},\;v_{i},\;h_{a}$ drawn according to Eq.(\ref{eq:B.2a})-Eq.(\ref{eq:B.2c})) using the Bethe approximation message-passing (MP) mean-field algorithm of Huang and Toyoizumi \cite{HuangToyoizumi2015}. Huang and Toyoizumi \cite{HuangToyoizumi2015} have already shown that the MP estimation of RBM free energy is accurate by comparing to exact enumeration of states for smaller scale RBMs and also by comparing to estimates obtained from Gibbs sampling methods \cite{Hinton2006}. Figure \ref{fig:Figure1} shows a comparison of the leading order replica-symmetry based approximation to $\mathbb{E}\left ( \log Z\right )$ and ${\rm Var}\left ( \log Z\right )$ over a range of values of $N$. In Fig.\ref{fig:Figure1} the replica-symmetry based estimates are compared to the same quantities evaluated over 5000 runs of the MP algorithm. Here we have set $\tilde{\sigma}^{2}_{J}=10$, $\mu_{v} = \mu_{h} = 0$ and $\sigma^{2}_{v} = \sigma^{2}_{h}=0.1$. From Fig.\ref{fig:Figure1} we can see that the leading order replica-symmetry based approximation of $\mathbb{E}\left ( \log Z\right )$ is accurate to under 1\%, even at relatively small values of $N$. 

\begin{figure*}
	\begin{center}
		\scalebox{0.5}{%
			\includegraphics*[bb=0 0 432 288]{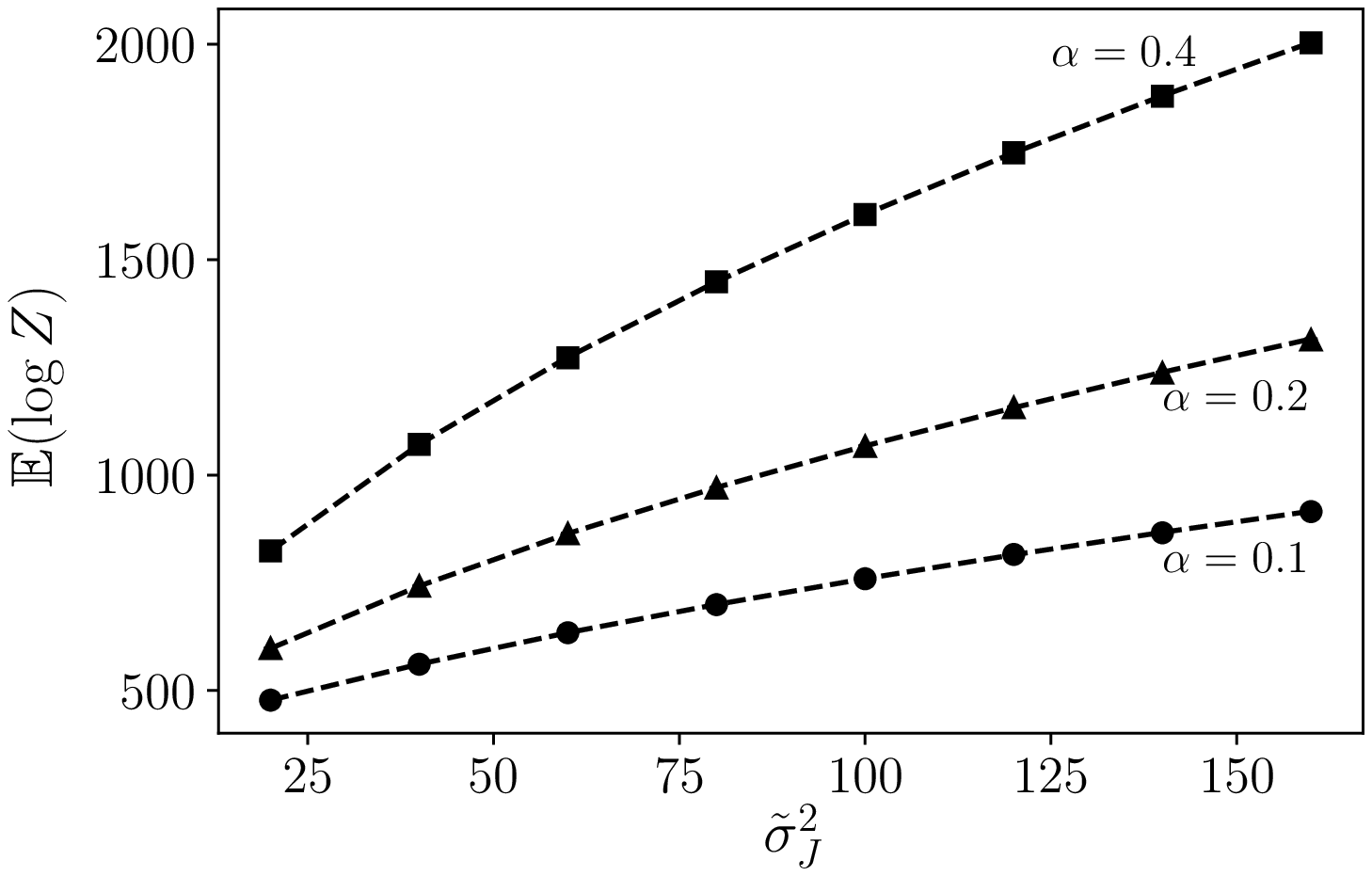},
			\includegraphics*[bb=0 0 432 288]{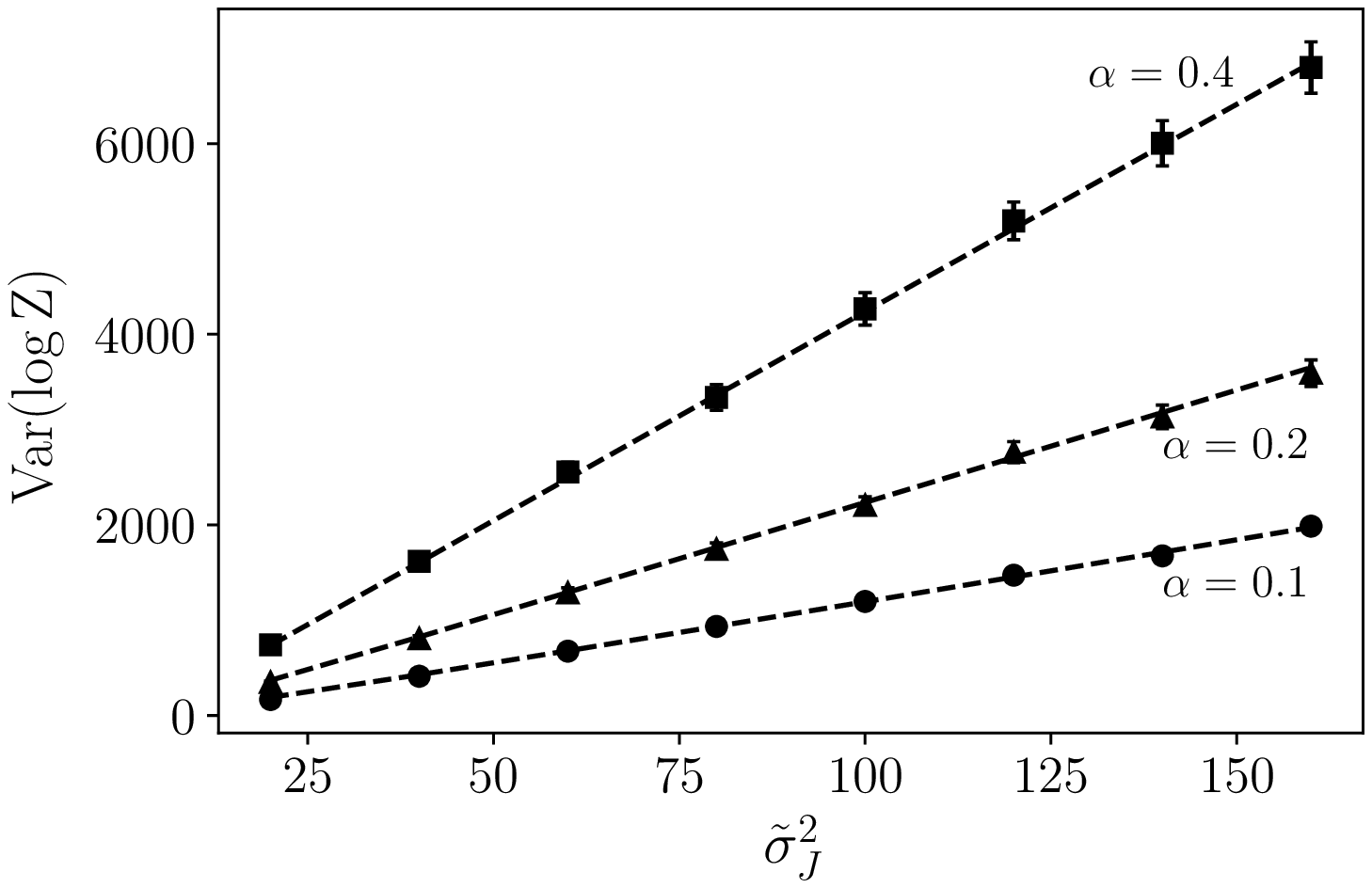}}
	\end{center}
    \caption{The left-hand plot shows estimates of $\mathbb{E}\left(\log Z\right)$ for different values of $\tilde{\sigma}^{2}_{J}$, whilst he right-hand plot shows estimates ${\rm Var}\left (\log Z \right)$ for different values of $\tilde{\sigma}^{2}_{J}$. In both plots the solid symbols show the simulation average (over 5000 RBMs) calculated using the Huang \& Toyozuimi message-passing algorithm, whilst the dashed lines show the corresponding replica-symmetric approximation. Error bars (typically less than the size of the symbols) correspond to 95\% confidence intervals of the simulation estimates.}
    \label{fig:Figure2}
\end{figure*}

Similarly, Figure \ref{fig:Figure2} shows the leading order replica-symmetry based approximation to $\mathbb{E}\left ( \log Z\right )$ and ${\rm Var}\left (\log Z\right )$ over a range of values of $\tilde{\sigma}^{2}_{J}$. Here we have set $N=500$, $\mu_{v} = \mu_{h} = 0$ and $\sigma^{2}_{v} = \sigma^{2}_{h}=0.1$. From Fig.\ref{fig:Figure2} we can see that the replica-symmetry based approximations are accurate across a wide range of values of $\tilde{\sigma}^{2}_{J}$. 

\begin{figure*}
	\begin{center}
		\scalebox{0.5}{%
			\includegraphics*[bb=0 0 432 288]{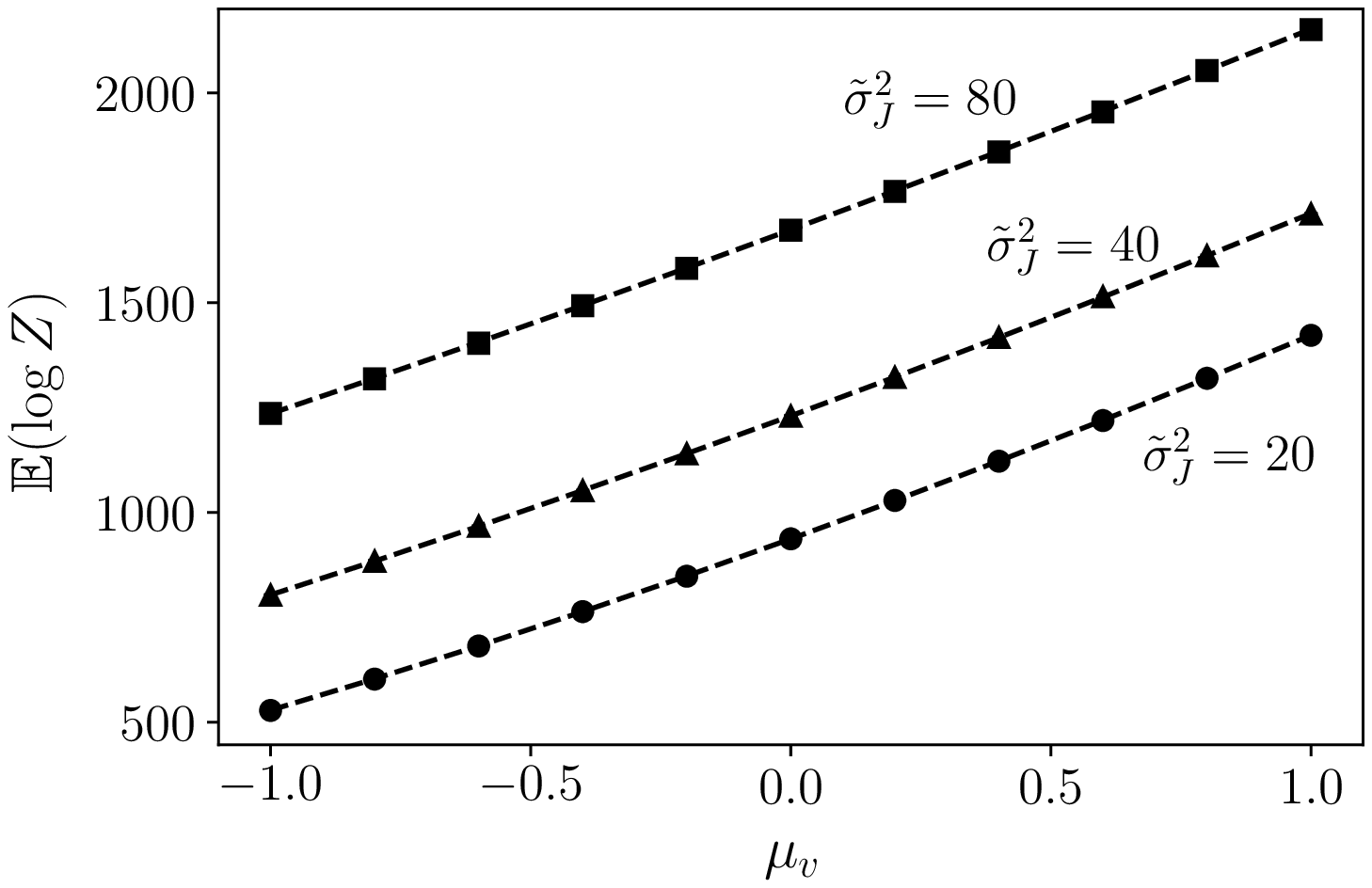},
			\includegraphics*[bb=0 0 432 288]{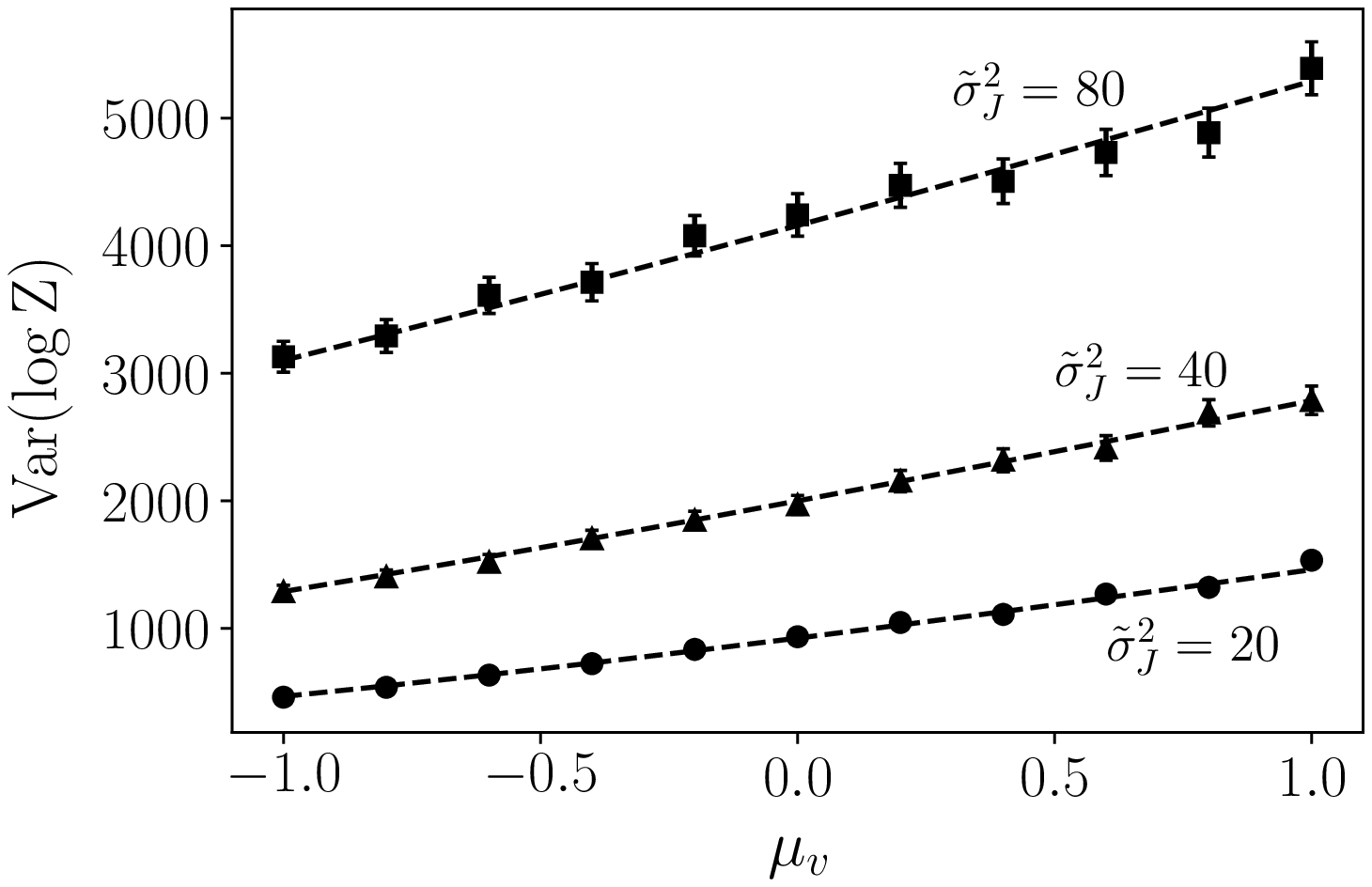}}
	\end{center}
	\caption{The left-hand plot shows estimates of $\mathbb{E}\left(\log Z\right)$ for different values of $\mu_{v}$ and $\tilde{\sigma}^{2}_{J}$, whilst the right-hand plot shows estimates of ${\rm Var}\left (\log Z \right)$ for different values of $\mu_{v}$ and $\tilde{\sigma}^{2}_{J}$. Simulation set up is similar to that for Fig.\ref{fig:Figure1} and Fig.\ref{fig:Figure2} - see main text for details.}
	\label{fig:Figure3}
\end{figure*}

\begin{figure*}
	\begin{center}
		\scalebox{0.5}{%
			\includegraphics*[bb=0 0 432 288]{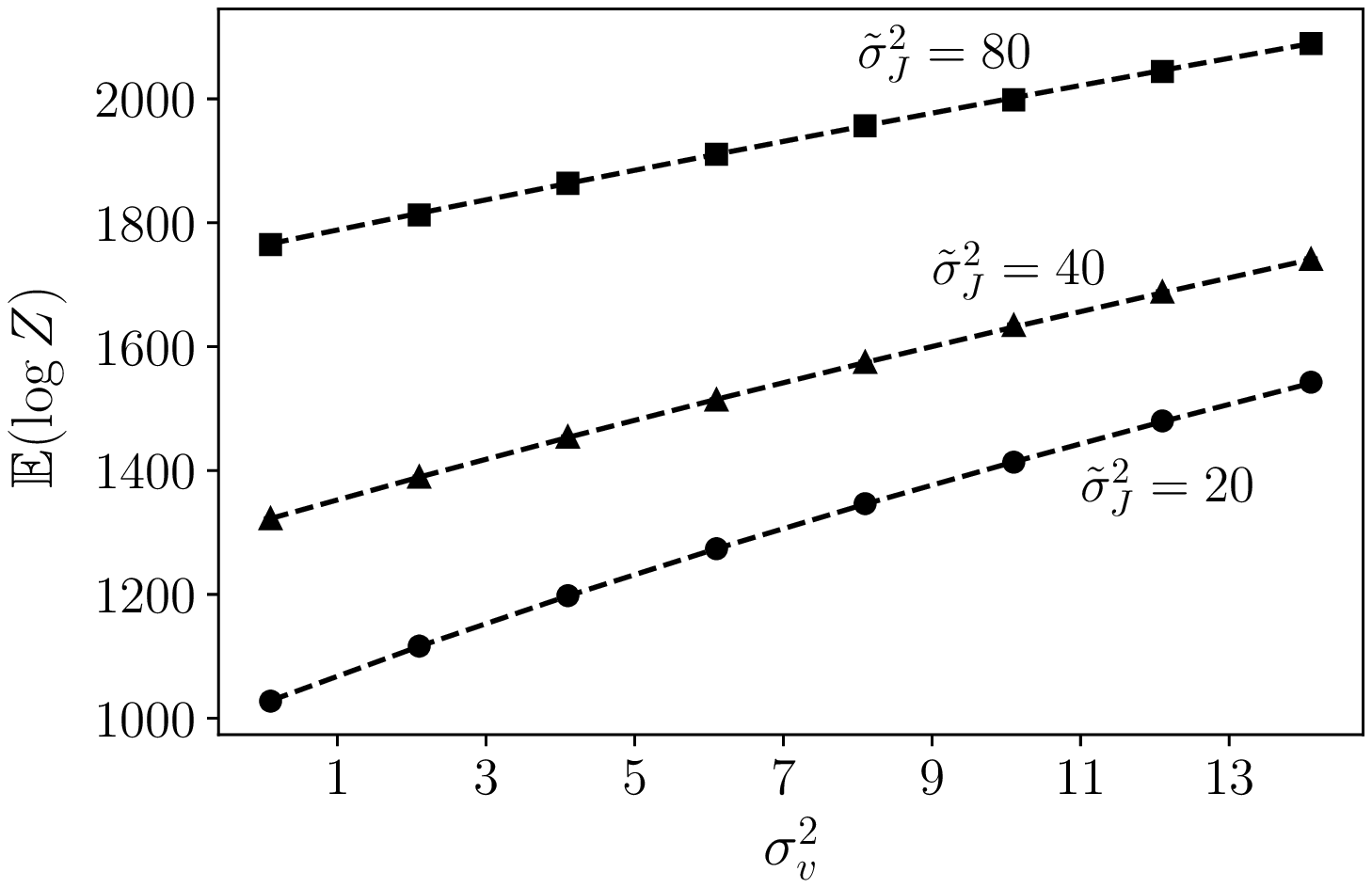},
			\includegraphics*[bb=0 0 432 288]{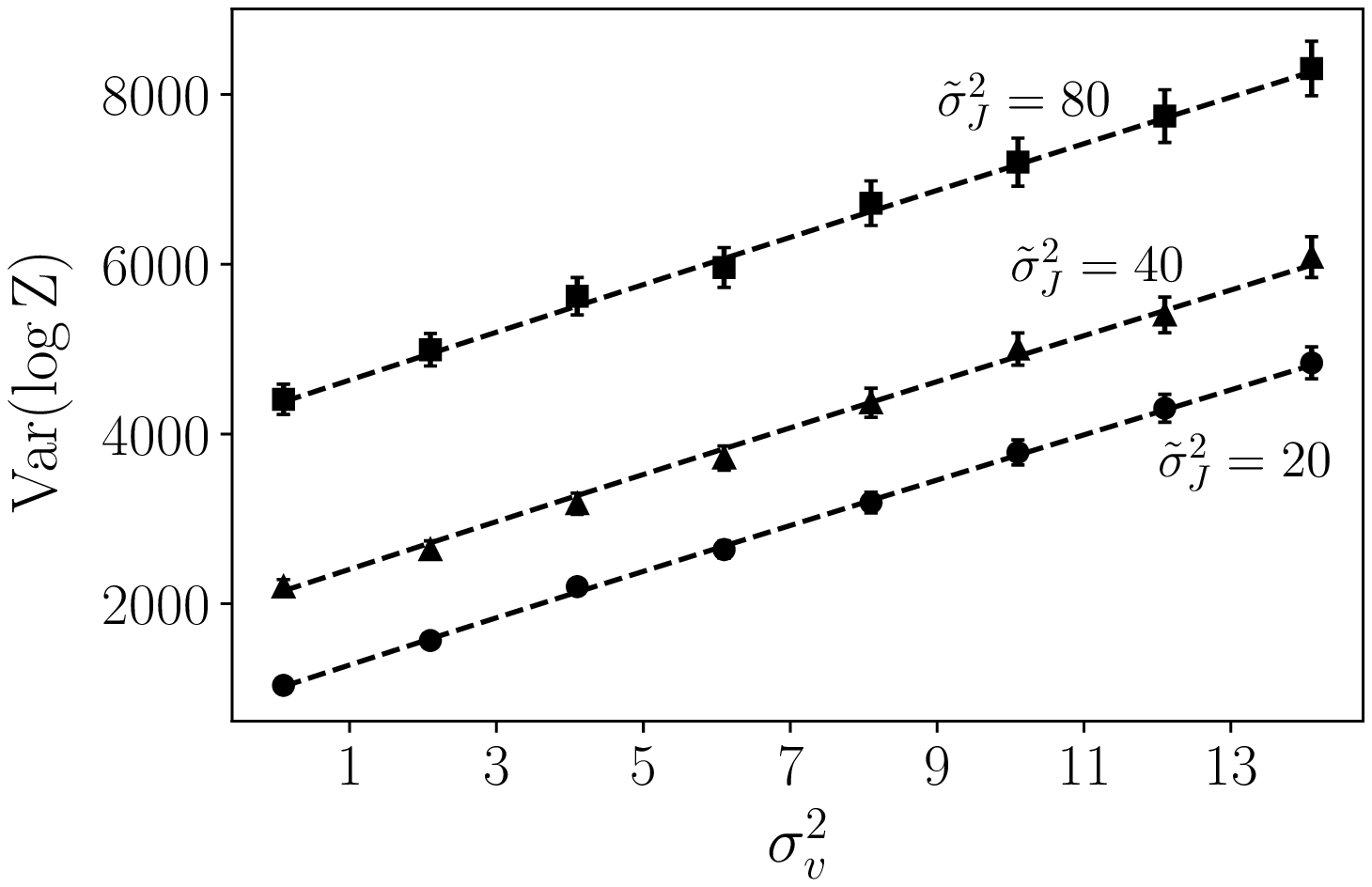}}
	\end{center}
	\caption{The left-hand plot shows estimates of $\mathbb{E}\left(\log Z\right)$ for different values of $\sigma^{2}_{v}$ and $\tilde{\sigma}^{2}_{J}$, whilst the right-hand plot shows estimates of ${\rm Var}\left (\log Z \right)$ for different values of $\sigma^{2}_{v}$ and $\tilde{\sigma}^{2}_{J}$. Simulation set up is similar to that for Fig.\ref{fig:Figure1} and Fig.\ref{fig:Figure2} - see main text for details.}
	\label{fig:Figure4}
\end{figure*}

The expressions for $\mathbb{E}\left(\log Z\right)$ in Eq.(\ref{eq:B.22}) and for ${\rm Var}\left (\log Z \right)$ in Eq.(\ref{eq:B.35}) have been derived for when we have bias and heterogeneity in the external field parameters of the lattice-gas RBM Hamiltonian and not just in the interaction couplings, that is when $\mu_{v}, \mu_{h} \neq 0$ and when $\sigma^{2}_{v}, \sigma^{2}_{h} > 0$. Therefore, we also test the accuracy of the expressions in Eq.(\ref{eq:B.22}) and Eq.(\ref{eq:B.35}) in these scenarios. Figure \ref{fig:Figure3} shows plots of $\mathbb{E}\left(\log Z\right)$ and ${\rm Var}\left (\log Z \right)$ for different values of $\mu_{v}$ and at three different values of $\tilde{\sigma}^{2}_{J}$. In Fig.\ref{fig:Figure3} we have set $N=500, M=250$ and $\sigma^{2}_{v} = \sigma^{2}_{h} = 0.1$. For all the simulations in Fig.\ref{fig:Figure3} we have set $\mu_{h} =\mu_{v}$, solely to simplify the number of combinations of different distribution parameter values tested with the simulations. Similarly, Figure \ref{fig:Figure4} shows plots of $\mathbb{E}\left(\log Z\right)$ and ${\rm Var}\left (\log Z \right)$ for different values of $\sigma_{v}^{2}$ and the same three values of $\tilde{\sigma}^{2}_{J}$. In Fig.\ref{fig:Figure4} we again set $N=500, M=250$, but have set $\mu_{v} = \mu_{h} = 0.2$ and now set $\sigma^{2}_{h} = \sigma^{2}_{v}$. As with Fig.\ref{fig:Figure1} and Fig.\ref{fig:Figure2}, the solid symbols in Fig.\ref{fig:Figure3} and Fig.\ref{fig:Figure4} show the simulation average (over 5000 RBMs) calculated using the Huang \& Toyozuimi message-passing algorithm, whilst the dashed lines show the corresponding replica-symmetric approximation. The error bars (typically less than the size of the symbols) correspond to 95\% confidence intervals of the simulation estimates. As we can see from Fig.\ref{fig:Figure3} and Fig.\ref{fig:Figure4}, the agreement between the simulation averages and the expressions in Eq.(\ref{eq:B.22}) and Eq.(\ref{eq:B.35}) are within those 95\% confidence intervals, confirming the accuracy of the leading order expressions for $\mathbb{E}\left(\log Z\right)$ and ${\rm Var}\left (\log Z \right)$ when we have heterogeneity in the RBM Hamiltonian external field parameters.


\section{Behaviour of the replica symmetric solution \label{sec:phase_behaviour}}

The derivation of $\mathbb{E}\left (\log Z \right )$ in Eq.(\ref{eq:B.22}) for the lattice-gas RBM yields a non-trivially different formula compared to the equivalent result for the Ising RBM, which is given in \ref{sec:AppendixIsing} in Eq.(\ref{eq:X.0}). This difference is due to the fact that despite the lattice-gas and Ising RBM Hamiltonian forms being isomorphic to each other and therefore equivalent up to a global constant, the two ensembles of Hamiltonian parameters that we have averaged over, are not equivalent to each other. A diagonal covariance matrix in one representation is not equivalent to a diagonal covariance matrix in the other representation. The simulation results show that the expression for $\mathbb{E}\left ( \log Z\right )$ in Eq.(\ref{eq:B.22}) is numerically accurate and therefore there are numerical differences, beyond the global constant, between the lattice-gas expression for $\mathbb{E}\left ( \log Z\right )$ in Eq.(\ref{eq:B.22}) and the equivalent expression in Eq.(\ref{eq:X.0}) for the Ising RBM. The question we now ask is whether the \emph{qualitative behaviour} of the lattice-gas RBM saddle-point solutions is different to what we would expect from the equivalent Ising RBM. That is, within a fixed representation does changing from averaging with a diagonal parameter covariance matrix to averaging with a non-diagonal parameter covariance matrix induce qualitative changes in the behaviour of the RBM free energy and not just numerical changes. The qualitative behaviour of the Ising RBM quenched free energy averaged with a diagonal covariance matrix is encapsulated in its phase behaviour, and so we first recap some of the main features and characteristics of that phase behaviour.

Replica analysis of the standard Sherrington-Kirkpatrick (SK) spin-glass model \cite{SherringtonKirkpatrick1975} reveals a rich phase behaviour \cite{Nishimori2001}, including paramagnetic, spin-glass and ferromagnetic phases. This analysis has been extended to Ising-based RBMs in a number of directions with again paramagnetic, spin-glass, and ferromagnetic phases being uncovered \cite{Barra2017b}. For Ising RBMs averaged with a diagonal covariance matrix, when we have zero external field the expected global magnetizations (the quenched average values of the spin variables in the visible and hidden nodes) are zero, in both the paramagnetic and spin-glass phases  \cite{Barra2017b}. At these points of the phase diagram, the expected value of the replica-replica overlaps is zero for paramagnetic phase and non-zero for the spin-glass phase \cite{Barra2017b}. 

To investigate the phase behaviour of the RBM quenched free energy it is common to introduce the quantities,
\begin{eqnarray}
l & = & \frac{1}{nN}\mathbb{E}_{J,v,h}\left(
\sum_{i}\sum_{\nu}\langle n_{i}^{(\nu)}\rangle \right )\;\;,  \label{eq:Z.1} \\
p & = & \frac{1}{nM}\mathbb{E}_{J,v,h}\left( \sum_{a}\sum_{\nu}\langle m_{a}^{(\nu)}\rangle \right )\;\;, \label{eq:Z.2} \\
q & = & \frac{1}{n(n-1)N}\mathbb{E}_{J,v,h}\left( \sum_{i}\sum_{\nu, \nu'\neq\nu}\langle n_{i}^{(\nu)}n_{i}^{(\nu')}\rangle \right )\;\;, \label{eq:Z.3} \\
r & = & \frac{1}{n(n-1)M}\mathbb{E}_{J,v,h}\left( \sum_{a}\sum_{\nu, \nu'\neq\nu}\langle m_{a}^{(\nu)}m_{a}^{(\nu')}\rangle \right )\;\;. \label{eq:Z.4}
\end{eqnarray}

\noindent Here, the expectation $\mathbb{E}_{J,v,h}(\cdot)$ represents an expectation over the parameters of the Hamiltonian, whilst $\langle \cdot \rangle$ represents the expectation over the state of the lattice-gas occupancy variables at a fixed set of Hamiltonian parameters. The first two quantities in Eq.(\ref{eq:Z.1}) and Eq.(\ref{eq:Z.2}) clearly represent the average occupancy (magnetizations) within the visible and hidden layers, respectively. The last two quantities in Eq.(\ref{eq:Z.3}) and Eq.(\ref{eq:Z.4}) represent overlaps between different replicas, again within the visible and hidden layers respectively. For the Ising RBM, the equivalent quantities to $l,p,q$ and $r$ serve as order parameters characterizing the differences between paramagnetic, spin-glass and ferromagnetic phases.

For the lattice-gas RBM we can easily evaluate $l,p,q$ and $r$ by introducing additional source terms into the definition of $Z^{n}$. For example, to evaluate $q$ we can define,
\begin{equation}
	Z^{n}(g)\;=\;\sum_{{\bm n}^{(1)}, {\bm m}^{(1)}}\ldots \sum_{{\bm n}^{(n)}, {\bm m}^{(n)}}
	\exp \left [ -\sum_{\nu =1}^{n} H\left ( {\bm n}^{(\nu)}, {\bm m}^{(\nu)}\right ) \;+\;g\sum_{i=1}^{n}\sum_{\nu,\nu'}n_{i}^{(\nu)}n_{i}^{(\nu')}\right ]\;\; .
	\nonumber 
\end{equation}

\noindent From this definition of $Z^{n}(g)$ we can then evaluate,
\begin{equation}
	\mathbb{E}_{J,v,h}\left ( \sum_{i=1}^{n}\sum_{\nu,\nu'}\langle n_{i}^{(\nu)}n_{i}^{(\nu')}\rangle \right ) \;=\;
		\left . \frac{\partial}{\partial g} \mathbb{E}_{J,v,h}\left ( \log Z^{n}(g)\right ) \right |_{g=0}\;=\;	
	\left . \left . \frac{\partial}{\partial g}\frac{\partial}{\partial s} \mathbb{E}_{J,v,h}\left ( \left (Z^{n}(g)\right )^{s}\right )\right |_{s=0}\right |_{g=0} \;\; .
	\nonumber 
\end{equation}

\noindent Evaluation of $\mathbb{E}_{J,v,h}\left ( \left (Z^{n}(g)\right )^{s}\right )$ then proceeds in a similar fashion to the evaluation of $\mathbb{E}_{J,v,h}\left ( Z^{n}(g=0)\right )$ via Eq.(\ref{eq:B.7}) - Eq.(\ref{eq:B.21}), again using a replica symmetry ansatz. Doing so, and with similar expressions to help evaluate $l,p$ and $r$, we find in the thermodynamic limit $N, M\rightarrow\infty, M =\alpha N$ we have $l,p,q,r$ given by,
\begin{eqnarray}
l & = & I_{1}\left (\mu_{v},\sigma^{2}_{v},\frac{\alpha}{2}\tilde{\sigma}^{2}_{J}(r-p), \frac{\alpha}{2}\tilde{\sigma}^{2}_{J}r \right )\;\;, \label{eq:Z.5}\\
p & = & I_{1}\left (\mu_{h},\sigma^{2}_{h},\frac{1}{2}\tilde{\sigma}^{2}_{J}(q-l), \frac{1}{2}\tilde{\sigma}^{2}_{J}q \right )\;\;, \label{eq:Z.6}\\
q & = & I_{2}\left (\mu_{v},\sigma^{2}_{v},\frac{\alpha}{2}\tilde{\sigma}^{2}_{J}(r-p), \frac{\alpha}{2}\tilde{\sigma}^{2}_{J}r \right )\;\;, \label{eq:Z.7}\\
r & = & I_{2}\left (\mu_{h},\sigma^{2}_{h},\frac{1}{2}\tilde{\sigma}^{2}_{J}(q-l), \frac{1}{2}\tilde{\sigma}^{2}_{J}q \right )\;\;. \label{eq:Z.8}
\end{eqnarray}
\noindent With these relations we can write $\mathbb{E}(\log Z)$ in Eq.(\ref{eq:B.22}) in terms of $l,p,q,r$ as,
\begin{multline}
\lim_{\substack{N\rightarrow\infty\\M=\alpha N}} \frac{1}{N}\mathbb{E}\left (\log Z \right ) \; = \; \frac{\alpha}{2}\tilde{\sigma}^{2}_{J}\left ( qr\;-\; lp\right )
\;+\; I_{0,1}\left(\mu_{v}, \sigma^{2}_{v}, \frac{\alpha}{2}\tilde{\sigma}^{2}_{J}(r-p), \frac{\alpha}{2}\tilde{\sigma}^{2}_{J}r \right ) \\ 
+ \;\alpha I_{0,1}\left ( \mu_{h},\sigma^{2}_{h},\frac{1}{2}\tilde{\sigma}^{2}_{J}(q-l), \frac{1}{2}\tilde{\sigma}^{2}_{J}q\right)\;\;. \nonumber 
\end{multline}

\noindent with $l,p,q,r$ given by the solutions to Eq.(\ref{eq:Z.5})-Eq.(\ref{eq:Z.8}). Given the fact that the lattice-gas RBM Hamiltonian is equivalent to an Ising RBM Hamiltonian, we now ask if there are solutions to Eq.(\ref{eq:Z.5})-Eq.(\ref{eq:Z.8}) that have the same characteristics as the paramagnetic and spin-glass phases seen in the Ising RBM when averaged with a diagonal covariance? Specifically, we ask the questions, i) are there solutions to Eq.(\ref{eq:Z.5}) - Eq.(\ref{eq:Z.8}) which are equivalent to having zero global magnetizations and zero expectation for the Ising replica-replica overlaps, and, ii) are there solutions to Eq.(\ref{eq:Z.5}) - Eq.(\ref{eq:Z.8}) which are equivalent to having zero global magnetizations and non-zero expectation for the Ising replica-replica overlaps?

In the lattice-gas representation, the equivalent of zero global magnetization in the Ising representation would require $l=p=\frac{1}{2}$. From the form of the integrand defining $I_{1}(\mu, \sigma^{2}, x, y)$ in Eq.(\ref{eq:B.21}), we have from Eq.(\ref{eq:Z.5}) and Eq.(\ref{eq:Z.6}) that the criteria $l=p=\frac{1}{2}$ are equivalent to,
\begin{equation}
\mu_{v}\;-\;\frac{1}{4}\alpha\tilde{\sigma}^{2}_{J}(2r-1) \;= \; 0\;\;\;,\;\;\;
\mu_{h}\;-\;\frac{1}{4}\tilde{\sigma}^{2}_{J}(2q-1) \; = \; 0 \;\;.\label{eq:Z.10}
\end{equation}

\noindent We can interpret these constraints as defining the zero effective external field condition, as these are by construction the conditions under which we have zero global magnetization of the equivalent Ising RBM. It is interesting to note that only in the absence of heterogeneity of the spin-couplings or if the replica overlaps $q=r=\frac{1}{2}$, would the effective zero field conditions correspond to the conditions $\mu_{v} = \mu_{h} = 0$, which we would have naively anticipated from the mappings in Eq.(\ref{eq:B.1c}) (for the $\mu_{J}=0$ case).

Clearly, if the conditions in Eq.(\ref{eq:Z.10}) are met then we have zero global magnetization solutions to Eq.(\ref{eq:Z.5}) and Eq.(\ref{eq:Z.6}). To address whether we can also have zero expectation of the Ising replica-replica overlaps we note that under the lattice-gas to Ising isomorphism this corresponds to having $q=r=\frac{1}{4}$. Inserting the relations in Eq.(\ref{eq:Z.10}) into Eq.(\ref{eq:Z.7}) and also setting $q=r=\frac{1}{4}$ we find this then requires satisfaction of the constraint,
\begin{equation}
0\;=\;\int_{-\infty}^{\infty}dt\,\exp \left ( -\frac{t^{2}}{2\left (\sigma^{2}_{v}\;+\;\frac{\alpha}{4}\tilde{\sigma}^{2}_{J}\right)}\right )\tanh^{2} \left ( \frac{t}{2} \right)\;\;.
\label{eq:Z.11}
\end{equation}

\noindent If $\sigma^{2}_{v} > 0$ and $\tilde{\sigma}^{2}_{J} > 0 $, then the right-hand-side of Eq.(\ref{eq:Z.11}) is strictly positive and it is clear that the constraint in Eq.(\ref{eq:Z.11}) cannot be met. A similar constraint that is obtained from Eq.(\ref{eq:Z.10}) and Eq.(\ref{eq:Z.8}) and again setting $q=r=\frac{1}{4}$, also cannot be satisfied if $\sigma^{2}_{h} > 0$ and $\tilde{\sigma}^{2}_{J} > 0 $. From this we conclude that if there is heterogeneity in either the external field values or the spin-couplings, then there is no saddle-point solution for which $l=p=\frac{1}{2}$ and $q=r=\frac{1}{4}$. Conversely, if the external fields and the spin-couplings are all homogeneous, i.e., $\sigma^{2}_{v} = \sigma^{2}_{h} = \tilde{\sigma}^{2}_{J} =0$, then Eq.(\ref{eq:Z.11}) is satisfied. In contrast, analysis in \ref{sec:AppendixIsing} of the binary Ising RBM averaged with a diagonal covariance matrix reveals that zero global magnetization and zero expected Ising replica-replica overlaps is possible when $\tilde{\sigma}^{2}_{J} > 0$ provided there is no heterogeneity in the external field values. 

Similarly, we can investigate the existence of a solution for which the global magnetization in the equivalent Ising RBM is still zero, i.e., $l=p=\frac{1}{2}$, but the Ising replica-replica overlaps are positive, i.e., $q > \frac{1}{4}$ and $r > \frac{1}{4}$. The requirement that $l=p=\frac{1}{2}$ means that the constraints in Eq.(\ref{eq:Z.10}) would still apply. From Eq.(\ref{eq:Z.10}) we can obtain expressions for $q$ and $r$ and equate them to the expressions in Eq.(\ref{eq:Z.7}) and Eq.(\ref{eq:Z.8}). This defines a 3-dimensional surface within the space $\left (\mu_{v},\sigma^{2}_{v}, \mu_{h}, \sigma^{2}_{h}, \tilde{\sigma}^{2}_{J}\right )$ given by,
\begin{equation}
\frac{2\mu_{v}}{\alpha \tilde{\sigma}^{2}_{J}}\;+\;\frac{1}{2} \; = \; I_{2}\left ( 0,  \sigma^{2}_{h}, 0, \mu_{h}\;+\;\frac{1}{4}\tilde{\sigma}^{2}_{J}\right ) \;\;,\;\;
\frac{2\mu_{h}}{\tilde{\sigma}^{2}_{J}}\;+\;\frac{1}{2} \; = \; I_{2}\left (0,  \sigma^{2}_{v}, 0, \mu_{v}\;+\;\frac{1}{4}\alpha\tilde{\sigma}^{2}_{J}\right ) \label{eq:Z.12}\;\;,
\end{equation}

\begin{equation}
-\alpha\tilde{\sigma}^{2}_{J}\;\leq\; 4\mu_{v}\; \leq\; \alpha\tilde{\sigma}^{2}_{J} \;\;,\;\;
-\tilde{\sigma}^{2}_{J}\; \leq\; 4\mu_{h}\; \leq\; \tilde{\sigma}^{2}_{J}\;\;.
\label{eq:Z.13}
\end{equation}

\noindent Only on the 3-dimensional surface defined by Eq.(\ref{eq:Z.12}) and within the region defined by Eq.(\ref{eq:Z.13}) are we able to observe solutions for which $l=p=\frac{1}{2}$ and $p,q > \frac{1}{4}$. Similar conclusions are reached when $\mu_{J}\neq 0$. The derivation of the equations governing $l,p,q,r$ when $\mu_{J}\neq 0$ is given in \ref{sec:AppendixA1}. Again in contrast, analysis in \ref{sec:AppendixIsing} shows that for the Ising RBM with diagonal parameter covariance, zero global magnetization and zero expected Ising replica-replica overlaps are possible with the only restriction being that the external field values are zero on average.

Further differences between the behaviour of the lattice-gas RBM with diagonal covariance and the Ising RBM with diagonal covariance are highlighted when we look at the effective field, $\sum_{i=1}^{N}J_{ai}\langle n_{i}\rangle$, that acts upon the hidden node $a$. In the Ising RBM formulation (with diagonal covariance structure) the presence of a spin-glass phase is associated with a vanishing of the effective field \cite{Barra2017b}. We can examine the behaviour of the effective field by evaluating the expectation value $\mathbb{E}_{J,v,h}\left ( \sum_{i=1}^{N}J_{ai}\langle n_{i}\rangle \right )$, where $\mathbb{E}_{J,v,h}(\cdot)$ represents the expectation over the lattice-gas RBM Hamiltonian parameters. This is done in \ref{sec:AppendixA3} and we find,
\begin{equation}
\mathbb{E}_{J,v,h}\left (\sum_{i=1}^{N}J_{ai}\langle n_{i}\rangle \right ) \; = \; 
\tilde{\sigma}^{2}_{J}(l - q)p \label{eq:Z.14}\;\;.
\end{equation}

\noindent For finite, positive values of $\tilde{\sigma}^{2}_{J}, \sigma^{2}_{v}$ and $\sigma^{2}_{h}$ we have $l > q$ and $p > 0$, and so the expected value of the effective field is always positive (whilst the replica-symmetric ansatz is valid). Again, this behaviour can be contrasted with that of the binary Ising RBM with diagonal covariance where vanishing of the effective field is possible when we have heterogeneity in the spin-spin couplings - see \ref{sec:AppendixIsing}. 

Overall, from all of these observations just discussed we would conclude that the presence of parameter heterogeneity in conjunction with the correlations between Hamiltonian parameters has qualitatively changed the behaviour of the equivalent Ising RBM when compared to using a diagonal covariance ensemble for an Ising RBM.

Do we expect to see any other behaviour in the lattice-gas RBM free energy that we do not see in the Ising RBM? Analysis by Russo \cite{Russo1998} of the low-temperature behaviour of the lattice-gas version of the SK model reveals a first order phase transition, driven by the external field, from a low occupancy phase to a higher occupancy phase. At zero temperature the low occupancy phase becomes a zero-occupancy phase. From the definitions given in Eq.(\ref{eq:Z.5}) - Eq.(\ref{eq:Z.8}) it is clear that we cannot obtain a zero-occupancy phase $l=p=q=r=0$ within the replica-symmetric theory for any finite values of $\mu_{v}, \mu_{h}, \sigma^{2}_{v}, \sigma^{2}_{h}, \sigma^{2}_{J}$. However, the equivalent here to the zero temperature limit would be to simultaneously scale $|\mu_{v}|, |\mu_{h}|, \sigma^{2}_{v}, \sigma^{2}_{h}$ and $\sigma^{2}_{J}$ to $+\infty$. Such a limit is unlikely to represent a realistic scenario we would encounter when using a lattice-gas RBM. Consequently, whilst we suspect that the lattice-gas RBM would display a correspondingly similar phase behaviour to that identified by Russo in the lattice-gas SK model, we do not analyse that aspect of the lattice-gas RBM behaviour in this paper.

Of course, the validity of our analysis of the lattice-gas RBM behaviour and the expression in Eq.(\ref{eq:Z.14}) is dependent upon the validity of the replica symmetric ansatz. Where the replica-symmetric saddle-points are unstable we would expect to see replica symmetry breaking. The stability of the replica-symmetric saddle-points we examine in the Section \ref{sec:stability}.

\section{Asymptotic expansion of the saddle-point solution}
The leading order replica-symmetric calculation would appear to provide an accurate approximation to the expectation of $\log Z$ for the lattice-gas RBM. However, we obviously expect the replica-symmetry approximation to break down, particularly as the level or disorder, quantified by the variance $\tilde{\sigma}^{2}_{J}$, increases. Therefore, it is helpful to study the behaviour of the replica-symmetric saddle-point as $\tilde{\sigma}^{2}_{J}$ increases. Again, for simplicity, we consider only the scenario where $\mu_{J} = 0$. For large values of $\tilde{\sigma}^{2}_{J}$ the form of the equations in Eq.(\ref{eq:B.20d}) suggests asymptotic expansions, as $\tilde{\sigma}^{2}_{J}\rightarrow\infty$, of the form,
\begin{eqnarray}
\frac{\Delta z + \Delta \hat{z}}{\tilde{\sigma}^{2}_{J}} & = & \frac{c_{1}^{(1)}}{\tilde{\sigma}_{J}}\;+\;\frac{c_{1}^{(2)}}{\tilde{\sigma}^{2}_{J}}
\;+\;{\mathcal O}\left (\tilde{\sigma}^{-3}_{J} \right )\;\;, \label{eq:C.13}\\
\frac{z_{1} + \hat{z}_{1}}{\tilde{\sigma}^{2}_{J}} & = & c_{2}^{(0)}\;+\;\frac{c_{2}^{(1)}}{\tilde{\sigma}_{J}}\;+\;\frac{c_{2}^{(2)}}{\tilde{\sigma}^{2}_{J}}
\;+\;{\mathcal O}\left (\tilde{\sigma}^{-3}_{J} \right )\;\;, \label{eq:C.14}\\
\frac{\Delta z - \Delta \hat{z}}{\tilde{\sigma}^{2}_{J}} & = & \frac{c_{3}^{(1)}}{\tilde{\sigma}_{J}}\;+\;\frac{c_{3}^{(2)}}{\tilde{\sigma}^{2}_{J}}
\;+\;{\mathcal O}\left (\tilde{\sigma}^{-3}_{J} \right )\;\;, \label{eq:C.15}\\
\frac{z_{1} - \hat{z}_{1}}{\tilde{\sigma}^{2}_{J}} & = & c_{4}^{(0)}\;+\;\frac{c_{4}^{(1)}}{\tilde{\sigma}_{J}}\;+\;\frac{c_{4}^{(2)}}{\tilde{\sigma}^{2}_{J}}
\;+\;{\mathcal O}\left (\tilde{\sigma}^{-3}_{J} \right )\;\;. \label{eq:C.16}
\end{eqnarray}

\noindent Using these asymptotic forms with the relations in Eq.(\ref{eq:B.20d}) we find a set of self-consistent equations for the leading order expansion coefficients, $c_{1}^{(1)}, c_{2}^{(0)},c_{3}^{(1)}, c_{4}^{(0)}$,
\begin{equation}
c_{1}^{(1)} \; = \; -\frac{1}{2}\frac{\exp \left ( -\left( c_{3}^{(1)}\right )^{2}/4c_{4}^{(0)}\right )}{\sqrt{4\pi c_{4}^{(0)}}}\;\;\;,\;\;\;
c_{2}^{(0)} \; = \; \frac{1}{4} \erfc \left( \frac{1}{2} \frac{c_{3}^{(1)}}{\sqrt{c_{4}^{(0)}}}\right) \;\;, \label{eq:C.17a}
\end{equation}
\begin{equation}
c_{3}^{(1)} \; = \; -\frac{\alpha}{2}\frac{\exp \left ( -\left( c_{1}^{(1)}\right )^{2}/4c_{2}^{(0)}\right )}{\sqrt{4\pi c_{2}^{(0)}}}\;\;\;,\;\;\;
c_{4}^{(0)} \; = \; \frac{\alpha}{4} \erfc \left( \frac{1}{2} \frac{c_{1}^{(1)}}{\sqrt{c_{2}^{(0)}}}\right)\;\;.\label{eq:C.17b}
\end{equation}

\noindent These equations are easily solved numerically, via iteration. The next-to-leading order expansion coefficients are then obtained via the matrix equation,

\begin{equation}
\begin{bmatrix}  c_{2}^{(1)} \\ c_{4}^{(1)} \\ c_{1}^{(2)} \\ c_{3}^{(2)}\end{bmatrix}
\;=\; 
{\bm W} ^{-1} 
\begin{bmatrix} 
c_{1}^{(1)}(1 - \mu_{v}) \\ c_{3}^{(1)}(1 - \mu_{h}) \\ \frac{1}{2}c_{1}^{(1)}c_{3}^{(1)}\mu_{v}/c_{4}^{(0)} \\ \frac{1}{2}c_{1}^{(1)}c_{3}^{(1)}\mu_{h}/c_{2}^{(0)}
\end{bmatrix}\;\;,
\label{eq:C.19}
\end{equation}

\noindent where the matrix ${\bm W}$ is given by,
\begin{equation}
{\bm W}
\;=\; 
  \begin{bmatrix}  1 & \frac{c_{1}^{(1)}c_{3}^{(1)}}{2c_{4}^{(0)}} & 0 & -c_{1}^{(1)} \\
                   \frac{c_{1}^{(1)}c_{3}^{(1)}}{2c_{2}^{(0)}} & 1 & -c_{3}^{(1)} & 0 \\
                   0 & c_{1}^{(1)}\left ( \frac{1}{2c_{4}^{(0)}} - \frac{\left ( c_{3}^{(1)}\right )^{2}}{4\left ( c_{4}^{(0)}\right )^{2}} \right ) & 1 & \frac{c_{1}^{(1)}c_{3}^{(1)}}{2c_{4}^{(0)}} \\
                   c_{3}^{(1)}\left ( \frac{1}{2c_{2}^{(0)}} - \frac{\left ( c_{1}^{(1)}\right )^{2}}{4\left ( c_{2}^{(0)}\right )^{2}} \right ) & 0 & \frac{c_{1}^{(1)}c_{3}^{(1)}}{2c_{2}^{(0)}} & 1 
  \end{bmatrix} \;\;.
\nonumber
\end{equation}

\noindent Expressions for higher-order expansion coefficients can similarly be obtained in terms of the lower order coefficients, although we have not done so. Note also that we considered fixed values for $\mu_{v}, \mu_{h}, \sigma^{2}_{v}, \sigma^{2}_{h}$ - that is, they do not scale with $\tilde{\sigma}^{2}_{J}$ in any way. Given the expansion coefficients in Eq.(\ref{eq:C.17a})-Eq.(\ref{eq:C.19}), it is possible to provide asymptotic expansion of other quantities of interest. For example, the replica approximation to $\mathbb{E}\left ( \log Z\right)$ can be written as,

\begin{equation}
\lim_{\substack{N\rightarrow\infty\\M=\alpha N}} \frac{1}{N}\mathbb{E}\left (\log Z \right )
\;=\; -\left [ 4\tilde{\sigma}_{J}\left ( c_{4}^{(0)}c_{1}^{(1)}\;+\;c_{2}^{(0)}c_{3}^{(1)}\right )
 \;-\;\left ( 2\mu_{v}c_{2}^{(0)} \;+\; 2\mu_{h}c_{4}^{(0)}\;+\;c_{1}^{(1)}c_{3}^{(1)}\right )\;+\;{\mathcal O}(\tilde{\sigma}^{-1}_{J})\right ]\;\;,
\label{eq:C.20}
\end{equation}

\noindent from which we can see that the expectation of $\log Z$ is (to leading order) linear in $\tilde{\sigma}_{J}$. Interestingly, due to cancellation of a number of terms, the next to leading order contribution in Eq.(\ref{eq:C.20}) still only depends upon the leading order expansion coefficients $c_{2}^{(0)}, c_{4}^{(0)}, c_{1}^{(1)}$ and $c_{3}^{(1)}$. The expression in Eq.(\ref{eq:C.20}) provides a means to quickly approximate the partition function of single lattice-gas RBM instance once suitable values for $\alpha,\tilde{\sigma}^{2}_{J},\mu_{v}, \mu_{h}$ have been calculated for that single instance.

\section{Replica symmetry stability criterion \label{sec:stability}}
The replica-symmetric saddle-point becomes unstable when any of the eigenvalues of the Hessian, ${\bm H}$, of the action $S$ in Eq.(\ref{eq:B.14}) evaluated at the replica-symmetric saddle-point, become positive. The Hessian ${\bm H}$ is a $2n^{2}\times 2n^{2}$ matrix and so there are $2n^{2}$ eigenvalues, although they are highly degenerate so there are only a few distinct values. Calculation of the Hessian ${\bm H}$, its eigenvectors and corresponding eigenvalues is given in \ref{sec:AppendixB}. From the analysis in \ref{sec:AppendixB} we find that the Hessian ${\bm H}$ is negative definite only if, 
\begin{multline}
1 \; > \; \frac{\alpha}{256}\frac{\tilde{\sigma}^{4}_{J}}{2\pi}
\left (\frac{1}{\sqrt{\sigma^{2}_{v}+2z_{1}-2\hat{z}_{1}}}\int_{-\infty}^{\infty}dx\,
\exp\left ( -\frac{(x-\mu_{v}+\Delta z-\Delta \hat{z})^{2}}{2(\sigma^{2}_{v} + 2z_{1} - 2\hat{z}_{1})}\right )\sech^{4}\left (\frac{x}{2} \right )\right ) \\
\times
\left (\frac{1}{\sqrt{\sigma^{2}_{h}+2z_{1}+2\hat{z}_{1}}}\int_{-\infty}^{\infty}dx\,
\exp\left ( -\frac{(x-\mu_{h}+\Delta z+\Delta \hat{z})^{2}}{2(\sigma^{2}_{h} + 2z_{1} + 2\hat{z}_{1})}\right )\sech^{4}\left (\frac{x}{2} \right )\right )\;\;.
\label{eq:E.1}
\end{multline}

\noindent Unsurprisingly, the stability in Eq.(\ref{eq:E.1}) is very similar in mathematical form to the criterion determining the location of the Almeida-Thouless line in the SK model \cite{AlmeidaThouless1978}, that indicates a breakdown of replica-symmetry in the SK model as we move into a spin-glass phase. The stability criterion in Eq.(\ref{eq:E.1}) can be solved, say, for $\tilde{\sigma}^{2}_{J}$ as a function of $\alpha$ to determine the point at which the level of disorder in the interactions $J_{ai}$ is such that replica-symmetry is broken. From the of the saddle-point equations in Eq.(\ref{eq:B.20d}) it is trivial to verify that the right-hand-side of Eq.(\ref{eq:E.1}) is invariant under the transformations $\alpha\rightarrow\alpha^{-1}\;,\;\tilde{\sigma}^{2}_{J}\rightarrow\alpha\tilde{\sigma}^{2}_{J}$. Denoting the value $\tilde{\sigma}^{2}_{J}$ at the stability threshold by $\tilde{\sigma}^{2}_{J,\star}$, we then have,
\begin{equation}
\tilde{\sigma}^{2}_{J,\star}(\alpha^{-1})\;=\;\alpha\tilde{\sigma}^{2}_{J,\star}(\alpha)\;\;.
\label{eq:E.3}
\end{equation} 

\noindent On first impression, Eq.(\ref{eq:E.1}) above suggests that the right-hand-side will increase, and the stability criterion exceeded, as $\tilde{\sigma}^{2}_{J}\rightarrow\infty$. This suggests using the the asymptotic expansions of the saddle-point location given in Eq.(\ref{eq:C.13})-Eq.(\ref{eq:C.16}) to approximately locate the stability threshold $\tilde{\sigma}^{2}_{J,\star}$. Doing so we obtain the replica stability criterion as,
\begin{equation}
1 \;=\; \frac{\tilde{\sigma}^{2}_{J,\star}}{64} c_{1}^{(1)}c_{3}^{(1)}
\times \left (\int_{-\infty}^{\infty}dx\,\sech^{4}\left(\frac{x}{2} \right ) \right )^{2}\times \left (1+{\mathcal O}\left (\tilde{\sigma}^{-1}_{J,\star} \right ) \right )\;\;,
\nonumber 
\end{equation}

\noindent and so to first approximation, we find that the limit of stability of replica symmetry is located (for $\mu_{J}=0$) at,
\begin{equation}
\tilde{\sigma}^{2}_{J,\star} \; \simeq \; \frac{9}{c_{1}^{(1)}c_{3}^{(1)}}\;\;.
\label{eq:E.5}
\end{equation}

\begin{figure*}
	\begin{center}		
		\scalebox{0.7}{%
			\includegraphics*[bb=0 0 432 288]{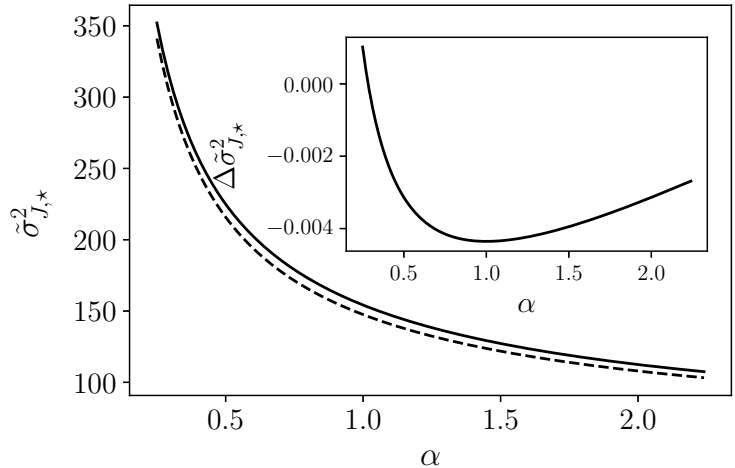}}
		\caption{Plot of the estimates of the replica symmetry stability threshold $\tilde{\sigma}^{2}_{J,\star}$ as a function of $\alpha$. The main plot shows the leading order (solid line) and next-to-leading order (dashed line) asymptotic estimates (as $\tilde{\sigma}^{2}_{J,\star}\rightarrow\infty$). The inset plot shows the difference, $\Delta \tilde{\sigma}^{2}_{J,\star}$, between exact numerical solution of the stability criterion in Eq.(\ref{eq:E.1}) and the next-to-leading order estimate from Eq.(\ref{eq:E.6}) when $\mu_{v} = \mu_{h} = 0$.}
		\label{fig:Figure5}
	\end{center}
\end{figure*}

\noindent Note that from Eq.(\ref{eq:C.17a}) and  Eq.(\ref{eq:C.17b}) the expansion coefficients $c_{1}^{(1)}$ and $c_{3}^{(1)}$ are functions only of $\alpha$, and so the right-hand-side of Eq.(\ref{eq:E.5}) is simply a function of $\alpha$. Again it is trivial to confirm that this leading order approximation to $\tilde{\sigma}^{2}_{J,\star}$ still satisfies the symmetry relation in Eq.(\ref{eq:E.3}). The plot in Figure \ref{fig:Figure5} shows how the leading order asymptotic estimate of $\tilde{\sigma}^{2}_{J,\star}$ varies with $\alpha$.

Continuing the expansion of the right-hand-side of Eq.(\ref{eq:E.1}) using the asymptotic expansions of the saddle-point location given in Eq.(\ref{eq:C.17a})-Eq.(\ref{eq:C.19}), we obtain the next-to-leading order approximation for $\tilde{\sigma}^{2}_{J,\star}$,
\begin{equation}
\tilde{\sigma}^{2}_{J,\star} \; \simeq \;
\frac{9}{c_{1}^{(1)}c_{3}^{(1)}\left [1+\left ( \frac{c_{1}^{(2)}}{c_{1}^{(1)}} + \frac{c_{3}^{(2)}}{c_{3}^{(1)}}\right )\tilde{\sigma}^{-1}_{J,\star} \right ]}\;\;.
\label{eq:E.6}
\end{equation}

\noindent Eq.(\ref{eq:E.6}) must be solved iteratively, but we have found that iteration converges quickly when starting from the value of $\tilde{\sigma}^{2}_{J,\star}$ calculated from Eq.(\ref{eq:E.5}). The extra expansion coefficients $c_{1}^{(2)}$ and $c_{3}^{(2)}$ in Eq.(\ref{eq:C.19}) are functions of $\alpha, \mu_{v}, \mu_{h}$ only. The next-to-leading order estimate of $\tilde{\sigma}^{2}_{J,\star}$, for $\mu_v = \mu_{h}=0$, is also shown in Fig.\ref{fig:Figure5}, as well as the difference (which we denote by $\Delta\tilde{\sigma}^{2}_{J,\star}$) between  $\tilde{\sigma}^{2}_{J,\star}$ obtained from exact numerical solution of Eq.(\ref{eq:E.1}) and the next-to-leading order estimate obtained from Eq.(\ref{eq:E.6}).

\section{Equivalence of replica stability threshold and Bethe approximation stability threshold}

We also expect instability to replica symmetry-breaking to impact any message-passing algorithm based on the Bethe approximation \cite{Rivoire2004}. For their message-passing process Huang and Toyoizumi \cite{HuangToyoizumi2015} also developed a stability measure, $S(t)$, given by,
\begin{equation}
S(t)\;=\;\sum_{(i,a)}{\mathcal V}_{i\rightarrow a}(t)\;\;,
\label{eq:E.8}
\end{equation}

\noindent where the sum in Eq.(\ref{eq:E.8}) is taken over all connected pairs of spins $(i,a)$. The messages ${\mathcal V}_{i\rightarrow a}$ are calculated from,
\begin{equation}
{\mathcal V}_{i\rightarrow a}(t) \; = \; \frac{(1-m_{i\rightarrow a}^{2}(t))^{2}}{4}\sum_{b\in \partial i\backslash a}{\mathcal P}_{b\rightarrow i}(t)\times \left [\tanh\left (\Gamma_{b\rightarrow i}(t)\right )\;-\;\tanh \left ( \Gamma_{b\rightarrow i}(t) -2w_{bi} \right )\right ]^{2}\;\;,
\label{eq:E.9a} 
\end{equation}
\begin{equation}
\Gamma_{b\rightarrow i}(t) \; \equiv \; \psi_{b}\;+\;G_{b\rightarrow i}(t)\;+\;w_{bi}\;\;\;,\;\;\;
{\mathcal P}_{b\rightarrow i}(t) \; \equiv \; \sum_{j\in\partial b\backslash i}w_{bj}^{2}{\mathcal V}_{j\rightarrow b}(t-1)\;\;,
\label{eq:E.9b} 
\end{equation}
\begin{equation}
G_{b\rightarrow i}(t) \; = \; \sum_{j\in\partial b\backslash i}w_{bj}m_{j\rightarrow b}(t-1)\;\;\;,\;\;\;
m_{i\rightarrow a}(t) \; = \; \tanh \left ( \phi_{i}\;+\;\sum_{b\in \partial i\backslash a}u_{b\rightarrow i}(t-1) \right )\;\;,
\label{eq:E.9c} 
\end{equation}
\begin{equation}
u_{b\rightarrow i}(t) \; = \; \frac{1}{2}\log \frac{\cosh \left ( \psi_{b}\;+\;G_{b\rightarrow i}(t-1)\;+\;w_{bi}\right )}{\cosh \left ( \psi_{b}\;+\;G_{b\rightarrow i}(t-1)\;-\;w_{bi}\right )}\;\;.
\label{eq:E.9d}
\end{equation}

\begin{figure*}
	\begin{center}		
		\scalebox{0.70}{%
			\includegraphics*[bb=0 0 432 288]{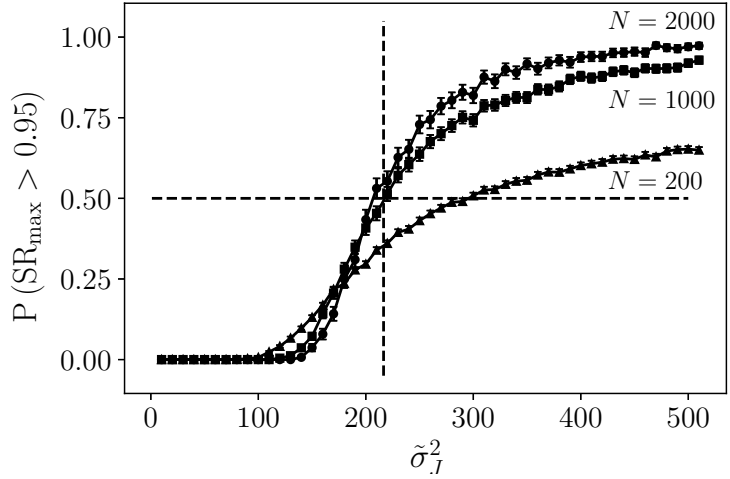}}
		\caption{Plots of the proportion of RBM instances for which ${\rm SR_{max}} > 0.95$, plotted against $\tilde{\sigma}_{J}^{2}$. The error bars plotted with each point represent estimates of the 95\% confidence intervals. The different plots are for different visible layer sizes; $N=200$ (triangle), $N=1000$ (square), $N=2000$ (circle). For all plots we have set $\alpha=0.5$. The vertical dashed line indicates the replica-symmetry stability threshold, $\tilde{\sigma}^{2}_{J,\star}$ estimated from Eq.(\ref{eq:E.6}). The horizontal dashed line is at $P({\rm SR_{max}}> 0.95) = 0.5$, and serves as a guide to the eye.}
		\label{fig:Figure6}
	\end{center}
\end{figure*}

\noindent Here $\partial i\backslash a$ denotes all the hidden layer nodes, except $a$, that are connected to the visible layer node $i$. Likewise, $\partial b\backslash i$ denotes all the visible layer nodes, except $i$, that are connected to the hidden layer node $b$. Stability of the message-passing process is indicated by convergence $S(t)\rightarrow 0$ as $t\rightarrow\infty$, whilst divergence $S(t)\rightarrow +\infty$ as $t\rightarrow\infty$ indicates instability. We might therefore expect to observe divergence in $S(t)$ for individual instances of the RBM Hamiltonian as we approach the threshold of stability for replica-symmetry. Similar to Huang and Toyoizumi we track a number of metrics derived from $S(t)$ as we iterate the message-passing algorithm when calculating $\log Z$ for a specific RBM instance. Ideally, we would like to evaluate $S(t)$ in the limit $t\rightarrow\infty$, but instead we iterate for a fixed, but large number of iterations which is sufficient for convergence to occur in the majority of instances, in this case $t=100$. To determine if divergence of $S(t)$ is occurring we compare the time-point within $t\in {1,\ldots,100}$ at which $S(t)$ reaches it largest observed value, and express it as a ratio to the number of the iterations, i.e., 100. We denote this ratio as ${\rm SR_{max}}$. For example, if ${\rm SR_{max}}=1$ this indicates that the maximum value of $S(t)$ was observed at the end of the sequence of 100 iterations, and so the sequence of values of $S(t)$ was still growing. In contrast a value of ${\rm SR_{max}}=0.1$ indicates that by $t=100$ convergence has essentially occurred. We then compare ${\rm SR_{max}}$ to a fixed threshold, e.g., 0.95, and evaluate the proportion of RBM instances that have $SR_{max}$ above this threshold. We regard the message-passing for this proportion of RBM instances to have essentially failed to converge. We study this proportion as a function of $\tilde{\sigma}^{2}_{J}$. We expect that this proportion will display the same qualitative behaviour as the stability of the replica-symmetry calculation, and so we computationally study the behaviour of this proportion in the neighbourhood of $\tilde{\sigma}^{2}_{J}=\tilde{\sigma}^{2}_{J,\star}$ for $\alpha=0.5, \mu_{v}=\mu_{h}=0$ and $\sigma^{2}_{v}=\sigma^{2}_{h}=0.1$, and where $\tilde{\sigma}^{2}_{J,\star}$ is calculated from Eq.(\ref{eq:E.6}). The choice of threshold (0.95 in this case), or the number of message-passing iterations is largely unimportant provided they are both sensibly large. Figure \ref{fig:Figure6} shows the proportion of RBM instances with ${\rm SR_{max}} > 0.95$, evaluated from running the message-passing algorithm of Huang and Toyoizumi over a large number of RBM instances. From Fig.\ref{fig:Figure6} we can see that the proportion of RBM instances that display instability (as defined by ${\rm SR_{max}} > 0.95$) increases markedly around $\tilde{\sigma}^{2}_{J} = \tilde{\sigma}^{2}_{J,\star}$. As we increase $N$, the width of the transition region from low values of $P({\rm SR_{max}} > 0.95)$ to high values of $P({\rm SR_{max}} > 0.95)$ narrows and appears to be centred close to $\tilde{\sigma}^{2}_{J,\star}$, providing numerical evidence that the limit of stability of the replica-symmetry saddle-point and the limit of stability of the Bethe approximation based message passing solution coincide.

We can formally establish the link between replica-symmetry breakdown in RBMs and divergence of $S(t)$ by determining the criterion for $S(t)$ divergence in the thermodynamic limit. Thus we consider the behaviour of the ratio $S(t)/S(t-1)$ in the limit $N\rightarrow\infty$ with $M = \alpha N$. From Eq.(\ref{eq:E.8}) and (\ref{eq:E.9a}) we can write,
\begin{equation}
	\frac{1}{MN}S(t)\;=\;\frac{M}{MN}\sum_{(i,a)} \frac{(1-m_{i\rightarrow a}^{2}(t))^{2}}{4}\frac{1}{M}\sum_{b\in \partial i\backslash a}{\mathcal P}_{b\rightarrow i}(t)\times \left [\tanh\left (\Gamma_{b\rightarrow i}(t)\right )\;-\;\tanh \left ( \Gamma_{b\rightarrow i}(t) -2w_{bi} \right )\right ]^{2}\;\;.
	\label{eq:E.10}
\end{equation}

\noindent The inner and outer summations in Eq.(\ref{eq:E.10}) are effectively sample means and so will converge in the thermodynamic limit to their expectation values, with vanishing variance. Here, the distribution of summand values, such as ${\mathcal P}_{b\rightarrow i}(t)\times \left [\tanh\left (\Gamma_{b\rightarrow i}(t)\right )\;-\;\tanh \left ( \Gamma_{b\rightarrow i}(t) -2w_{bi} \right )\right ]^{2}$ is across the population of different visible and hidden nodes $b$ and $i$, and arises because of the variation of the Hamiltonian parameters $w_{bi}, \phi_{i}$ and $\psi_{b}$ across those nodes. The quantity ${\mathcal P}_{b\rightarrow i}(t)/N$ itself represents a sample average, and so overall for large $N$ we can approximate to leading order,
\begin{multline}
	\frac{1}{MN}S(t) \;\simeq\; M \times \left [ \frac{1}{MN} \sum_{(i,a)} \frac{(1-m_{i\rightarrow a}^{2}(t))^{2}}{4}\right ] \times \left [ \frac{1}{MN} \sum_{(i,b)}{\mathcal P}_{b\rightarrow i}\right] \\ 
	\times \left [ \frac{1}{MN} \sum_{(i,b)}  \left [\tanh\left (\Gamma_{b\rightarrow i}(t)\right )\;-\;\tanh \left ( \Gamma_{b\rightarrow i}(t) -2w_{bi} \right )\right ]^{2} \right ]\;\;.
	\nonumber
\end{multline}

\noindent As we approach the thermodynamic limit we will have vanishing correlation between $w^{2}_{bj}$ and $\left ( 1 - m^{2}_{j\rightarrow b}(t-1)\right )^{2}$ and hence between $w^{2}_{bj}$ and ${\mathcal V}_{j\rightarrow b}(t-1)$, and so we can approximate to leading order,
\begin{flalign}
	\frac{1}{MN}\sum_{(i,b)}{\mathcal P}_{b\rightarrow i}\;=\;\frac{1}{MN}\sum_{(b,i)}\sum_{j\in\partial b\backslash i}w^{2}_{bj}{\mathcal V}_{j\rightarrow b}(t-1)  &\simeq \; N\times \frac{1}{MN}\sum_{(b,j)}w^{2}_{bj}\times \frac{1}{MN}\sum_{(b,j)} {\mathcal V}_{j\rightarrow b}(t-1) \nonumber \\
	&\simeq \; \tilde{\sigma}^{2}_{w}\frac{1}{MN}S(t-1) \;\;. \nonumber
\end{flalign}

\noindent Here we are only considering the scenario where $\mathbb{E}(w_{bi})=0$ and $\sigma^{2}_{w}$ is the variance of $w_{ai}$ for all $a$ and $i$. Similar to $\sigma^{2}_{J}$ we consider $\sigma^{2}_{w}$ to scale with $N$ as $\sigma^{2}_{w} = \tilde{\sigma}^{2}_{w}/N$. With this, we have,
\begin{equation}
\frac{S(t)}{S(t-1)}\;\simeq\; M \tilde{\sigma}^{2}_{w}\left [ \frac{1}{MN} 
\sum_{(i,a)} \frac{(1-m_{i\rightarrow a}^{2}(t))^{2}}{4} \right ] \times \left [\frac{1}{MN}\sum_{(i,b)}\left [\tanh\left (\Gamma_{b\rightarrow i}(t)\right )\;-\;\tanh \left ( \Gamma_{b\rightarrow i}(t) -2w_{bi} \right )\right ]^{2}\right ]\;\;.
\label{eq:E.13}
\end{equation}

\noindent With $\sigma^{2}_{w}\rightarrow 0^{+}$ as $N\rightarrow\infty$, values of $w_{bi}$ will typically be small, and so we can replace the last summand in Eq.(\ref{eq:E.13}) by its leading order Taylor expansion. If we also use the identity $\sech^2(x)=1-\tanh^{2}(x)$, then we obtain,
\begin{multline}
\frac{S(t)}{S(t-1)}\;\simeq\; M \tilde{\sigma}^{2}_{w}\left [\frac{1}{MN}
\sum_{(i,a)} \sech^{4}\left( \phi_{i}\;+\;\sum_{b'\in\partial i\backslash a}u_{b'\rightarrow i}(t-1)\right) \right ] \\
\times \left [ \frac{1}{MN}\sum_{(i,b)}\left ( w^{2}_{bi}\sech^{4}\left (\psi_{b}\;+\; G_{b\rightarrow i}(t)\right )\;+\;{\mathcal O}\left ( w_{bi}^{4}\right ) \right )\right ] \;\;.
\label{eq:E.14}
\end{multline}

\noindent As $G_{b\rightarrow i}(t)$ contains messages $m_{j\rightarrow b}$ except from the edge between $b$ and $i$, we expect any correlation between $w_{bi}$ and $G_{b\rightarrow i}(t)$ to be vanishingly small as $N\rightarrow\infty$. We therefore, to leading order, further approximate,
\begin{equation}
	\frac{1}{MN}\sum_{(i,b)}\left ( w^{2}_{bi}\sech^{4}\left (\psi_{b}\;+\; G_{b\rightarrow i}(t)\right ) \right ) \; \simeq \;
	\tilde{\sigma}^{2}_{J} \frac{1}{MN}\sum_{(i,b)}\sech^{4}\left (\psi_{b}\;+\; G_{b\rightarrow i}(t)\right )\;\;. \nonumber 
\end{equation}
 
On passing to the thermodynamic limit we can replace the main summations in Eq.(\ref{eq:E.14}) by corresponding integrals. From the Central Limit Theorem, we can also replace $\phi_{i} + \sum_{b'\in\partial i\backslash a}u_{b'\rightarrow i}$ and $\psi_{b} + G_{b\rightarrow i}$ by Gaussian random variables. Here we assume the distributions of those values are self-averaging in the thermodynamic limit, in that the distributions of values over nodes $i,b$ for a single RBM instance tend to the distributions of values over RBM instances for fixed choice of $i$ and $b$. From this we obtain,
\begin{equation}
\lim_{\substack{N\rightarrow\infty\\M=\alpha N}}\frac{S(t)}{S(t-1)} \; = \; \alpha\tilde{\sigma}^{4}_{w}\int_{-\infty}^{\infty}\sech^{4}(x)P(x)dx \; \times \; \int_{-\infty}^{\infty}\sech^{4}(y)P(y)dy\;\;,
\nonumber
\end{equation}
\begin{equation}
x \;\sim\; {\mathcal N}\left ( \mu_{x}, \sigma^{2}_{x}\right ) \;,\; y \;\sim\; {\mathcal N}\left(\mu_{y}, \sigma^{2}_{y}\right )\;\;, \nonumber
\end{equation}
\noindent with $\mu_{x}, \mu_{y}, \sigma^{2}_{x}, \sigma^{2}_{y}$ given by,
\begin{equation}
\mu_{x} \; = \; \lim_{\substack{N\rightarrow\infty\\M=\alpha N}}\mathbb{E}_{w,\phi,\psi}\left( \phi_{i}\;+\;\sum_{b\in\partial i\backslash a} u_{b\rightarrow i}\right) \;,\;
\sigma^{2}_{x} \; = \; \lim_{\substack{N\rightarrow\infty\\M=\alpha N}}{\rm Var}_{w,\phi,\psi}\left( \phi_{i}\;+\;\sum_{b\in\partial i\backslash a} u_{b\rightarrow i}\right)\;\;\;\forall i,a\;\;,
\label{eq:E.18}
\end{equation}
\begin{equation}
\mu_{y} \; = \; \lim_{\substack{N\rightarrow\infty\\M=\alpha N}}\mathbb{E}_{w,\phi,\psi}\left( \psi_{b}\;+\;\sum_{j\in\partial b\backslash i} w_{bj}m_{j\rightarrow b}\right) \;,\;
\sigma^{2}_{y} \; = \; \lim_{\substack{N\rightarrow\infty\\M=\alpha N}}{\rm Var}_{w,\phi,\psi}\left( \psi_{b}\;+\;\sum_{j\in\partial b\backslash i} w_{bj}m_{j\rightarrow b}\right)\;\;\;\forall i,b\;\;.
\label{eq:E.20}
\end{equation}

\noindent Eq.(\ref{eq:E.18}) and Eq.(\ref{eq:E.20}) tell us that the means and variances of $x$ and $y$ are inter-dependent. It is those inter-dependencies we will now evaluate. We can simplify the arguments to the expectation operators in Eq.(\ref{eq:E.18}) and Eq.(\ref{eq:E.20}) using the relations in Eq.(\ref{eq:E.9b}) - Eq.(\ref{eq:E.9d}), and in particular expanding $u_{b\rightarrow i}$ in powers of $w_{bi}$. If we then denote,
\begin{eqnarray}
x_{ia} & = & \phi_{i}\;+\;\sum_{b\in\partial i\backslash a} w_{bi}\tanh (y_{bi})\;\;,\nonumber \\
y_{bi} & = & \psi_{b}\;+\;\sum_{j\in\partial b\backslash i}w_{bj}\tanh(x_{jb})\;\;,\nonumber
\end{eqnarray}

\noindent we obviously have in the thermodynamic limit, $\mu_{x} = \mathbb{E}_{w,\phi,\psi}(x_{ia})$, for any choice of $i$ and $a$. Similar expressions for $\mu_{x}, \sigma^{2}_{x}, \sigma^{2}_{y}$ in terms of $x_{ia}$ and $y_{bi}$ can be written. Substituting $w_{bi}$, $\psi_{b}$ and $\phi_{i}$ by their lattice-gas equivalents, $x_{ia}$ and $y_{bi}$ become,
\begin{eqnarray}
x_{ia} & = & \frac{1}{2}v_{i}\;+\;\frac{1}{4}J_{ai}\;+\;\frac{1}{2}\sum_{b\in\partial i\backslash a}J_{bi}r(y_{bi})\;\equiv\; \tilde{x}_{ia}\;+\;\frac{1}{4}J_{ai}\;\;,\label{eq:E.23} \\
y_{bi} & = & \frac{1}{2}h_{b}\;+\;\frac{1}{4}J_{bi}\;+\;\frac{1}{2}\sum_{j\in\partial b\backslash i}J_{bj}r(x_{jb})\;\equiv\; \tilde{y}_{bi}\;+\;\frac{1}{4}J_{bi}\;\;,\label{eq:E.24}
\end{eqnarray}

\noindent where $r(t) = e^{2t}/(1+e^{2t})$, and the right-hand-side of Eq.(\ref{eq:E.23}) and Eq.(\ref{eq:E.24}) are used to define $\tilde{x}_{ia}$ and $\tilde{y}_{bi}$ respectively. From these definitions, we have, 
\begin{equation}
\mu_{x} \;=\;\lim_{\substack{N\rightarrow\infty\\M=\alpha N}}\mathbb{E}_{J,v,h}(\tilde{x}_{ia})\;\;\;\;,\;\;\;\;
\sigma^{2}_{x} \;=\;
\lim_{\substack{N\rightarrow\infty\\M=\alpha N}} {\rm Var}_{J,v,h}(\tilde{x}_{ia})\;\;. \nonumber
\end{equation}

\noindent The expectation $\mathbb{E}_{J,v,h}(\tilde{x}_{ia})$ can be evaluated by,
\begin{eqnarray}
\mathbb{E}_{J,v,h}(\tilde{x}_{ia}) &=& \frac{1}{2}\mu_{v}\;+\;\frac{1}{2}\sum_{b\in\partial i\backslash a}\mathbb{E}_{J,\tilde{y}}\left (J_{bi}r(\tilde{y}_{bi}\;+\;\frac{1}{4}J_{bi}) \right )\nonumber \\
& = & \frac{1}{2}\mu_{v}\;+\;\frac{1}{2}\sum_{b\in\partial i\backslash a}\mathbb{E}_{J,\tilde{y}}\left (J_{bi}r(\tilde{y}_{bi})\;+\;\frac{1}{4}J^{2}_{bi}r'(\tilde{y}_{bi})\;+\;{\mathcal O}(J^{3}) \right )\nonumber \\
& \rightarrow &  \frac{1}{2}\mu_{v}\;+\;\frac{\alpha}{8}\tilde{\sigma}^{2}_{J}\lim_{\substack{N\rightarrow\infty\\M=\alpha N}} \mathbb{E}_{\tilde{y}}\left ( r'(\tilde{y}_{bi})\right )\;\;,\; {\rm as}\; N\rightarrow\infty\;\;,\;\forall i,b\;\;.\nonumber 
\end{eqnarray}

\noindent A similar expression can be derived for $\mathbb{E}_{J,v,h}(\tilde{y}_{bi})$. Expressions for ${\rm Var}_{J,v,h}(\tilde{x}_{ia})$ and ${\rm Var}_{J,v,h}(\tilde{y}_{bi})$ are more easily obtained, and we arrive at,
\begin{equation}
\lim_{\substack{N\rightarrow\infty\\M=\alpha N}} \frac{S(t)}{S(t-1)} \; = \; \alpha \tilde{\sigma}_{w}^{4}
\mathbb{E}_{\tilde{x}}\left (\sech^{4}(\tilde{x}) \right )\mathbb{E}_{\tilde{y}}\left (\sech^{4}(\tilde{y}) \right )\;\;,\label{eq:E.27}
\end{equation}
\begin{equation}
\tilde{x}\;\sim\; {\mathcal N}\left ( \mathbb{E}(\tilde{x}), {\rm Var}(\tilde{x})\right )\;\;,\;\;
\tilde{y}\;\sim\; {\mathcal N}\left ( \mathbb{E}(\tilde{y}), {\rm Var}(\tilde{y})\right )\;\;,
\label{eq:E.28}
\end{equation}
\begin{equation}
\mathbb{E}(\tilde{x}) \; = \;  \frac{1}{2}\mu_{v}\;+\;\frac{\alpha}{8}\tilde{\sigma}^{2}_{J}\mathbb{E}_{\tilde{y}}\left ( r'(\tilde{y})\right )\;\;,\;\;
{\rm Var}(\tilde{x}) \; = \; \frac{1}{4}\sigma^{2}_{v}\;+\;\frac{\alpha}{4}\tilde{\sigma}^{2}_{J}\mathbb{E}_{\tilde{y}}\left ( r^{2}(\tilde{y})\right )\;\;,
\label{eq:E.29} 
\end{equation}
\begin{equation}
\mathbb{E}(\tilde{y}) \; = \;  \frac{1}{2}\mu_{h}\;+\;\frac{1}{8}\tilde{\sigma}^{2}_{J}\mathbb{E}_{\tilde{x}}\left ( r'(\tilde{x})\right )\;\;,\;\;
{\rm Var}(\tilde{y}) \; = \; \frac{1}{4}\sigma^{2}_{h}\;+\;\frac{1}{4}\tilde{\sigma}^{2}_{J}\mathbb{E}_{\tilde{x}}\left ( r^{2}(\tilde{x})\right )\;\;.\label{eq:E.30}
\end{equation}

\noindent Finally, substituting $\tilde{x}=\frac{1}{2}\hat{x}$, $\tilde{y}=\frac{1}{2}\hat{y}$ and identifying,
\begin{eqnarray}
\mathbb{E}(\hat{x}) -\mu_{v}\; = \; -\left ( \Delta z -\Delta \hat{z}\right )\;\;,\;\;
\mathbb{E}(\hat{y}) -\mu_{h}& = & -\left ( \Delta z +\Delta \hat{z}\right )\;\;, \nonumber \\
{\rm Var}(\hat{x}) -\sigma^{2}_{v}\; = \; 2\left ( z_{1} - \hat{z}_{1}\right )\;\;,\;\;
{\rm Var}(\hat{y}) -\sigma^{2}_{h} & = & 2\left ( z_{1} + \hat{z}_{1}\right )\;\;, \nonumber 
\end{eqnarray}

\noindent we recover from Eq.(\ref{eq:E.27})-Eq.(\ref{eq:E.30}) the replica-symmetric saddle-point equations in Eq.(\ref{eq:B.20d}) and the replica symmetry stability condition in Eq.(\ref{eq:E.1}). Thus, in the thermodynamic limit $N\rightarrow\infty\;,\; M=\alpha N$, the message passing convergence threshold $\lim_{t\rightarrow\infty} S(t)/S(t-1)\;=\;1$ is equivalent to the stability threshold of the replica-symmetric saddle-point. A similar equivalence between the stability criterion of a message passing algorithm and the stability of replica-symmetry has also been shown for a number of other systems \cite{Kabashima2003a, Kabashima2003b, Zhao2022}.

\section{Discussion and Conclusions}
Message passing algorithms, such as that of Huang and Toyoizumi \cite{HuangToyoizumi2015}, provide efficient methods for obtaining computational estimates of the free energy of any individual RBM. However, to make analytical progress in understanding the properties of RBMs requires studying their properties under expectation, that is, averaging over the parameters of the RBM Hamiltonian. The replica trick is a widely used technique for performing that averaging, and has been used to study both spin-glass models and RBMs - see for example the tutorial by Montanari and Sen \cite{MontanariSen2022} for a recent and accessible introduction. The averaging over parameters is done assuming a particular form, usually diagonal, for the parameter-parameter covariance matrix. Although the mapping from the Ising form of the RBM Hamiltonian to the lattice-gas form is simple, the mapping changes a diagonal covariance matrix of parameters in one form to a non-diagonal covariance matrix in the other form. Consequently, replica trick results obtained from analysis of the Ising form of RBM Hamiltonian with a diagonal covariance are not necessarily applicable to analysis for the lattice-gas RBM Hamiltonian with a diagonal covariance. This point has already been made in the context of the SK model \cite{Russo1998}. Use of the lattice-gas form is common within the machine learning literature, where the log RBM partition function is a key quantity of interest. Therefore, it is important to understand, from an analytic perspective, the properties of the lattice-gas RBM log partition function.

We have derived the leading-order (tree-level) and one-loop finite-size correction (in \ref{sec:AppendixB}) for the replica-symmetric approximation to the lattice-gas RBM log partition function when there is heterogeneity in all the parameters of the lattice-gas RBM Hamiltonian. We have also, for completeness, derived in \ref{sec:AppendixIsing} a number of the corresponding results for the binary-to-binary Ising RBM Hamiltonian with a diagonal covariance matrix, although a number of variants of Ising RBMs have already been extensively studied by others \cite{Barra2011, Barra2017, Barra2017b, Agliari2019, DecelleFurtlehner2021}. For the lattice-gas RBM, below the replica-symmetry instability point the agreement between the replica-symmetric approximations and the message-passing algorithm of Huang and Toyoizumi is good, showing that the replica-symmetric approximation is very accurate, even at moderate system sizes, and when we have heterogeneity in both the external fields and spin-spin couplings. We have also shown that the replica-symmetry instability point (above which we expect replica symmetry breaking) coincides with the point at which the Huang and Toyoizumi Bethe approximation algorithm fails to converge, which is typically related to divergence of the spin-glass susceptibility \cite{Rivoire2004} and emergence of a spin-glass phase where we would expect replica symmetry-breaking. We have confirmed that the log partition function and other properties display the expected behaviour when interchanging the hidden layer characteristics and the visible layer characteristics along with the mapping $\alpha\rightarrow\alpha^{-1}$. From the replica-symmetric theory we have also derived a simplified approximation to the replica-symmetric saddle-point equations which is appropriate when the variance in the visible-to-hidden layer couplings is large. This allows for simplified approximations to the log partition function and the replica-symmetry instability point to be derived.

The asymptotic expansion results in Eq.(\ref{eq:C.19}) and Eq.(\ref{eq:C.20}), as $\tilde{\sigma}^{2}_{J} \rightarrow\infty$, suggest that the quantitative influence of heterogeneity in the external fields is lesser than the influence of the biases in the external fields, as only the biases enter the next-to-leading order terms in the expansion. To some extent this can be seen in the simulation results in Fig.\ref{fig:Figure4}, where the change in the value of $\mathbb{E}\left ( \log Z\right )$ over a wide range of $\sigma^{2}_{v}$ values is considerably less than the change in $\mathbb{E}\left ( \log Z\right )$ that occurs in Fig.\ref{fig:Figure3} when the mean $\mu_{v}$ of the external field $v$ is changed, despite the changes in $\mu_{v}$ seen in Fig.\ref{fig:Figure3} being smaller than the changes in, $\sigma_{v}$, the standard deviation of the external field $v$, that we see in Fig.\ref{fig:Figure4}. The weakening effect of $\sigma^{2}_{v}$ as $\tilde{\sigma}^{2}_{J}$ increases is also clearly apparent in Fig.\ref{fig:Figure4}. 

By analyzing $\mathbb{E}\left ( \log Z\right )$ for both a binary-to-binary lattice-gas RBM with diagonal covariance and for a binary-to-binary Ising RBM with diagonal covariance (see \ref{sec:AppendixIsing}) we have also shown that the behaviour of the replica-symmetric solution changes between these two forms of RBMs, even though the lattice-gas binary RBM Hamiltonian is precisely equivalent to a binary Ising RBM Hamiltonian. Therefore, these differences are due to the difference in the parameter ensembles averaged over when calculating the quenched averages. Specifically, this is due to the presence of the off-diagonal elements in the disorder covariance matrix induced in going from the lattice-gas representation to the Ising representation of the RBM Hamiltonian, and so indicates that taking into account parameter-parameter correlations is important when studying RBMs.

Although we have demonstrated the impact of using a lattice-gas representation instead of an Ising representation, the standard diagonal covariance structure we have used corresponds to an coupling matrix between the hidden and visible layers that is isotropic, and so this coupling does not contain any explicit structure or signal. From a machine learning perspective of using RBMs to represent meaningful structure in a training dataset, it may be more pertinent to ask when and how the hidden layer learns to represent that structure. This question was addressed by Tubiana and Monasson \cite{TubianaMonasson2017} who studied the conditions required for a compositional phase to be present, in which visible layer patterns can be represented by a small number of strongly activated hidden nodes. Decelle and co-workers \cite{Decelle2017, Decelle2018, DecelleFurtlehner2021} have also studied the question of how binary-to-binary RBMs learn the structure in a dataset by explicitly introducing non-isotropic structure into the spin-spin couplings using a ‘spiked’ spin-spin coupling population covariance matrix. The work of Decelle et al \cite{Decelle2017, Decelle2018, DecelleFurtlehner2021} focused on the Ising RBM Hamiltonian form but, in principle, the analysis presented here could be extended to spiked spin-coupling covariance matrices. We would expect to reveal similar issues as the analysis of the isotropic case has – namely that changes in covariance structure induced by going from a lattice-gas representation to an Ising representation (or vice-versa) lead to material differences in the quenched free energy. However, the isotropic covariance matrices studied in this paper have been sufficient to illustrate that point, and so we leave the replica analysis of the lattice-gas binary-to-binary RBM with a spiked parameter covariance matrix for further work.

Finally, an immediate question that also arises from a practical perspective, is when should one use theoretical results derived from the Ising form of the Hamiltonian to understand the behaviour of a particular instance of an RBM, and when should one use theoretical results derived from the lattice-gas form? In principle, this can be determined by testing whether the diagonal covariance structure or the non-diagonal form gives a higher likelihood for the estimated Hamiltonian parameters. Having decided which form is most appropriate for the given set of RBM parameters, the corresponding theoretical results and approximations can be applied. Alternatively, if a diagonal prior on the RBM Hamiltonian parameters has been included as part of the parameter estimation process, then the form (Ising or lattice-gas) of the Hamiltonian specified for that estimation should determine which theoretical analysis is appropriate. 

\appendix

\section{Replica-symmetric saddle-point equations for lattice-gas form RBMs when $\mu_{J}\;\neq\;0$\label{sec:AppendixA1}}
If we consider a non-zero value for the mean, $\mu_{J}$, of the interactions $J_{ai}$, the equivalent expression to Eq.(\ref{eq:B.7}) becomes,
\begin{multline}
\mathbb{E}\left ( Z^{n}\right) \; = \; \sum_{{\bm n}^{(1)}, {\bm m}^{(1)}}\ldots\sum_{{\bm n}^{(n)}, {\bm m}^{(n)}} \left [ \prod_{i}\exp \left ( \mu_{v}\sum_{\nu}n^{(\nu)}_{i}\;+\;\frac{\sigma_{v}^{2}}{2}\left ( \sum_{\nu}n^{(\nu)}_{i}\right )^{2}\right ) \right .  \\
\times \left . \prod_{a}\exp \left ( \mu_{h}\sum_{\nu}m^{(\nu)}_{a}\;+\;\frac{\sigma_{h}^{2}}{2}\left ( \sum_{\nu}m^{(\nu)}_{a}\right )^{2}\right ) \times \prod_{i,a} \exp \left (  \sum_{\nu, \nu'}n^{(\nu)}_{i}n^{(\nu')}_{i}m^{(\nu)}_{a}m^{(\nu')}_{a} \left ( \frac{\sigma_{J}^{2}}{2} + \mu_{J}\delta_{\nu\nu'}\right ) \right ) \right ]\;\;.
\label{eq:AppA2.1}
\end{multline}

\noindent The last factor in Eq.(\ref{eq:AppA2.1}) we can write as,
\begin{equation}
\prod_{i,a} \exp \left (  \sum_{\nu, \nu'}n^{(\nu)}_{i}n^{(\nu')}_{i}m^{(\nu)}_{a}m^{(\nu')}_{a} \left ( \frac{\sigma_{J}^{2}}{2} + \mu_{J}\delta_{\nu\nu'}\right ) \right ) \;=\;
\exp \left (  \sum_{\nu, \nu'} x_{\nu\nu'}y_{\nu\nu'}\left ( \frac{1}{2}\sigma^{2}_{J} + \mu_{J}\delta_{\nu\nu'}\right )\right )\;\;,
\nonumber 
\end{equation}

\noindent with $x_{\nu\nu'}$ and $y_{\nu\nu'}$ defined as before as $x_{\nu\nu'}=\sum_{i}n^{(\nu)}_{i}n^{(\nu')}_{i}$ and $y_{\nu\nu'}=\sum_{a}m^{(\nu)}_{a}m^{(\nu')}_{a}$. Similar to Eq.(\ref{eq:B.11}) we can then represent,
\begin{multline} 
\exp \left ( x_{\nu\nu'}y_{\nu\nu'}\left ( \frac{1}{2}\sigma^{2}_{J} + \mu_{J}\delta_{\nu\nu'}\right )\right )
\; = \; 
\int_{-\infty}^{\infty}dz_{\nu\nu'}\int_{-\infty}^{\infty}d\tilde{z}_{\nu\nu'}\left [ \exp \left (-2\sigma^{-2}_{J}\left (1+2\mu_{J}\sigma^{-2}_{J}\right )^{-1}\left ( z^{2}_{\nu\nu'}\;+\;\tilde{z}^{2}_{\nu\nu'}\right)\right ) 
\right . \\
 \times  \exp \left (z_{\nu\nu'}\left( x_{\nu\nu'} \;+\; y_{\nu\nu'}\right)\right ) \exp \left (\mathrm{i}\tilde{z}_{\nu\nu'}\left( x_{\nu\nu'} \;-\; y_{\nu\nu'}\right)\right ) \Big ]
\; \times\; \frac{2}{\pi}\sigma^{-2}_{J} \left (1+2\mu_{J}\sigma^{-2}_{J}\right )^{-1}\;\;.
\nonumber 
\end{multline}

\noindent The action in Eq.(\ref{eq:B.14}) now becomes,
\begin{equation}
\exp\left ( N\left [ \log Z_{1}\;+\;\alpha\log Z_{2}\;-\;2\tilde{\sigma}^{-2}_{J}\sum_{\nu,\nu'}\left ( z^{2}_{\nu\nu'}\;+\;\tilde{z}^{2}_{\nu\nu'}\right ) \left ( 1 + 2\mu_{J}\sigma^{-2}_{J} \right )^{-1}\right ]\right )\;\;.
\nonumber 
\end{equation}

\noindent The replica symmetric saddle-point equations in Eq.(\ref{eq:B.20d}) are modified to become,
\begin{flalign}
z_{0}  &= \; \frac{1}{4}\tilde{\sigma}^{2}_{J} (1 + 2\mu_{J}\sigma^{-2}_{J} ) I_{1}(\mu_{v}, \sigma^{2}_{v}, \Delta z - \Delta \hat{z}, z_{1} - \hat{z}_{1})\;+\;\frac{\alpha}{4}\tilde{\sigma}^{2}_{J} (1 + 2\mu_{J}\sigma^{-2}_{J} ) I_{1}(\mu_{h}, \sigma^{2}_{h}, \Delta z + \Delta \hat{z}, z_{1} + \hat{z}_{1})\;\;, \nonumber \\
z_{1} & = \; \frac{1}{4}\tilde{\sigma}^{2}_{J}I_{2}(\mu_{v}, \sigma^{2}_{v}, \Delta z - \Delta \hat{z}, z_{1} - \hat{z}_{1})\;+\;\frac{\alpha}{4}\tilde{\sigma}^{2}_{J} I_{2}(\mu_{h}, \sigma^{2}_{h}, \Delta z + \Delta \hat{z}, z_{1} + \hat{z}_{1})\;\;, \nonumber \\
\hat{z}_{0} &= \; \frac{1}{4}\tilde{\sigma}^{2}_{J} (1 + 2\mu_{J}\sigma^{-2}_{J} ) I_{1}(\mu_{v}, \sigma^{2}_{v}, \Delta z - \Delta \hat{z}, z_{1} - \hat{z}_{1})\;-\;\frac{\alpha}{4}\tilde{\sigma}^{2}_{J} (1 + 2\mu_{J}\sigma^{-2}_{J} ) I_{1}(\mu_{h}, \sigma^{2}_{h}, \Delta z + \Delta \hat{z}, z_{1} + \hat{z}_{1})\;\;, \nonumber \\
\hat{z}_{1} &= \; \frac{1}{4}\tilde{\sigma}^{2}_{J}I_{2}(\mu_{v}, \sigma^{2}_{v}, \Delta z - \Delta \hat{z}, z_{1} - \hat{z}_{1})\;-\;\frac{\alpha}{4}\tilde{\sigma}^{2}_{J} I_{2}(\mu_{h}, \sigma^{2}_{h}, \Delta z + \Delta \hat{z}, z_{1} + \hat{z}_{1})\;\;.\nonumber 
\end{flalign}

\noindent The value of $\mathbb{E}\left ( \log Z\right )$ in the thermodynamic limit in terms of the saddle-point solution is then,
\begin{multline}
\lim_{\substack{N\rightarrow\infty\\M=\alpha N}} \frac{1}{N}\mathbb{E}\left (\log Z \right ) \; = \; \frac{2}{\tilde{\sigma^{2}_{J}}}\left (1 + 2\mu_{J}\sigma^{-2}_{J} \right )^{-1}\left( \hat{z}_{0}^{2}-z_{0}^{2} \right )
\;-\; \frac{2}{\tilde{\sigma^{2}_{J}}} \left ( \hat{z}_{1}^{2} - z_{1}^{2}\right ) \\
 + \; I_{0,1}\left( \mu_{v},\sigma^{2}_{v},\Delta z -\Delta \hat{z}, z_{1} -\hat{z}_{1}\right )
\; + \; \alpha I_{0,1}\left ( \mu_{h}, \sigma^{2}_{h}, \Delta z + \Delta \hat{z}, z_{1} + \hat{z}_{1} \right)\;\;.
\nonumber 
\end{multline}

\noindent This expression for $\mathbb{E}\left ( \log Z\right )$ can also be expressed in terms of the quantities $l,p,q,r$ defined in Eq.(\ref{eq:Z.1}) - Eq.(\ref{eq:Z.4}) as,

\begin{multline}
\lim_{\substack{N\rightarrow\infty\\M=\alpha N}} \frac{1}{N}\mathbb{E}\left (\log Z \right ) \; = \; \frac{\alpha}{2}{\tilde{\sigma}^{2}_{J}}qr\;-\; \frac{\alpha}{2}{\tilde{\sigma}^{2}_{J}}\left (1 + 2\mu_{J}\sigma^{-2}_{J} \right )lp
\; + \; I_{0,1}\left( \mu_{v},\sigma^{2}_{v},\frac{\alpha}{2}\tilde{\sigma}^{2}_{J}\left (r-(1+2\mu_{J}\sigma^{-2}_{J})p \right), \frac{\alpha}{2}\tilde{\sigma}^{2}_{J}r \right ) \\
\; + \; \alpha I_{0,1}\left ( \mu_{h}, \sigma^{2}_{h}, \frac{1}{2}\tilde{\sigma}^{2}_{J}\left (q-(1+2\mu_{J}\sigma^{-2}_{J})l \right), \frac{1}{2}\tilde{\sigma}^{2}_{J}q \right)\;\;.
\nonumber 
\end{multline}

\noindent In the thermodynamic limit and in the limit $n\rightarrow 0$ the replica-symmetric values of $l,p,q,r$ are related to each other via, 
\begin{eqnarray}
l & = & I_{1}\left (\mu_{v},\sigma^{2}_{v},\frac{\alpha}{2}\tilde{\sigma}^{2}_{J}\left (r-(1+2\mu_{J}\sigma^{-2}_{J})p \right), \frac{\alpha}{2}\tilde{\sigma}^{2}_{J}r \right )\;\;, \nonumber \\
p & = & I_{1}\left (\mu_{h},\sigma^{2}_{h},\frac{1}{2}\tilde{\sigma}^{2}_{J}\left (q-(1+2\mu_{J}\sigma^{-2}_{J})l \right), \frac{1}{2}\tilde{\sigma}^{2}_{J}q \right )\;\;, \nonumber \\
q & = & I_{2}\left (\mu_{v},\sigma^{2}_{v},\frac{\alpha}{2}\tilde{\sigma}^{2}_{J}\left (r-(1+2\mu_{J}\sigma^{-2}_{J})p\right ), \frac{\alpha}{2}\tilde{\sigma}^{2}_{J}r \right )\;\;, \nonumber \\
r & = & I_{2}\left (\mu_{h},\sigma^{2}_{h},\frac{1}{2}\tilde{\sigma}^{2}_{J}\left (q-(1+2\mu_{J}\sigma^{-2}_{J})l\right), \frac{1}{2}\tilde{\sigma}^{2}_{J}q \right )\;\;. \nonumber
\end{eqnarray}

\section{Derivation of the expected effective field \label{sec:AppendixA3}}

We wish to evaluate $\mathbb{E}_{J,v,h}\left ( \sum_{i=1}^{N}J_{ai}\langle n_{i}\rangle \right )$, where $\mathbb{E}_{J,v,h}(\cdot)$ represents the expectation over the Hamiltonian parameters. To aid this we introduce source fields $\chi_{a}$. That is, we introduce a term of the form,
\begin{equation}
-\sum_{a} \chi_{a}\sum_{i=1}^{N}J_{ai}\sum_{\nu}n^{(\nu)}_{i}\;\;,
\nonumber 
\end{equation}

\noindent into the replica Hamiltonian, $\sum_{\nu} H\left ( {\bf n}^{(\nu)}, {\bf m}^{(\nu)} \right )$ in Eq.(\ref{eq:B.6}). The analysis now follows similar steps to the derivation of equations Eq.(\ref{eq:B.7}) to Eq.(\ref{eq:B.19}). After integrating over the distribution of the Hamiltonian parameters, the last factor on the right-hand-side of Eq.(\ref{eq:B.7}) now becomes,
\begin{equation}
\prod_{i,a} \exp \left ( \frac{\sigma_{J}^{2}}{2}\left ( \sum_{\nu}n^{(\nu)}_{i}\left (m^{(\nu)}_{a} + \chi_{a}\right )\right )^{2}\right ) \;=\;
\exp \left ( \frac{1}{2}\sigma^{2}_{J}\sum_{\nu, \nu'}x_{\nu\nu'}w_{\nu\nu'}\right )\;\;,
\nonumber 
\end{equation}

\noindent where we have defined $w_{\nu\nu'}\;=\;\sum_{a}\left ( m_{a}^{(\nu)} + \chi_{a}\right )\left ( m_{a}^{(\nu')} + \chi_{a}\right )$. Similar to Eq.(\ref{eq:B.11}), we represent,
\begin{multline}
\exp \left ( \frac{1}{2}\sigma^{2}_{J} x_{\nu\nu'}w_{\nu\nu'}\right ) \; = \; 
\frac{2}{\pi}\sigma^{-2}_{J}\int_{-\infty}^{\infty}dz_{\nu\nu'}\int_{-\infty}^{\infty}d\tilde{z}_{\nu\nu'}\big [ \exp \left (-2\sigma^{-2}_{J}\left ( z^{2}_{\nu\nu'}\;+\;\tilde{z}^{2}_{\nu\nu'}\right)\right ) \\
\times\; \exp \left (z_{\nu\nu'}\left( x_{\nu\nu'} \;+\; w_{\nu\nu'}\right)\right ) \exp \left (\mathrm{i}\tilde{z}_{\nu\nu'}\left( x_{\nu\nu'} \;-\; w_{\nu\nu'}\right)\right ) \big ]\;\;,
\nonumber 
\end{multline}

\noindent and so $\mathbb{E}\left (Z^{n} \right )$ becomes,
\begin{equation}
\mathbb{E}\left ( Z^{n}\right) \; = \; \left ( \frac{2}{\pi\sigma^{2}_{J}}\right )^{n^{2}}
\int_{-\infty}^{\infty} \prod _{\nu,\nu'} dz_{\nu\nu'}
\int_{-\infty}^{\infty} \prod _{\nu,\nu'} d\tilde{z}_{\nu\nu'}\,\exp \left (-2\sigma^{-2}_{j}\left ( z^{2}_{\nu\nu'}\;+\;\tilde{z}^{2}_{\nu\nu'}\right)\right ) Z_{1}^{N}\prod_{a=1}^{M}Z_{2}\left ( \chi_{a}\right )\;\;,
\nonumber 
\end{equation}

\noindent with $Z_{2}\left ( \chi_{a}\right )$ given by,
\begin{equation}
Z_{2}\left ( \chi_{a}\right ) \; = \; \sum_{m_{1},\ldots,m_{n}\in \{0,1\}} \exp\left ( \mu_{h}\sum_{\nu}m_{\nu}\;+\;\sum_{\nu,\nu'}\left( \frac{1}{2}\sigma^{2}_{h} m_{\nu}m_{\nu'}\;+\;\left ( z_{\nu\nu'}-\mathrm{i}\tilde{z}_{\nu\nu'}\right )\left (m_{\nu} + \chi_{a}\right ) \left ( m_{\nu'} + \chi_{a}\right ) \right) \right )\;\;.
\nonumber 
\end{equation}

Under replica symmetry $Z_{2}\left ( \chi_{a}\right )$ becomes,
\begin{equation}
Z_{2}\left ( \chi_{a}\right )\;=\; \exp \left (\chi^{2}_{a}\left (-n(\Delta z + \Delta \hat{z}) + n^{2}(z_{1} + \hat{z}_{1}) \right ) \right ) \times Z_{2}^{(RS)}\left ( \mu_{h} - 2\chi_{a}(\Delta z + \Delta\hat{z} ) + 2n\chi_{a}(z_{1} + \hat{z}_{1}), \sigma^{2}_{h}\right )\;\;,
\nonumber
\end{equation}

\noindent where $Z_{2}^{(RS)}\left (\mu_{h}, \sigma^{2}_{h} \right )$ is the replica-symmetric expression for $Z_{2}\left (\mu_{h} \right )$ and so is given by the representation in Eq.(\ref{eq:B.17}). From this we find,
\begin{equation}
\log Z_{2}\left (\chi_{a} \right )\;=\; -n\chi^{2}_{a}\left ( \Delta z + \Delta\hat{z} \right )\;+\;
nI_{0,1}\left ( \mu_{h} - 2\chi_{a}(\Delta z + \Delta\hat{z}), \sigma^{2}_{h}, \Delta z + \Delta\hat{z}, z_{1} + \hat{z}_{1} \right ) \; + \; {\mathcal O}\left ( n^{2} \right )\;\;.
\nonumber 
\end{equation}

\noindent Finally taking derivatives with respect to $\chi_{a}$ we obtain,
\begin{eqnarray}
\mathbb{E}_{J,v,h}\left (\sum_{i=1}^{N}J_{ai}\langle n_{i}\rangle \right ) & = & \lim_{n\rightarrow 0}\frac{1}{n} 
\left . \frac{\partial \mathbb{E}_{J,v,h}\left ( Z^{n}\right )}{\partial \chi_{a}} \right |_{{\bm \chi}={\bf 0}} \nonumber \\
& = & -2\left ( \Delta z + \Delta\hat{z}\right ) \frac{\partial}{\partial \mu_{h}}I_{0,1}\left ( \mu_{h}, \sigma^{2}_{h}, \Delta z + \Delta\hat{z}, z_{1} + \hat{z}_{1}\right ) \nonumber \\
& =& \tilde{\sigma}^{2}_{J}(l - q)p \;\;. \nonumber 
\end{eqnarray}

\noindent With $p\;>\;0$ and $l\;>\;q$ we have that the effective field is non-zero and positive when $\tilde{\sigma}^{2}_{J} > 0$.

\section{Eigen-decomposition of the replica symmetric saddle-point Hessian \label{sec:AppendixB}}
To evaluate the Hessian at the replica-symmetric saddle-point we organize the matrix elements into blocks,
\begin{equation}
{\rm Hessian}\;=\; {\bm H}\;=\;
\renewcommand\arraystretch{2}
\begin{bmatrix}
\frac{\partial^{2}S}{\partial z \partial z} & \frac{\partial^{2}S}{\partial z \partial \tilde{z}} \\
\frac{\partial^{2}S}{\partial \tilde{z} \partial z} & \frac{\partial^{2}S}{\partial \tilde{z} \partial \tilde{z}}
\end{bmatrix}\;\;,
\nonumber 
\end{equation}

\noindent where $S$ is the action as defined by Eq.(\ref{eq:B.14}). From this we find the Hessian, ${\bm H}$, at the replica-symmetric saddle-point takes the form,
\begin{eqnarray}
{\bm H} & = & N
\renewcommand\arraystretch{1.5}
\setlength\arraycolsep{10pt}
\begin{bmatrix} 
{\bm A}^{(1)}+\alpha{\bm A}^{(2)} & \mathrm{i}{\bm A}^{(1)}-\mathrm{i}\alpha{\bm A}^{(2)}\\
\mathrm{i}{\bm A}^{(1)}-\mathrm{i}\alpha{\bm A}^{(2)} & -{\bm A}^{(1)}-\alpha{\bm A}^{(2)}
\end{bmatrix} \;-\;N\frac{4}{\tilde{\sigma}^{2}_{J}}{\bm I}_{2n^{2}} \nonumber \\
& \equiv & N{\bm A}\;-\;N\frac{4}{\tilde{\sigma}^{2}_{J}}{\bm I}_{2n^{2}}\;\;.
\label{eq:AppB.3}
\end{eqnarray}

\noindent Determining the eigenvectors and eigenvalues of ${\bm H}$ is trivial once we have determined the eigenvectors and eigenvalues of ${\bm A}$. The matrices ${\bm A}^{(1)}$ and ${\bm A}^{(2)}$ are both of size $n^{2}\times n^{2}$, with their matrix elements being indexed by four replica indices, for which we use $\alpha,\beta,\gamma,\delta$. That is ${\bm A}^{(1)}$ has elements $A^{(1)}_{\alpha\beta,\gamma\delta}$, and ${\bm A}^{(2)}$ has elements $A^{(2)}_{\alpha\beta,\gamma\delta}$, with $\alpha,\beta,\gamma,\delta \in \{1,\ldots,n\}$. The superscripts $(1)$ and $(2)$ on the matrices ${\bm A}^{(1)}$ and ${\bm A}^{(2)}$ are used to denote contributions to the Hessian arising, respectively, from second partial derivatives of $\log Z_{1}$ and $\log Z_{2}$ in Eq.(\ref{eq:B.14}). The matrices ${\bm A}^{(1)}$ and ${\bm A}^{(2)}$ are given terms of the block matrices ${\bm B}^{(1)}, {\bm B}^{(2)}, {\bm C}^{(1)}, {\bm C}^{(2)}, {\bm D}^{(1)}, {\bm D}^{(2)}$ as,

\begin{equation} {\bm A}^{(1)}\;=\;
\begin{bmatrix}
{\bm B}^{(1)} & {\bm C}^{(1)}  \\
{{\bm C}^{(1)}}^{\top} & {\bm D^{(1)}}
\end{bmatrix}\;\;,\;\;
{\bm A}^{(2)}\;=\;
\begin{bmatrix}
{\bm B}^{(2)} & {\bm C}^{(2)}  \\
{{\bm C}^{(2)}}^{\top} & {\bm D^{(2)}}
\end{bmatrix}\;\;.
\nonumber 
\end{equation}

The matrices ${\bm B}^{(i)}, {\bm C}^{(i)}, {\bm D}^{(i)},\;i=1,2$, represent the different blocks of ${\bm A}^{(i)}$, according to how many contractions we have between the matrix element indices $\alpha,\beta,\gamma,\delta$, and so the matrices ${\bm B}^{(i)}, {\bm C}^{(i)}, {\bm D}^{(i)}$ are of sizes $n\times n$, $n\times n(n-1)$ and $n(n-1)\times n(n-1)$ respectively. To evaluate the matrix elements of ${\bm B}^{(i)}, {\bm C}^{(i)}, {\bm D}^{(i)}$ we have to evaluate, at the replica-symmetric saddle-point, second partial derivatives of $\log Z_{1}$, and $\log Z_{2}$ with respect to $z_{\alpha\beta}$ and $\tilde{z}_{\gamma\delta}$. Expressions for $Z_{1}$ and $Z_{2}$ are given in Eq.(\ref{eq:B.13a}) and Eq.(\ref{eq:B.13b}). Their second partial derivatives are easily evaluated and we find it convenient to introduce expressions, 
\begin{equation}
R_{k}(n,\mu,\sigma^{2}, x, y)\;=\; \ddfrac{\int_{-\infty}^{\infty}dt \left (\frac{e^{t}}{1 + e^{t}} \right )^{k} \left (1+e^{t} \right )^{n} e^{-\frac{(t\;-\;\mu\;+\;x)^{2}}{2(\sigma^{2}+y)}}}{\int_{-\infty}^{\infty}dt \left (1+e^{t} \right )^{n} e^{-\frac{(t\;-\;\mu\;+\;x)^{2}}{2(\sigma^{2}+y)}}}\;\;.
\nonumber 
\end{equation}

For brevity we will use a shorthand notation,
\begin{eqnarray}
R_{k}^{(1)} & = & R_{k}(n,\mu_{v}, \sigma^{2}_{v}, \Delta z\;-\;\Delta \hat{z}, z_{1}\;-\;\hat{z}_{1}) \;\;,\nonumber \\
R_{k}^{(2)} & = & R_{k}(n, \mu_{h}, \sigma^{2}_{h}, \Delta z\;+\;\Delta \hat{z}, z_{1}\;+\;\hat{z}_{1})\;\;. \nonumber 
\end{eqnarray}

\noindent In terms of the quantities $R_{k}^{(1)}, R_{k}^{(2)}$, the second partial derivatives of $\log Z_{1}$ and $\log Z_{2}$ are trivially evaluated and we find expressions for the block matrices ${\bm B}^{(i)}, {\bm C}^{(i)}, {\bm D}^{(i)}$ as follows,
\begin{eqnarray}
{\bm B}^{(i)}_{\alpha \gamma} & = & \left (R_{2}^{(i)}\;-\;{R_{1}^{(i)}}^{2} \right )\;+\; \left (R_{1}^{(i)}\;-\;R_{2}^{(i)} \right )\delta_{\alpha \gamma}\;\;\; i=1,2\;\;, \nonumber \\[2pt]
{\bm C}^{(i)}_{\alpha\alpha, \gamma\neq\delta} & = & \left (R_{3}^{(i)}\;-\;R_{2}^{(i)}R_{1}^{(i)} \right )\;+\; \left (R_{2}^{(i)}\;-\;R_{3}^{(i)} \right )\left (\delta_{\alpha \gamma}\;+\;\delta_{\alpha\delta}\right )\;\;\; i=1,2\;\;, \nonumber \\[2pt]
{\bm D}^{(i)}_{\alpha\neq\beta,\gamma\neq\delta} & = & \left (R_{4}^{(i)}\;-\;{R_{2}^{(i)}}^{2} \right )\;+\; \left (R_{3}^{(i)}\;-\;R_{4}^{(i)} \right )\left(\delta_{\alpha\gamma}\;+\;\delta_{\beta\gamma}\;+\;\delta_{\alpha\delta}\;+\;\delta_{\beta\delta}\right ) \nonumber \\
 & + & \left( R_{4}^{(i)}\;-\; 2R_{3}^{(i)}\;+\; R_{2}^{(i)} \right )\left( \delta_{\alpha\gamma}\delta_{\beta\delta}\;+\;\delta_{\alpha\delta}\delta_{\beta\gamma} \right)\;\;\; i=1,2\;\;. \nonumber 
\end{eqnarray}

\noindent In all of the above $\delta_{\alpha\beta}$ denotes the Kronecker-delta function. Note that in the limit $n\rightarrow 0$ we have $R_{k}(n,\mu,\sigma^{2}, x, y)\rightarrow I_{k}(\mu,\sigma^{2}, x, y)$ as given by Eq.(\ref{eq:B.21}).

The form of ${\bm A}$ in Eq.(\ref{eq:AppB.3}) suggests that its eigenvectors can be easily constructed from the eigenvectors of ${\bm A}^{(1)}$ and ${\bm A}^{(2)}$. The eigenvectors and eigenvalues of ${\bm A}^{(1)}$ can be constructed from the eigenvectors and eigenvalues of ${\bm B}^{(1)}$ and ${\bm D}^{(1)}$, and the singular vectors and singular values of ${\bm C}^{(1)}$. Similarly, we can construct the eigen-decomposition of ${\bm A}^{(2)}$ from the eigen-decompositions of ${\bm B}^{(2)}$ and ${\bm D}^{(2)}$ and the singular-value-decomposition of ${\bm C}^{(2)}$. Unsurprisingly, the matrices ${\bm B}^{(i)}$, ${\bm C}^{(i)}$ and ${\bm D}^{(i)}$ are similar in form to block matrices that arise in the stability analysis of the replica solution of the Ising perceptron \cite{GardnerDerrida1988, EngelVanDenBroeck2001} and also the SK model \cite{AlmeidaThouless1978, Nishimori2001}, and so we can directly utilise those analyses to obtain the decompositions of ${\bm B}^{(i)}$, ${\bm C}^{(i)}$ and ${\bm D}^{(i)}$. Details on the Hessian eigenvalues and eigenvectors from the replica analysis of the Ising perceptron are given by Engel and Van den Broeck \cite{EngelVanDenBroeck2001}. For brevity we simply state the corresponding results for ${\bm B}^{(i)}$, ${\bm C}^{(i)}$ and ${\bm D}^{(i)}$. The matrices ${\bm B}^{(i)}$  have 2 distinct eigenvalues, whilst the matrices  ${\bm D}^{(i)}$ have 4 distinct eigenvalues. The matrices ${\bm C}^{(i)}$ have 2 distinct singular values. We index the distinct eigenvalues and singular values by $s\in\{1,2,3,4\}$. The distinct eigenvalues, $\lambda^{(s)}$, are given by,
\begin{flalign}
\lambda^{(1)} ({\bm B}^{(i)}) & = \; R_{1}^{(i)}\;-\;R_{2}^{(i)}\;+\;n\left ( R_{2}^{(i)} \;-\; {R_{1}^{(i)}}^{2} \right)\;\;,\;\;{\rm degeneracy}\;=\;1\;\;, \nonumber \\
\lambda^{(2)} ({\bm B}^{(i)}) & = \; R_{1}^{(i)}\;-\;R_{2}^{(i)}\;\;,\;\;{\rm degeneracy}\;=\;n-1\;\;, \nonumber \\
\lambda^{(1)} ({\bm D}^{(i)}) & = \; 6R_{4}^{(i)}\;-8R_{3}^{(i)}\;+\;2R_{2}^{(i)}\;+\;n\left ( 4R_{3}^{(i)}\;-\;5R_{4}^{(i)}\;+\;{R_{2}^{(i)}}^{2}\right )\;+\;n^{2}\left ( R_{4}^{(i)}\;-\;{R_{2}^{(i)}}^{2}\right )\;\;, \nonumber \\
&\, \hspace{350pt} {\rm degeneracy}\;=\;1\;\;, \nonumber \\
\lambda^{(2)} ({\bm D}^{(i)}) & = \; 2\left (3R_{4}^{(i)}\;-\;4R_{3}^{(i)}\;+\;R_{2}^{(i)}\;+\;n\left(R_{3}^{(i)}\;-\;R_{4}^{(i)} \right ) \right )\;\;,\;\;{\rm degeneracy}\;=\;n-1\;\;, \nonumber \\
\lambda^{(3)} ({\bm D}^{(i)}) & = \; 2\left (R_{4}^{(i)}\;-\;2R_{3}^{(i)}\;+\;R_{2}^{(i)}\right)\;\;,\;\;{\rm degeneracy}\;=\;\frac{n}{2}(n-3)\;\;, \nonumber \\
\lambda^{(4)} ({\bm D}^{(i)}) & = \; 0\;\;,\;\;{\rm degeneracy}\;=\;\frac{n}{2}(n-1)\;\;. \nonumber
\end{flalign}

\noindent The corresponding eigenvectors, indexed by $m$, we denote by ${\bm v}^{(s)}_{m}({\bm B}^{(i)})$ and ${\bm v}^{(s)}_{m}({\bm D}^{(i)})$. For $s=1$ and $s=2$, these eigenvectors are also singular vectors of ${\bm C}^{(i)}$, with corresponding singular values,
\begin{eqnarray}
\Lambda^{(1)}({\bm C}^{(i)}) & = & \sqrt{n-1}\left ( n\left( R_{3}^{(i)}\;-\;R_{2}^{(i)}R_{1}^{(i)}\right )\;+\;2R_{2}^{(i)}\;-\;2R_{3}^{(i)}\right )\;\;,\;\;{\rm degeneracy}\;=\;1\;\;, \nonumber \\
\Lambda^{(2)}({\bm C}^{(i)}) & = & \sqrt{2(n-2)}\left ( R_{2}^{(i)}\;-\;R_{3}^{(i)} \right )\;\;,\;\;{\rm degeneracy}\;=\;n-1\;\;. \nonumber
\end{eqnarray}

\noindent For $s=1$ and $s=2$ we have ${\bm v}^{(s)}_{m}({\bm B}^{(1)}) = {\bm v}^{(s)}_{m}({\bm B}^{(2)})$, that is, the eigenvectors are the same irrespective of whether the matrix ${\bm B}$ is derived from $\log Z_{1}$ or from $\log Z_{2}$. Similarly, the eigenvectors ${\bm v}^{(s)}_{m}({\bm D}^{(i)})$ all take the form,
\begin{equation}
	{\bm v}^{(s)}_{m}\left ({\bm D}^{(i)}\right ) \;=\;
	\begin{bmatrix}
		{\bm u}_{m}^{(s)}\\
		\pm {\bm u}_{m}^{(s)}
	\end{bmatrix}\;\;,
	\nonumber
\end{equation}

\noindent where ${\bm u}_{m}^{(s)}$ is a $\frac{1}{2}n(n-1)$-dimensional vector, and so for all values of $s$ the eigenvector ${\bm v}^{(s)}_{m}({\bm D}^{(i=1)})$ is also an eigenvector of ${\bm v}^{(s)}_{m}({\bm D}^{(i=2)})$.  From this, we find that possible eigenvectors of ${\bm A}^{(i)}$ take one of the following forms,
\begin{eqnarray}
&&  \underbrace{
        \begin{bmatrix}
	        {\bm v}^{(s=1)}_{m}({\bm B}^{(i)})\\
	        \sigma_{1,i} {\bm u}_{m}^{(s=1)}\\
	        \sigma_{1,i} {\bm u}_{m}^{(s=1)}
        \end{bmatrix}}_{\text{m=1}}\;\;\;,\;\;\;
    \underbrace{
    	\begin{bmatrix}
    		{\bm v}^{(s=2)}_{m}({\bm B}^{(i)})\\
    		\sigma_{2,i} {\bm u}_{m}^{(s=2)}\\
    		\sigma_{2,i} {\bm u}_{m}^{(s=2)}
        \end{bmatrix}}_{\text{m=1,\ldots,n-1}}\;\;\;,\;\;\;
    \overbrace{
    \underbrace{
	    \begin{bmatrix}
		   {\bm 0}\\
		   {\bm u}_{m}^{(s=1)}\\
		  -{\bm u}_{m}^{(s=1)}
        \end{bmatrix}}_{\text{m=1}}}^{\text{Eigenvalue=0}}\;\;\;,\;\;\;
    \overbrace{
    \underbrace{
       \begin{bmatrix}
	      {\bm 0}\\
	      {\bm u}_{m}^{(s=2)}\\
	     -{\bm u}_{m}^{(s=2)}
       \end{bmatrix}}_{\text{m=1,\ldots,n-1}}}^{\text{Eigenvalue=0}}\;\;\;,\;\;\; \nonumber \\
&& \underbrace{
       \begin{bmatrix}
	       {\bm 0}\\
           {\bm u}_{m}^{(s=3)}\\
           {\bm u}_{m}^{(s=3)}
       \end{bmatrix}}_{\text{m=1,\ldots,n(n-3)/2}}\;\;\;,\;\;\;
   \overbrace{
   \underbrace{
       \begin{bmatrix}
	       {\bm 0}\\
		   {\bm u}_{m}^{(s=3)}\\
		  -{\bm u}_{m}^{(s=3)}
       \end{bmatrix}
   }_{\text{m=1,\ldots,n(n-3)/2}}}^{\text{Eigenvalue=0}}\;\;\;.
   \label{eq:AppB.21}
\end{eqnarray}

\noindent The multiplier $\sigma_{s,i}$ in Eq.(\ref{eq:AppB.21}) is given by,
\begin{equation}
		\sigma_{s,i}\;=\;\frac{1}{2\Lambda^{(s)}({\bm C}^{(i)})}\left [\lambda^{(s)}( {\bm D}^{(i)}) - \lambda^{(s)}( {\bm B}^{(i)}) \pm \sqrt{\left ( \lambda^{(s)}( {\bm D}^{(i)}) - \lambda^{(s)}( {\bm B}^{(i)}) \right)^{2} + 4 {\Lambda^{(s)}({\bm C}^{(i)})}^{2}}\right ]\;\;.
		\nonumber 
\end{equation}

\noindent Now to construct the eigenvectors and eigenvalues of ${\bm A}$ we note that, where an eigenvector, ${\bm v}$ of ${\bm A}^{(1)}$ (with eigenvalue $\lambda_{1}$) is also an eigenvector of ${\bm A}^{(2)}$ (with eigenvalue $\lambda_{2}$) we can construct eigenvectors of ${\bm A}$ of the form, 
\begin{equation}
   \begin{bmatrix}
   	{\bm v}\\
   	\phi {\bm v}
   \end{bmatrix}\;\;\;,\;\;\;\phi\;=\;\frac{\mathrm{i}}{\lambda_{1}+\lambda_{2}}\left [(\lambda_{1}-\lambda_{2}) \pm \sqrt{\lambda_{1}\lambda_{2}} \right ]\;\;,
   \nonumber
\end{equation}

\noindent with corresponding eigenvalues $\pm 2N\sqrt{\alpha\lambda_{1}\lambda_{2}}$. The last four of the eigenvector forms in Eq.(\ref{eq:AppB.21}) fall into this category as they are constructed from the vectors ${\bm u}_{m}^{(s)}$. For the remaining eigenvectors of ${\bm A}$ we can construct them from a (non-orthogonal) basis consisting of the component vectors of the first two eigenvector forms in Eq.(\ref{eq:AppB.21}). That is we seek eigenvectors of ${\bm A}$ of the form,
\begin{equation}
	\underbrace{
	    \begin{bmatrix}
		    t_{1,1}{\bm v}^{(s=1)}_{m}\left ( {\bm B}^{(1)} \right ) \\
		    t_{1,2}{\bm v}^{(s=1)}_{m}\left ( {\bm D}^{(1)} \right ) \\
		    t_{1,3}{\bm v}^{(s=1)}_{m}\left ( {\bm B}^{(1)} \right ) \\
		    t_{1,4}{\bm v}^{(s=1)}_{m}\left ( {\bm D}^{(1)} \right )
	    \end{bmatrix} 
    }_{\text{m=1}}\;\;\;,\;\;\;
    \underbrace{
	    \begin{bmatrix}
	        t_{2,1}{\bm v}^{(s=2)}_{m}\left ( {\bm B}^{(1)} \right ) \\
	        t_{2,2}{\bm v}^{(s=2)}_{m}\left ( {\bm D}^{(1)} \right ) \\
	        t_{2,3}{\bm v}^{(s=2)}_{m}\left ( {\bm B}^{(1)} \right ) \\
	        t_{2,4}{\bm v}^{(s=2)}_{m}\left ( {\bm D}^{(1)} \right )
        \end{bmatrix}
    }_{\text{m=1,\ldots,n-1}}\;\;.
\label{eq:AppB.24}
\end{equation}

\noindent Requiring that the forms in Eq.(\ref{eq:AppB.24}) are eigenvectors of ${\bm A}$ allows us to solve for the coefficients $t_{s,k}$. Doing so, we find the remaining eigenvalues of ${\bm A}$ are eigenvalues of the matrices,
\begin{equation}
{\bm M}(s)\;=\;N
	\begin{bmatrix}
		\left ( {\bm S}^{(s,1)}\;+\;\alpha{\bm S}^{(s,2)}\right ) &\;\; \mathrm{i}\left ( {\bm S}^{(s,1)}\;-\;\alpha{\bm S}^{(s,2)}\right )\\
		\mathrm{i}\left ( {\bm S}^{(s,1)}\;-\;\alpha{\bm S}^{(s,2)}\right ) &\;\; -\left ({\bm S}^{(s,1)}\;+\;\alpha{\bm S}^{(s,2)} \right)
	\end{bmatrix}\;\;,
	\nonumber 
\end{equation}

\noindent where the $2\times 2$ matrices ${\bm S}^{(s,i)}$ are given by,
\begin{equation}
	{\bm S}^{(s,i)}\;=\;
	\begin{bmatrix}
		\lambda^{(s)}\left ( {\bm B}^{(i)}\right )\; & \;\Lambda^{(s)}\left ( {\bm C}^{(i)}\right ) \\
		\Lambda^{(s)}\left ( {\bm C}^{(i)}\right )\; & \;\lambda^{(s)}\left ( {\bm D}^{(i)}\right )
	\end{bmatrix}\;\;.
	\nonumber 
\end{equation}

\noindent The matrices ${\bm S}^{(s,1)}$ and ${\bm S}^{(s,2)}$ are J-symmetric in the sense of Benner et al. \cite{Benner2018} and so their eigenvalues are either of the form $\lambda, -\lambda, \lambda^{\star}, -\lambda^{\star}$, where $\lambda^{\star}$ denotes the complex conjugate of $\lambda$, or are of the form $\pm \lambda_{a}, \pm \lambda_{b}$ where $\lambda_{a}, \lambda_{b}$ are real eigenvalues. In fact, by considering,
\begin{equation}
{\bm M}^{2}(s)\;=\; 4N^{2}\alpha
     \begin{bmatrix}
       {\bm S}^{(s,1)}{\bm S}^{(s,2)} & {\bm 0} \\
       {\bm 0} & {\bm S}^{(s,1)}{\bm S}^{(s,2)}
     \end{bmatrix}\;\;,
     \nonumber 
\end{equation} 

\noindent we see that the remaining eigenvalues of ${\bm A}$ are $\pm 2N\alpha^{\frac{1}{2}}$ times the square root of the eigenvalues of ${\bm S}^{(s,1)}{\bm S}^{(s,2)}, s=1,2$. Putting all the pieces together, we have that the eigenvalues of the Hessian ${\bm H}$ in Eq.(\ref{eq:AppB.3}) are,

\begin{flalign}
&-\frac{4N}{\tilde{\sigma}^{2}_{J}} \;\; ,\;\;{\rm degeneracy}\;=\; n(n-1)\;\;, \label{eq:AppB.28}
\end{flalign}

\begin{flalign}
&-\frac{4N}{\tilde{\sigma}^{2}_{J}} \;+\;2N\alpha^{\frac{1}{2}}\sqrt{\lambda^{(s=3)} \left ({\bm D}^{(1)}\right)\lambda^{(s=3)}\left ({\bm D}^{(2)} \right )} \;\; ,\;\;{\rm degeneracy}\;=\;\frac{n(n-3)}{2}\;\;, \nonumber \\
&-\frac{4N}{\tilde{\sigma}^{2}_{J}} \;-\;2N\alpha^{\frac{1}{2}}\sqrt{\lambda^{(s=3)} \left ({\bm D}^{(1)}\right)\lambda^{(s=3)}\left ({\bm D}^{(2)} \right )} \;\; ,\;\;{\rm degeneracy}\;=\;\frac{n(n-3)}{2}\;\;. \label{eq:AppB.29}
\end{flalign}

\begin{flalign}
&-\frac{4N}{\tilde{\sigma}^{2}_{J}}\;+\;2N\alpha^{\frac{1}{2}}\sqrt{ \lambda \left( {\bm S}^{(1,1)}{\bm S}^{(1,2)}\right)}\;\;,\;\;{\rm degeneracy}=1\;\;, \nonumber \\
&-\frac{4N}{\tilde{\sigma}^{2}_{J}}\;-\;2N\alpha^{\frac{1}{2}}\sqrt{ \lambda \left( {\bm S}^{(1,1)}{\bm S}^{(1,2)}\right)}\;\;,\;\;{\rm degeneracy}=1\;\;, \nonumber \\
&-\frac{4N}{\tilde{\sigma}^{2}_{J}}\;+\;2N\alpha^{\frac{1}{2}}\sqrt{ \lambda \left( {\bm S}^{(2,1)}{\bm S}^{(2,2)}\right)}\;\;,\;\;{\rm degeneracy}=n-1\;\;, \nonumber\\
&-\frac{4N}{\tilde{\sigma}^{2}_{J}}\;-\;2N\alpha^{\frac{1}{2}}\sqrt{ \lambda \left( {\bm S}^{(2,1)}{\bm S}^{(2,2)}\right)}\;\;,\;\;{\rm degeneracy}=n-1\;\;, \label{eq:AppB.30}
\end{flalign}

\noindent where again, we have used $\lambda \left( {\bm S}^{(s,1)}{\bm S}^{(s,2)}\right)$ to denote, in an obvious fashion, an eigenvalue of ${\bm S}^{(s,1)}{\bm S}^{(s,2)}$. 

We are now in a position to finally evaluate $\log \det \left (-{\bm H}\right )$ and identify which eigenvectors of ${\bm H}$ lead to an instability of the replica-symmetric saddle-point. Introducing a short-hand notation,
\begin{eqnarray}
I_{k,v} & = & I_{k}(\mu_{v}, \sigma^{2}_{v}, \Delta z -\Delta \hat{z}, z_{1} -\hat{z}_{1} )\;\;, \nonumber \\
I_{k,h} & = & I_{k}(\mu_{h}, \sigma^{2}_{h}, \Delta z +\Delta \hat{z}, z_{1} +\hat{z}_{1} )\;\;, \nonumber 
\end{eqnarray}

\noindent we find that,
\begin{equation}
\lim_{n\rightarrow 0}\frac{1}{n}\log \det \left ( -{\bm H}\right ) \; = \; \log \tilde{\sigma}^{2}_{J}\;+\;
\log \left ( 1\;-\;\frac{\alpha}{4} \tilde{\sigma}^{2}_{J}{\rm tr}{\bm Q}\;+\;\frac{\alpha^{2}}{16}\tilde{\sigma}^{4}_{J}\det {\bm Q}  \right ) \;
 + \; \log \left ( 1\;-\;\alpha \tilde{\sigma}^{4}_{J}K \right )\;\;,
\label{eq:AppB.33}
\end{equation}

\noindent where $K$ is given by,
\begin{flalign}
K & = \; \left (I_{4,v}\;-\;2I_{3,v}\;+\;I_{2,v} \right ) \left (I_{4,h}\;-\;2I_{3,h}\;+\;I_{2,h} \right ) \nonumber \\
& = \; \frac{1}{256}\frac{1}{2\pi}
\left (\frac{1}{\sqrt{\sigma^{2}_{v}+2z_{1}-2\hat{z}_{1}}}\int_{-\infty}^{\infty}dx\,
\exp\left ( -\frac{(x-\mu_{v}+\Delta z-\Delta \hat{z})^{2}}{2(\sigma^{2}_{v} + 2z_{1} - 2\hat{z}_{1})}\right )\sech^{4}\left (\frac{x}{2} \right )\right ) \nonumber \\
& \hspace{50pt} \times \;
\left (\frac{1}{\sqrt{\sigma^{2}_{h}+2z_{1}+2\hat{z}_{1}}}\int_{-\infty}^{\infty}dx\,
\exp\left ( -\frac{(x-\mu_{h}+\Delta z+\Delta \hat{z})^{2}}{2(\sigma^{2}_{h} + 2z_{1} + 2\hat{z}_{1})}\right )\sech^{4}\left (\frac{x}{2} \right )\right )\;\;.
\nonumber 
\end{flalign}

\noindent In Eq.(\ref{eq:AppB.33}) the matrix ${\bm Q}=\lim_{n\rightarrow 0}{\bm S}^{(1,1)}{\bm S}^{(1,2)} = \lim_{n\rightarrow 0}{\bm S}^{(2,1)}{\bm S}^{(2,2)}$, and has properties,
\begin{multline}
{\rm tr}{\bm Q} \;=\; \left ( I_{2,v}-I_{1,v}\right )\left ( I_{2,h}-I_{1,h}\right )\;+\;\left ( 6I_{4,v}\;-\;8I_{3,v}\;+\;2I_{2,v}\right )\left ( 6I_{4,h}\;-\;8I_{3,h}\;+\;2I_{2,h}\right ) \\
 + \; 8 \left (I_{2,v} \;-\;I_{3,v}\right )\left (I_{2,h} \;-\;I_{3,h}\right )\;\;, \nonumber
\end{multline}
\begin{multline}
\det {\bm Q} \;=\; \left [ \left ( I_{2,v}-I_{1,v}\right )\left ( 6I_{4,v}\;-\;8I_{3,v}\;+\;2I_{2,v}\right )\;+\;4\left (I_{2,v} \;-\;I_{3,v}\right )^{2}\right ] \\
 \times \; \left [ \left ( I_{2,h}-I_{1,h}\right )\left ( 6I_{4,h}\;-\;8I_{3,h}\;+\;2I_{2,h}\right )\;+\;4\left (I_{2,h} \;-\;I_{3,h}\right )^{2}\right ]\;\;.
\nonumber 
\end{multline}

\noindent The expression in Eq.(\ref{eq:AppB.33}) provides a means for calculating the next-to-leading order (1-loop) contribution to $\mathbb{E}\left( \log Z\right)/N$. 

Of all the eigenvectors of ${\bm H}$ it is only the eigenvectors corresponding to the eigenvalues in Eq.(\ref{eq:AppB.29}) that do not represent fluctuations that stay within the replica-symmetric subspace, and so it is the behaviour of only these eigenvalues that will determine the stability of the replica-symmetric saddle-point \cite{EngelVanDenBroeck2001}. Thus, the replica-symmetric saddle-point becomes unstable when either of these eigenvalues become positive. Consequently, we find the threshold for stability of replica-symmetry is given by,
\begin{equation}
1 \; = \; \alpha \tilde{\sigma}^{4}_{J}K\;\;.
\label{eq:AppB.37}
\end{equation}

\noindent If the right-hand-side of Eq.(\ref{eq:AppB.37}) exceeds 1 then the Hessian ${\bm H}$ is no longer negative-definite.

\section{Derivation of leading order replica-symmetric results for the Ising RBM \label{sec:AppendixIsing}}
Leading order replica-symmetric approximations for $\mathbb{E}\left (\log Z \right)$ for the Ising RBM can easily be derived in a similar manner to the derivations for the lattice-gas RBM. We will give those results here. For brevity, rather than detail every step we will make reference to the previous lattice-gas RBM derivations. To aid that comparison we will, just for this section, keep  the same notation as for the lattice-gas RBM and simply change the set of values that the visible and hidden node variables can take, from $\{0,1\}$ to $\{-1,1\}$. That is, we consider a Hamiltonian of the form,   
\begin{equation}
	H({\bm n}, {\bm m})  \;=\; -\sum_{i,a}n_{i}J_{ai}m_{a}\;-\;\sum_{i=1}^{N}n_{i}v_{i}\;-\;\sum_{a=1}^{M}m_{a}h_{a}\;\;.
	\nonumber 
\end{equation}

\noindent The visible node variables $n_{i}$ and hidden node variables $m_{a}$ take values in $\{-1,1\}$. The Hamiltonian parameters $v_{i}, h_{a}, J_{ai}$ are drawn from the distributions given in Eq.(\ref{eq:B.2a}) - Eq.(\ref{eq:B.2c}), i.e., a Gaussian multi-variate distribution with diagonal covariance matrix.

The derivation of $\mathbb{E}\left ( Z^{n}\right )$ proceeds as before up to Eq.(\ref{eq:B.13a}) and Eq.(\ref{eq:B.13b}). For the Ising RBM $Z_{1}$ and $Z_{2}$ are now given by,
\begin{equation}
	Z_{1} \; = \; \sum_{n_{1},\ldots,n_{n}\in \{-1,1\}} \exp\left ( \mu_{v}\sum_{\nu}n_{\nu}\;+\;\sum_{\nu,\nu'}\left( \frac{1}{2}\sigma^{2}_{v}+z_{\nu\nu'}+\mathrm{i}\tilde{z}_{\nu\nu'}\right)n_{\nu}n_{\nu'}\right)\;\;, \nonumber
\end{equation}
\begin{equation}
	Z_{2} \; = \; \sum_{m_{1},\ldots,m_{n}\in \{-1,1\}} \exp\left ( \mu_{h}\sum_{\nu}m_{\nu}\;+\;\sum_{\nu,\nu'}\left( \frac{1}{2}\sigma^{2}_{h}+z_{\nu\nu'}-\mathrm{i}\tilde{z}_{\nu\nu'}\right)m_{\nu}m_{\nu'}\right)\;\;. \nonumber
\end{equation}

\noindent The action $S$ is still given by Eq.(\ref{eq:B.14}) but with these changed expressions for $Z_{1}$ and $Z_{2}$. Under a replica-symmetric ansatz for the dominant saddle points of $S$, we obtain the following expressions,
\begin{equation}
\log Z_{1} \; = \; n\left ( z_{0} - \hat{z}_{0} -z_{1} + \hat{z}_{1}\;+\; \frac{1}{\sqrt{2\pi}}\int_{-\infty}^{\infty}dt\,e^{-\frac{1}{2}t^{2}} \log \left ( 2\cosh \left ( \mu_{v}\;+\;t\left ( \sigma^{2}_{v}+2z_{1}-2\hat{z}_{1}\right )^{\frac{1}{2}} \right ) \right ) \right ) \;+\; {\cal O}\left ( n^{2}\right ) \;\;,
\nonumber 
\end{equation}

\begin{equation}
	\log Z_{2} \; = \; n\left ( z_{0} + \hat{z}_{0} -z_{1} - \hat{z}_{1}\;+\; \frac{1}{\sqrt{2\pi}}\int_{-\infty}^{\infty}dt\,e^{-\frac{1}{2}t^{2}} \log \left ( 2\cosh \left ( \mu_{h}\;+\;t\left ( \sigma^{2}_{h}+2z_{1}+2\hat{z}_{1}\right )^{\frac{1}{2}} \right ) \right ) \right ) \;+\; {\cal O}\left ( n^{2}\right ) \;\;,
	\nonumber 
\end{equation}

\noindent where again $\tilde{z}_{0} = \mathrm{i}\hat{z}_{0}$ and $\tilde{z}_{1} = \mathrm{i}\hat{z}_{1}$. Here, $z_{0}, \tilde{z}_{0}, z_{1}, \tilde{z}_{1}$ are defined again as in Eq.(\ref{eq:B.15a}) and Eq.(\ref{eq:B.15b}). Saddle-point values for $z_{0}, \hat{z}_{0}, z_{1}$ and $\hat{z}_{1}$ are easily identified as,
\begin{flalign}
	z_{0} &= \; \frac{\tilde{\sigma}^{2}_{J}}{4} \left ( 1 + \alpha \right ) \;\;, \nonumber \\
	\hat{z}_{0} &= \; \frac{\tilde{\sigma}^{2}_{J}}{4} \left ( 1 - \alpha \right ) \;\;, \nonumber \\
	z_{1} &= \; \frac{\tilde{\sigma}^{2}_{J}}{4} \frac{1}{\sqrt{2\pi}}\int_{-\infty}^{\infty}dt\,e^{-\frac{1}{2}t^{2}} \left [ \tanh^{2} \left ( \mu_{v} + t\left ( \sigma^{2}_{v}+2z_{1}-2\hat{z}_{1}\right )^{\frac{1}{2}} \right ) \;+\;
	\alpha \tanh^{2} \left ( \mu_{h} + t\left ( \sigma^{2}_{h}+2z_{1}+2\hat{z}_{1}\right )^{\frac{1}{2}} \right )
	 \right ] \,, \nonumber \\
	\hat{z}_{1} &= \; \frac{\tilde{\sigma}^{2}_{J}}{4} \frac{1}{\sqrt{2\pi}}\int_{-\infty}^{\infty}dt\,e^{-\frac{1}{2}t^{2}} \left [ \tanh^{2} \left ( \mu_{v} + t\left ( \sigma^{2}_{v}+2z_{1}-2\hat{z}_{1}\right )^{\frac{1}{2}} \right ) \;-\;
	\alpha \tanh^{2} \left ( \mu_{h} + t\left ( \sigma^{2}_{h}+2z_{1}+2\hat{z}_{1}\right )^{\frac{1}{2}} \right )
	\right ] \,. \nonumber
\end{flalign}

\noindent The leading order approximation to $\mathbb{E}\left ( \log Z \right )$ can be expressed in terms of those saddle-point values as,
\begin{flalign}
	\lim_{\substack{N\rightarrow\infty\\M=\alpha N}} \frac{1}{N}\mathbb{E}\left (\log Z \right ) &= \;  \frac{2}{\tilde{\sigma}^{2}_{J}}\left ( z_{1}^{2}\;-\;\hat{z}_{1}^{2}\right ) \;+\; \hat{z}_{1}\left ( 1 - \alpha\right )\;-\;z_{1}\left ( 1 + \alpha\right )\;+\;\frac{\alpha}{2}\tilde{\sigma}^{2}_{J}\;+\;\left ( 1 + \alpha \right )\log 2 \nonumber \\
	 &+ \; \frac{1}{\sqrt{2\pi}}\int_{-\infty}^{\infty}dt\,e^{-\frac{1}{2}t^{2}} \log \cosh \left ( \mu_{v}\;+\;t\left ( \sigma^{2}_{v}+2z_{1} - 2\hat{z}_{1}\right )^{\frac{1}{2}} \right ) \nonumber \\
	&+ \; \frac{\alpha}{\sqrt{2\pi}}\int_{-\infty}^{\infty}dt\,e^{-\frac{1}{2}t^{2}} \log \cosh \left ( \mu_{h}\;+\;t\left ( \sigma^{2}_{h}+2z_{1}+2\hat{z}_{1}\right )^{\frac{1}{2}} \right )\;\;.
	\label{eq:X.0}
\end{flalign}

\noindent Similarly, to leading order ${\rm Var}\left ( \log Z\right )$ is given by,
\begin{flalign}
	\lim_{\substack{N\rightarrow\infty\\M=\alpha N}} \frac{1}{N}{\rm Var}\left (\log Z \right ) &= \;
	\frac{4}{\tilde{\sigma}^{2}_{J}}\left ( \hat{z}_{1}^{2} - z_{1}^{2} \right ) 
	\; + \; \frac{1}{\sqrt{2\pi}}\int_{-\infty}^{\infty}dt\, e^{-\frac{1}{2}t^{2}} \left ( \log \cosh \left ( \mu_{v}\;+\;t\left ( \sigma^{2}_{v}+2z_{1}-2\hat{z}_{1}\right )^{\frac{1}{2}} \right )\right )^{2} \nonumber \\
	&-\; \left ( \frac{1}{\sqrt{2\pi}}\int_{-\infty}^{\infty}dt\, e^{-\frac{1}{2}t^{2}}  \log \cosh \left ( \mu_{v}\;+\;t\left ( \sigma^{2}_{v}+2z_{1}-2\hat{z}_{1}\right )^{\frac{1}{2}} \right ) \right )^{2} \nonumber \\
	&+\; \frac{\alpha}{\sqrt{2\pi}}\int_{-\infty}^{\infty}dt\, e^{-\frac{1}{2}t^{2}} \left ( \log \cosh \left ( \mu_{h}\;+\;t\left ( \sigma^{2}_{h}+2z_{1}+2\hat{z}_{1}\right )^{\frac{1}{2}} \right )\right )^{2} \nonumber \\
	&-\; \alpha\left ( \frac{1}{\sqrt{2\pi}}\int_{-\infty}^{\infty}dt\, e^{-\frac{1}{2}t^{2}} \log \cosh \left ( \mu_{h}\;+\;t\left ( \sigma^{2}_{h}+2z_{1}+2\hat{z}_{1}\right )^{\frac{1}{2}} \right ) \right )^{2}\;\;. \nonumber 
\end{flalign}

\noindent Introducing quantities $l,p,q$ and $r$ defined as before in Eq.(\ref{eq:Z.1}) - Eq.(\ref{eq:Z.4}) we find (after repeating the derivations of Section \ref{sec:phase_behaviour}) that $l,p,q,r$ are given by,
\begin{eqnarray}
	l & = & \frac{1}{\sqrt{2\pi}}\int_{-\infty}^{\infty}dt\,e^{-\frac{1}{2}t^{2}} \tanh \left ( \mu_{v}\;+\;t\left ( \sigma^{2}_{v}+2z_{1}-2\hat{z}_{1}\right )^{\frac{1}{2}} \right ) \;\;, \nonumber \\
	p & = & \frac{1}{\sqrt{2\pi}}\int_{-\infty}^{\infty}dt\,e^{-\frac{1}{2}t^{2}} \tanh \left ( \mu_{h}\;+\;t\left ( \sigma^{2}_{h}+2z_{1}+2\hat{z}_{1}\right )^{\frac{1}{2}} \right ) \;\;, \nonumber \\
	q & = & \frac{1}{\sqrt{2\pi}}\int_{-\infty}^{\infty}dt\,e^{-\frac{1}{2}t^{2}} \tanh^{2} \left ( \mu_{v}\;+\;t\left ( \sigma^{2}_{v}+2z_{1}-2\hat{z}_{1}\right )^{\frac{1}{2}} \right ) \;\;, \nonumber \\
	r & = & \frac{1}{\sqrt{2\pi}}\int_{-\infty}^{\infty}dt\,e^{-\frac{1}{2}t^{2}} \tanh^{2} \left ( \mu_{h}\;+\;t\left ( \sigma^{2}_{h}+2z_{1}+2\hat{z}_{1}\right )^{\frac{1}{2}} \right ) \;\;. \nonumber 
\end{eqnarray}

\noindent From this we can express the saddle-point values of $z_{1}$ and $\hat{z}_{1}$ as,
\begin{equation}
	z_{1} \; = \; \frac{\tilde{\sigma}^{2}_{J}}{4}\left ( q\;+\;\alpha r\right ) \;\;\;\;,\;\;\;\;
	\hat{z}_{1} \; = \; \frac{\tilde{\sigma}^{2}_{J}}{4}\left ( q\;-\;\alpha r\right ) \;\;,\nonumber 
\end{equation}

\noindent and from this we can write $l,p,q,r$ as satisfying the self-consistent relations,
\begin{eqnarray}
	l & = & \frac{1}{\sqrt{2\pi}}\int_{-\infty}^{\infty}dt\,e^{-\frac{1}{2}t^{2}} \tanh \left ( \mu_{v}\;+\;t\left ( \sigma^{2}_{v}+\alpha \tilde{\sigma}^{2}_{J}r\right )^{\frac{1}{2}} \right ) \;\;, \label{eq:X.1} \\
	p & = & \frac{1}{\sqrt{2\pi}}\int_{-\infty}^{\infty}dt\,e^{-\frac{1}{2}t^{2}} \tanh \left ( \mu_{h}\;+\;t\left ( \sigma^{2}_{h}+\tilde{\sigma}^{2}_{J}q\right )^{\frac{1}{2}} \right ) \;\;, \label{eq:X.2} \\
	q & = & \frac{1}{\sqrt{2\pi}}\int_{-\infty}^{\infty}dt\,e^{-\frac{1}{2}t^{2}} \tanh^{2} \left ( \mu_{v}\;+\;t\left ( \sigma^{2}_{v}+\alpha \tilde{\sigma}^{2}_{J}r\right )^{\frac{1}{2}} \right ) \;\;, \label{eq:X.3} \\
	r & = & \frac{1}{\sqrt{2\pi}}\int_{-\infty}^{\infty}dt\,e^{-\frac{1}{2}t^{2}} \tanh^{2} \left ( \mu_{h}\;+\;t\left ( \sigma^{2}_{h}+\tilde{\sigma}^{2}_{J}q\right )^{\frac{1}{2}} \right ) \;\;. \label{eq:X.4}
\end{eqnarray}

\noindent The leading order replica-symmetric approximation to $\mathbb{E}\left (\log Z\right )$ is then given by,
\begin{flalign}
	\lim_{\substack{N\rightarrow\infty\\M=\alpha N}} \frac{1}{N}\mathbb{E}\left (\log Z \right ) &= \;  \frac{\alpha}{2}\tilde{\sigma}^{2}_{J}\left ( 1 - q\right )\left( 1 - r\right) \;+\; \left ( 1 + \alpha \right )\log 2 \nonumber \\
	&+\; \frac{1}{\sqrt{2\pi}}\int_{-\infty}^{\infty}dt\,e^{-\frac{1}{2}t^{2}} \log \cosh \left ( \mu_{v}\;+\;t\left ( \sigma^{2}_{v}+\alpha \tilde{\sigma}^{2}_{J}r\right )^{\frac{1}{2}} \right ) \nonumber \\
	&+\; \frac{\alpha}{\sqrt{2\pi}}\int_{-\infty}^{\infty}dt\,e^{-\frac{1}{2}t^{2}} \log \cosh \left ( \mu_{h}\;+\;t\left ( \sigma^{2}_{h}+\tilde{\sigma}^{2}_{J}q\right )^{\frac{1}{2}} \right )\;\;, \label{eq:X.5}
\end{flalign}

\noindent where the values of $l,p,q$ and $r$ are given by the relations in Eq.(\ref{eq:X.1}) - Eq.(\ref{eq:X.4}).\newline

Previous studies of other variants the Ising RBMs have revealed paramagnetic, spin-glass, and ferromagnetic phases \cite{Barra2017, Barra2017b}. These studies found that in zero external field both the paramagnetic and spin-glass phases have zero global magnetization values, whilst the paramagnetic phase also has a zero expectation value for the replica-replica overlaps and the spin-glass phase has a non-zero replica-replica overlap expectation. To confirm that the expressions in Eq.(\ref{eq:X.1}) - Eq.(\ref{eq:X.4}) can yield these same characteristics we consider the conditions on $\mu_{v}, \mu_{h}, \sigma^{2}_{v}, \sigma^{2}_{h}$ necessary for, i) zero global magnetizations and zero expected replica-replica overlaps, ii) zero global magnetizations with non-zero expected replica-replica overlaps.\newline

$\bullet$ For zero global magnetizations and zero expected replica-replica overlaps we would require $l = p = q = r = 0$. From Eq.(\ref{eq:X.1}) and Eq.(\ref{eq:X.2}) we can easily see from the integrand that the condition $l=0$ requires $\mu_{v} = 0$, whilst the condition $p=0$ requires $\mu_{h} = 0$. From the integrands in Eq.(\ref{eq:X.3}) and Eq.(\ref{eq:X.4}), which are non-negative, we can then see that $q=0$ can only be satisfied by having $\sigma^{2}_{v} = 0$ and $r=0$. Similarly, we can see that $r=0$ can only be satisfied by having $\sigma^{2}_{h} = 0$ and $q=0$. So $l = p = q = r = 0$ is possible only if $\mu_{v} = \mu_{h} = \sigma^{2}_{v} = \sigma^{2}_{h} = 0$, which for this Ising RBM corresponds to zero external field.\newline

$\bullet$ For zero global magnetizations and non-zero expected replica-replica overlaps we need $l = p = 0$ and $q > 0, r > 0$. Clearly the conditions $l = p = 0$ require $\mu_{v} = \mu_{h} = 0$ as before. The conditions $q > 0, r > 0$ are automatically met as the integrands are always positive and non-zero once $q > 0$ and $r >0$. So $l = p = 0$ and $q > 0, r > 0$ requires $\mu_{v} = \mu_{h} = 0$, which for this Ising RBM corresponds only to zero \emph{expected} external field.\newline

Whilst we have identified conditions on $\mu_{v}, \mu_{h}, \sigma^{2}_{v}, \sigma^{2}_{h}$ for zero global magnetizations and zero/non-zero expected values for the replica-replica overlaps, we have not determined any phase boundaries and transitions - nor shall we - as the phase behaviour of other variants of the Ising RBM has already been extensively studied by Barra et al. \cite{Barra2017, Barra2017b}.\newline

We can also repeat a similar calculation to that in \ref{sec:AppendixA3} to derive the expected value of the effective field acting on the spin variable at a hidden node. Doing so we find,
\begin{equation}
\mathbb{E}_{J,v,h}\left ( \sum_{i=1}^{N}J_{ai}\langle n_{i}\rangle \right ) \;=\;\tilde{\sigma}^{2}_{J} p \left ( 1 - q\right )\;\;.\nonumber
\end{equation}

\section*{References}
\bibliography{Hoyle_LatticeGasRBM_arxiv_bibliography}
\bibliographystyle{unsrt}

\end{document}